\pdfoutput=1
\documentclass[preprint,superscriptaddress,nofootinbib]{revtex4}
\usepackage{amssymb}
\usepackage{amsmath}
\usepackage{epsfig}
\usepackage{graphicx}
\usepackage{here}
\usepackage{color}

\def\lapproxeq{\lower .7ex\hbox{$\;\stackrel{\textstyle <}{\sim}\;$}}
\def\gapproxeq{\lower .7ex\hbox{$\;\stackrel{\textstyle >}{\sim}\;$}}

\begin{document} 

\title{\Large\bfseries $R(D^{(*)})$ in a general two Higgs doublet model}
\author{Syuhei Iguro}
\affiliation{\it Department of Physics, Nagoya University, Nagoya, 464-8602, Japan}
\author{Kazuhiro Tobe}
\affiliation{\it Department of Physics, Nagoya University, Nagoya, 464-8602, Japan}
\affiliation{\it Kobayashi-Maskawa Institute for the Origin of Particles and the Universe, Nagoya University, Nagoya, 464-8602, Japan}

\begin{abstract}
  Motivated by an anomaly in $R(D^{(*)})={\rm BR}(\bar{B}\rightarrow D^{(*)} \tau^-\bar{\nu})
  /{\rm BR}(\bar{B}\rightarrow D^{(*)} l^-\bar{\nu})$
  reported by BaBar, Belle and LHCb, we study $R(D^{(*)})$ in a general two Higgs doublet model (2HDM).
  Although it has been suggested that it is difficult for the 2HDM to explain the current world average for $R(D^{(*)})$,
  it would be important to clarify how large deviations from the standard model predictions for $R(D^{(*)})$
  are possible in the 2HDM.
  We investigate possible corrections to $R(D^{(*)})$ in the 2HDM by taking into account various
  flavor physics constraints (such as $B_c^-\rightarrow \tau^- \bar{\nu}$,
  $b\rightarrow s\gamma$, $b\rightarrow s l^+l^-$, $\Delta m_{B_{d,s}}$, $B_s\rightarrow \mu^+\mu^-$ and $\tau^+\tau^-$, and
  $B^-\rightarrow \tau^- \bar{\nu}$), and find that it would be possible (impossible)
  to accommodate the 1$\sigma$ region suggested by the Belle's result when we adopt a constraint ${\rm BR}(B_c^-\rightarrow \tau^- \bar{\nu})\le30~\%$
  (${\rm BR}(B_c^-\rightarrow \tau^- \bar{\nu})\le10~\%$).
  We also study productions and decays of heavy neutral and charged Higgs bosons at the Large Hadron Collider (LHC)
  experiment and discuss the constraints and implications at the LHC. We show that in addition to well-studied production
  modes $bg\rightarrow tH^-$ and $gg\rightarrow H/A$, exotic productions of heavy Higgs bosons such as
  $cg\rightarrow bH^+,t+H/A$ and $c\bar{b}\rightarrow H^+$
  would be significantly large, and the search for their exotic decay modes such as
  $H/A\rightarrow t\bar{c}+c\bar{t},~\mu^\pm\tau^\mp$ and $H^+\rightarrow c\bar{b}$ as well as $H/A\rightarrow \tau^+\tau^-$
  and $H^+\rightarrow \tau^+\nu$ would be important to probe the interesting parameter regions for $R(D^{(*)})$.
\end{abstract}

\maketitle



\section{Introduction}
The standard model of elementary particles has been very successful to explain phenomena in nature.
Recently, however, there are several observables in experiments, which may be suggesting the existence
of new physics. For example, measurements of $b\rightarrow c$ transition processes $\bar{B}\rightarrow D^{(*)} l^-\bar{\nu}$
at BaBar~\cite{Lees2012xj,Lees2013uzd}, Belle~\cite{Huschle2015rga,Sato2016svk,Hirose2016wfn}
and LHCb~\cite{Aaij2015yra,FPCP2017LHCb} indicate a discrepancy between the experimental and theoretical
values of
\begin{align}
  R(D^{(*)})=\frac{{\rm BR}(\bar{B}\rightarrow D^{(*)} \tau^- \bar{\nu})}
  {{\rm BR}(\bar{B}\rightarrow D^{(*)} l^-\bar{\nu})},
\end{align}
where $l=e,~{\rm or}~\mu$.
Measurements of $b\rightarrow s$ transition process $B\rightarrow K^{*}\mu^+ \mu^-$ at Belle~\cite{P5pBelle},
ATLAS~\cite{P5pATLAS}, CMS~\cite{P5pCMS},BaBar~\cite{P5pBaBar} and LHCb~\cite{P5pLHCb} also
suggest an anomaly in the angular observable and moreover the LHCb has reported 
deviations from the SM predictions in $R_K^{(*)}={\rm BR}(B\rightarrow K^{(*)} \mu^+\mu^-)/
{\rm BR}(B\rightarrow K^{(*)} e^+ e^-)$~\cite{RKLHCb}. A discrepancy between experimental and theoretical values
of the muon anomalous magnetic moment has been also well-recognized~\cite{Davier2017zfy}. These anomalies may indicate some hints of
new flavor structure beyond the standard model.

One of the simplest extensions of the standard model is a two Higgs doublet model (2HDM).
When both Higgs doublets couple to all fermions,
it is notorious that the Higgs bosons have flavor violating interactions even at tree level, and
it tends to induce the large flavor violating phenomena.
Therefore, without any experimental supports, it would be difficult to justify this kind of extension.
As we mentioned above, however, currently there are several indications which may support the extra flavor
violation beyond the standard model, and hence we seriously consider a general type of two Higgs doublet
model as a possibility to explain some of the anomalies mentioned above.
In Refs.~\cite{Omura2015nja,Omura2015xcg,Tobe2016qhz}, it has been pointed out that
$\mu-\tau$ flavor violations in the general 2HDM
may be able to explain the muon $g-2$ anomaly.\footnote{Although recently the CMS collaboration reported a stronger constraint
  on the lepton flavor violating Higgs boson decay $h\rightarrow \mu\tau$~\cite{LFVmutau}, the $\mu-\tau$ flavor violations
  in the general 2HDM can still explain the muon $g-2$ anomaly, which is consistent with the recent CMS constraint,
  as seen in Refs.~\cite{Omura2015xcg,Tobe2016qhz}.}
In this paper,  we concentrate on 
$R(D^{(*)})$ in the general 2HDM.

The current status of $R(D^{(*)})$ measurements
is summarized in Table~\ref{RD_status}. Although results are not conclusive,
the current measured average values deviate from the SM prediction
about $4\sigma$.
\begin{table}[h]
  \begin{center}
    \begin{tabular}{|c|c|c|c|}
      \hline
      Experiment & $R(D^*)$ & $R(D)$ & References\\
\hline \hline
BaBar  & $0.332\pm 0.024\pm 0.018$ & $0.440\pm 0.058 \pm 0.042$ & \cite{Lees2012xj,Lees2013uzd} \\
\hline
Belle & $0.293\pm 0.038\pm 0.015$ & $0.375\pm 0.064\pm 0.026$ &\cite{Huschle2015rga}\\
& $0.302\pm 0.030\pm 0.011$ & $-$ & \cite{Sato2016svk}\\
& $0.270\pm 0.035^{+0.028}_{-0.025}$ & $-$  &\cite{Hirose2016wfn}\\
\hline
LHCb & $0.336\pm 0.027\pm 0.030$ & $-$ & \cite{Aaij2015yra}\\
& $0.285\pm 0.019\pm 0.029$ & & \cite{FPCP2017LHCb}\\
\hline \hline
Average & $0.304\pm 0.013 \pm 0.007$ & $0.407\pm 0.039\pm 0.024$ & \cite{Amhis2016xyh}\\
\hline
    \end{tabular}
  \end{center}
  \label{RD_status}
  \caption{Current status of experimental measurements of $R(D^{(*)})$. The SM prediction for
    $R(D^*)$ is $R(D^*)=0.252\pm 0.003$~\cite{Fajfer2012vx},
    The SM predictions for
    $R(D)$ based on recent Lattice calculations are
    $R(D)=0.299\pm 0.011$~\cite{Lattice2015rga}, $R(D)=0.300\pm 0.008$~\cite{Na2015kha} and $R(D)=0.299\pm 0.003$\cite{Bigi2016mdz}.
  See also Refs.~\cite{Bigi2017jbd,Jaiswal:2017rve}.}
\end{table}

In order to study the anomaly, a model-independent effective operator
approach~\cite{Tanaka2012nw,Biancofiore:2013ki,Freytsis2015qca,Celis2016azn,Bardhan:2016uhr,Choudhury:2017qyt} is very useful to
understand what type of interaction is relevant to the anomaly. On the other hand, to identify
the new physics model and to know constraints from other processes and possible correlation
among other phenomena, it is necessary to study the
anomaly in a model-dependent way. Furthermore, the anomaly would be explained by the new particles
whose masses are of $O(1)$ TeV scale. The current Large Hadron Collider (LHC) experiment
is powerful to investigate such $O(1)$ TeV scale physics as stressed in Refs.~\cite{highpt,Altmannshofer2017poe}.
Therefore the interplay between the flavor
physics and LHC physics would be also very helpful to probe a source of the anomaly.

Several studies on $R(D^{(*)})$ have been done in the general 2HDM
(for example, see Refs.~\cite{Crivellin2012ye,Celis:2012dk,Tanaka2012nw,Crivellin:2013wna,Crivellin:2015hha,Cline:2015lqp,
Alonso2016oyd}), and
it seems difficult for the 2HDM to explain the anomaly of $R(D^{(*)})$ within $1\sigma$ of the current world
average. However, it has not been clear how large deviations from the SM predictions of $R(D^{(*)})$ are possible
when the various constraints from other processes are taken into account, and hence we would like
to clarify the predictions of $R(D^{(*)})$ in the general 2HDM. Since the current experimental results and SM predictions are 
subject to improve in future, it is important to understand the possible
predicted values of $R(D^{(*)})$ in the general 2HDM. Furthermore,
we would like to show how the LHC experiment can probe the source for the anomaly
of $R(D^{(*)})$, and thus the interplay between flavor physics and LHC physics would be
very important to test the new physics model.

This paper is organized as follows: In section II, we briefly review a general 2HDM. In section III, 
we summarize a formula for $\bar{B}\rightarrow D^{(*)}l^-\bar{\nu}~(l=e,\mu,\tau)$ in a general 2HDM. In section IV,
to study $R(D^{(*)})$ in the general 2HDM, we consider three typical scenarios and show the allowed regions
of $R(D^{(*)})$ which are consistent with various flavor physics constraints such as $B_c^-\rightarrow \tau^-\bar{\nu}$,
$B_{d,s}-\bar{B}_{d,s}$ mixing, $b\rightarrow s\gamma$, $b\rightarrow s l^+ l^-$, $B_{d,s}\rightarrow \mu^+\mu^-,~\tau^+\tau^-$
and $B^-\rightarrow \tau^-\bar{\nu}$ in each scenario. We also show the result of the type II 2HDM as a comparison.
In section V, we study the productions and decays of heavy Higgs bosons at the LHC, and discuss the constraints from the LHC results
and implications at the LHC searches. In section VI, we summarize our studies.

\section{General two Higgs doublet model}
In a two Higgs doublet model, when the Higgs potential is minimized in the SM-like vacuum,
both neutral components of Higgs doublets get vacuum expectation values (vevs) in general.
Taking a certain linear combination, we can always consider a basis (so called Higgs basis or Georgi basis~\cite{Georgi:1978ri,Donoghue:1978cj},
and see also, for example, \cite{Lavoura:1994fv,Lavoura:1994yu,Botella:1994cs,Branco:1999fs,Davidson:2005cw})
where only one of the Higgs doublets
has the vev as follows:
\begin{eqnarray}
  H_1 =\left(
  \begin{array}{c}
    G^+\\
    \frac{v+\phi_1+iG}{\sqrt{2}}
  \end{array}
  \right),~~~
  H_2=\left(
  \begin{array}{c}
    H^+\\
    \frac{\phi_2+iA}{\sqrt{2}}
  \end{array}
  \right),
\label{HiggsBasis}
\end{eqnarray}
where $G^+$ and $G$ are Nambu-Goldstone bosons, and $H^+$ and $A$ are a charged Higgs boson and a CP-odd
Higgs boson, respectively.
CP-even neutral Higgs bosons $\phi_1$ and $\phi_2$ can mix and form mass
eigenstates, $h$ and $H$ ($m_H>m_h$),
\begin{eqnarray}
  \left(
  \begin{array}{c}
    \phi_1\\
    \phi_2
  \end{array}
  \right)=\left(
  \begin{array}{cc}
    \cos\theta_{\beta \alpha} & \sin\theta_{\beta \alpha}\\
    -\sin\theta_{\beta \alpha} & \cos\theta_{\beta \alpha}
  \end{array}
  \right)\left(
  \begin{array}{c}
    H\\
    h
  \end{array}
  \right).
\end{eqnarray}
Here $\theta_{\beta \alpha}$ is the mixing angle. Note that when $\cos\theta_{\beta \alpha}\rightarrow 0$
  ($\sin\theta_{\beta\alpha}\rightarrow 1$), the interactions of $\phi_1$ approach to those of the SM Higgs boson.
In this paper, we adopt Higgs basis. In Appendix A, we summarize a relation between the Higgs basis
and the general basis.

If any discrete symmetries are not imposed, both Higgs doublets can couple to all fermions.\footnote{
Sometimes, this is called the Type III two Higgs doublet model. See, for example,
Refs.~\cite{Liu:1987ng,Cheng:1987rs,Savage:1991qh,Hou:1991un,Antaramian:1992ya,Hall:1993ca,Luke:1993cy,Atwood:1995ud,Atwood:1996vj,Aoki:2009ha} for the model
and its phenomenological studies.
However, sometimes the Type III 2HDM is referred to as a different type of 2HDM~\cite{Barger:1989fj}, and hence
we simply call it general 2HDM.}
In the mass eigenbasis for the fermions, the Yukawa interactions are expressed by
\begin{eqnarray}
  {\cal L} &=&-\bar{Q}_L^i H_1 y_d^i d_R^i -\bar{Q}_L^i H_2 \rho_d^{ij} d_R^j-\bar{Q}_L^i (V^\dagger)^{ij} \tilde{H}_1 y_u^j u_R^j
  -\bar{Q}_L^i (V^\dagger)^{ij} \tilde{H}_2 \rho_u^{jk}u_R^k \nonumber\\
  &&-\bar{L}_L^i H_1 y^i_e e_R^i -\bar{L}_L^i H_2 \rho^{ij}_e e_R^j.
\label{yukawas}
\end{eqnarray}
Here $i,j$ represent flavor indices, $Q=(V^\dagger u_L,d_L)^T$ and $L=(V_{\rm MNS} \nu_L, e_L)^T$,
where $V$ and $V_{\rm MNS}$ are Cabbibo-Kobayashi-Maskawa (CKM) and the Maki-Nakagawa-Sakata (MNS) matrices,
respectively. Here we have assumed the neutrino masses are explained by the seesaw mechanism introducing
super-heavy right-handed neutrinos, so that in the low-energy effective theory, the left-handed neutrinos have
a $3\times 3$ Majorana mass matrix, which is diagonalized by the $V_{\rm MNS}$ matrix.
Note that Yukawa couplings $y_f$ are expressed by the fermion masses $m_f$, 
$y_f=\sqrt{2}m_f/v$, on the other hand, Yukawa couplings $\rho_f^{ij}$ are unknown general $3\times3$ complex matrices and
can be sources of the Higgs-mediated flavor violation.

In mass eigenstates of Higgs bosons, the Yukawa interactions are given by
\begin{align}
  {\cal L}&=-\sum_{f=u,d,e}\sum_{\phi=h,H,A} y^f_{\phi i j}\bar{f}_{Li} \phi f_{Rj}+{\rm h.c.}
  \nonumber\\
  &\quad -\bar{\nu}_{Li} (V_{\rm MNS}^\dagger \rho_e)^{ij}  H^+ e_{Rj}
  -\bar{u}_i(V\rho_d P_R-\rho_u^\dagger V P_L)^{ij} H^+d_j+{\rm h.c.},
\end{align}
where
\begin{align}
  y^f_{hij}&=\frac{m_{f}^i}{v}s_{\beta\alpha}\delta_{ij}+\frac{\rho_{f}^{ij}}{\sqrt{2}} c_{\beta\alpha},
  \nonumber\\
  y^f_{Hij}&=\frac{m_f^i}{v} c_{\beta \alpha}\delta_{ij}-\frac{\rho_f^{ij}}{\sqrt{2}} s_{\beta\alpha},
  \nonumber \\
  y^f_{Aij}&=
  \left\{
  \begin{array}{c}
    -\frac{i\rho_f^{ij}}{\sqrt{2}}~({\rm for}~f=u),\\
    \frac{i\rho_f^{ij}}{\sqrt{2}}~({\rm for}~f=d,~e),
  \end{array}
  \right.
  \label{yukawa}
\end{align}
where $c_{\beta\alpha}\equiv \cos\theta_{\beta\alpha}$ and $s_{\beta\alpha}\equiv\sin\theta_{\beta\alpha}$.
Note that when $c_{\beta\alpha}$ is small, the Yukawa interactions of $h$ are almost equal to those of
the SM Higgs boson, however, there are small flavor-violating interactions $\rho_f^{ij}$ which are suppressed
by $c_{\beta\alpha}$. On the other hand, the Yukawa interactions of heavy Higgs bosons ($H$, $A$, and $H^+$)
mainly come from the $\rho_f$ couplings. Therefore, for the SM-like Higgs boson, the tree level flavor
violation can be suppressed by the small mixing $c_{\beta\alpha}$ and for the heavy Higgs bosons,
it would be suppressed by their heaviness and/or the smallness of the extra flavor violation.

Here we also stress that the interactions of the charged Higgs boson are simply parameterized
by $\rho_f$ Yukawa couplings in the Higgs basis. In order to analyze the effects on $R(D^{(*)})$, we adopt
the Higgs basis in our analysis because it is convenient to effectively understand how large deviation
from the SM prediction on the $R(D^{(*)})$ is possible in the 2HDM, which is the aim of this paper.
To understand the effects on $R(D^{(*)})$, we can consider some simple flavor violation in the Higgs basis.
On the other hand, if we consider the simple flavor violation
in the original basis, it may correspond to the complex flavor violation in the Higgs basis (as shown in
Appendix A), and hence
it induces the effects not only on $R(D^{(*)})$ but also on other processes, which may generate strong
constraints in the model. In that sense, we expect that our approach is conservative to see the possibly
large effects on $R(D^{(*)})$, but consistent with other constraints.
Therefore, in our analysis, using the Higgs basis, we try to clarify how large deviations are possible
within the framework of the 2HDM in general.

\subsection{Higgs mass spectrum}
A scalar potential in the general 2HDM is given by
\begin{align}
  V&=M_{11}^2 H_1^\dagger H_1+M_{22}^2 H_2^\dagger H_2-\left(M_{12}^2H_1^\dagger H_2+{\rm h.c.}
  \right)\nonumber \\
&+\frac{\lambda_1}{2}(H_1^\dagger H_1)^2+\frac{\lambda_2}{2}(H_2^\dagger H_2)^2+\lambda_3(H_1^\dagger H_1)(H_2^\dagger H_2)
+\lambda_4 (H_1^\dagger H_2)(H_2^\dagger H_1)\nonumber \\
&+
\frac{\lambda_5}{2}(H_1^\dagger H_2)^2+\left\{
\lambda_6 (H_1^\dagger H_1)+\lambda_7 (H_2^\dagger H_2)\right\} (H_1^\dagger H_2)+{\rm h.c.}.
\end{align}
In the basis shown in Eq. (\ref{HiggsBasis}), Higgs boson masses are related as follows:
\begin{align}
m_{H^+}^2&=M_{22}^2+\frac{v^2}{2}\lambda_3, \nonumber \\
m_A^2-m_{H^+}^2&=-\frac{v^2}{2}(\lambda_5-\lambda_4),\nonumber \\
(m_H^2-m_h^2)^2&=\left\{m_A^2+(\lambda_5-\lambda_1)v^2\right\}^2+4\lambda_6^2v^4,\nonumber \\
\sin 2\theta_{\beta\alpha}&=-\frac{2\lambda_6 v^2}{m_H^2-m_h^2}.
\end{align}
Especially, when $c_{\beta \alpha}$ is close to zero (or $\lambda_6 \sim 0$), we approximately
get the following expressions for the Higgs boson masses:
\begin{align}
  m_h^2& \simeq \lambda_1 v^2,\nonumber \\
  m_H^2& \simeq m_A^2+\lambda_5 v^2,\nonumber \\
  m_{H^+}^2 &= m_A^2-\frac{\lambda_4-\lambda_5}{2} v^2,\nonumber \\
  m_A^2 & = M_{22}^2+\frac{\lambda_3+\lambda_4-\lambda_5}{2} v^2.
  \label{Higgs_spectrum2}
\end{align}
Note that fixing the couplings $\lambda_i$, the heavy Higgs boson masses
are expressed by the CP-odd Higgs boson mass $m_A$.
We also note that a dangerous contribution to Peskin-Takeuchi's T-parameter are suppressed by the
degeneracy between $m_A$ and $m_{H^+}$ or $m_A$ and $m_H$ as well as the small Higgs mixing parameter $c_{\beta \alpha}$.

\section{$\bar{B}\rightarrow D^{(*)} l^- \bar{\nu}~(l=e,~\mu,~\tau)$ in a general 2HDM}
\subsection{$\bar{B}\rightarrow D  l^-\bar{\nu}$}
In the 2HDM, the charged Higgs boson generates a new contribution to $B\rightarrow Dl^-\bar{\nu}$.
The relevant hadronic matrix elements are parameterized by the form factors $f_0(q^2)$ and $f_+(q^2)$
as follows~\cite{Caprini1997mu,FormfactorsforRDRDs}:
\begin{align}
  \langle D(p_D)| \bar{c}\gamma^\mu b | \bar{B}(p_B)\rangle & =
  \left(p_B^\mu+p_D^\mu -\frac{m_B^2-m_D^2}{q^2} q^\mu \right) f_+(q^2)+\frac{m_B^2-m_D^2}{q^2}q^\mu f_0(q^2),\\
  \langle D(p_D)| \bar{c} b |\bar{B}(p_B)\rangle & =
  \frac{m_B^2-m_D^2}{m_b-m_c} f_0(q^2).
\end{align}
where $p_B$ and $p_D$ ($m_B$ and $m_D$) are momenta (masses) of $B$ and $D$ mesons, respectively, and
$q$ is a momentum transfer $q=p_B-p_D$ ($m_l^2\le q^2 \le (m_B-m_D)^2$), and $m_b$ and $m_c$ are $b$ and $c$ quark masses,
respectively. In Appendix B, we summarize an information on form factors 
and in Appendix D, we list numerical values of various parameters we use in our numerical analysis.

Then the differential decay rate in the 2HDM is given by
\begin{align}
  &\frac{d\Gamma(\bar{B}\rightarrow D l^- \bar{\nu})}{dq^2}=\frac{G_{\rm F}^2 |V_{cb}|^2}{192\pi^3 m_B^3}
  \left(1-\frac{m_l^2}{q^2}\right)^2\sqrt{\lambda_D} \nonumber \\
&\hspace{0.5cm}\times \left[\lambda_D \left(1+\frac{m_l^2}{2q^2}\right)^2 f_+(q^2) 
     +\frac{3m_l^2}{2q^2}(m_B^2-m_D^2)^2 f_0^2(q^2) \left(1+\delta^{Dl}_{H^+}(q^2)\right)\right],
\end{align}
where $\lambda_D=(m_B^2-m_D^2-q^2)^2-4m_D^2 q^2$.
The charged Higgs boson induces the corrections which are proportional to the form factor
$f_0^2(q^2)$, and the contributions $\delta^{Dl}_{H^+}$ are expressed by
\begin{align}
\delta^{Dl}_{H^+}(q^2) &= -\frac{q^2}{m_l(m_b-m_c)}
      \frac{{\rm Re}\left[\rho_e^{ll}V_{cb}(\rho_d^\dagger V^\dagger-V^\dagger \rho_u)_{bc}\right]}
           {\sqrt{2}G_{\rm F}m_{H^+}^2|V_{cb}|^2} \nonumber \\
           &\hspace{0.5cm} +\frac{q^4}{m_l^2(m_b-m_c)^2}
           \frac{(\rho_e^\dagger \rho_e)^{ll}\left|(\rho_d^\dagger V^\dagger -V^\dagger \rho_u)_{bc}\right|^2}
                {8G_{\rm F}^2 m_{H^+}^4|V_{cb}|^2},
\label{delta_D}
\end{align}
where $m_{H^+}$ is a charged Higgs boson mass, and $\rho_f$ $(f=e,~u,~d)$ are Yukawa couplings introduced in Eq.~(\ref{yukawas}).
The first term in Eq.~(\ref{delta_D}) is an interference between the SM W-boson and charged Higgs boson contributions suppressed
by $m_{H^+}^2$, and the second term comes from the charged Higgs boson contribution suppressed by $m_{H^+}^4$.

Notice that the charged Higgs boson contribution to $\Gamma(\bar{B}\rightarrow D l^- \bar{\nu})$ is proportional to
$\rho_e^{ll}$ or $(\rho_e^\dagger \rho_e)^{ll}$.
We also note that if $(\rho_d^\dagger V^\dagger-V^\dagger \rho_u)_{bc}$ is small, the charged Higgs boson contribution is suppressed
because the scalar coupling $\bar{b}c H^+$ is small. On the other hand, if $(\rho_d^\dagger V^\dagger-V^\dagger \rho_u)_{bc}$
is sizable, the charged Higgs contribution can be significant especially in the large $q^2$ region.

\subsection{$\bar{B}\rightarrow D^* l^- \bar{\nu}$}
The relevant hadronic matrix elements for this process are parameterized by the form factors $V(q^2)$ and
$A_i(q^2)~(i=1-3)$~\cite{Caprini1997mu,FormfactorsforRDRDs}:
\begin{align}
\langle D^*(p_{D^*},\epsilon)|\bar{c}\gamma_\mu b | \bar{B}(p_B)\rangle &=-i \epsilon_{\mu\nu\rho\sigma} \epsilon^{\nu *}
p_B^\rho p_{D^*}^\sigma \frac{2V(q^2)}{m_B+m_{D^*}} \nonumber\\
\langle D^*(p_{D^*},\epsilon)|\bar{c} \gamma_\mu \gamma_5 b | \bar{B}(p_B)\rangle
& = \epsilon^*_{\mu} (m_B+m_{D^*}) A_1(q^2)\nonumber \\
-(p_B+p_{D^*})_\mu (\epsilon^* \cdot q)&\frac{A_2(q^2)}{m_B+m_{D^*}} 
-q_\mu (\epsilon^*\cdot q)\frac{2m_{D^*}}{q^2}\left\{A_3(q^2)-A_0(q^2)\right\},\\
  \langle D^*(p_{D^*},\epsilon)|\bar{c}\gamma_5 b |\bar{B}(p_B)\rangle
  &=-\frac{1}{m_b+m_c} q_\mu \langle D^*(p_{D^*},\epsilon)|\bar{c}\gamma^\mu \gamma_5 b|
  \bar{B}(p_B)\rangle
\end{align}
where
\begin{align}
A_3(q^2)=\frac{m_B+m_{D^*}}{2m_{D^*}} A_1(q^2)-\frac{m_B-m_{D^*}}{2m_{D^*}} A_2(q^2).
\end{align}
Here $p_{D^*}$ and $m_{D^*}$ are momentum and mass of $D^*$, respectively, and $q$ is a momentum transfer
$q=p_B-p_{D^*}$. In Appendix B, we summarize an information on the form factors.

The differential decay rate in the 2HDM is given by
\begin{align}
  &\frac{d\Gamma(\bar{B}\rightarrow D^* l^- \bar{\nu})}{dq^2}=\frac{G_{\rm F}^2 |V_{cb}|^2}{192\pi^3 m_B^3}\left(1-\frac{m_l^2}{q^2}\right)
  \sqrt{\lambda_{D^*}}\nonumber \\
  &\hspace{0.5cm}\times \left[ \left( 1+\frac{m_l^2}{2q^2}\right)
    \left\{
    q^2(m_B+m_{D^*})^2 \left(2+\frac{(q^2-m_B^2+m_{D^*})^2}{4m_{D^*}^2 q^2}\right)
    A_1^2(q^2)\right.\right.\nonumber \\
    &\hspace{0.5cm}+\frac{\lambda_{D^*}^2}{4m_{D^*}^2(m_B+m_{D^*})^2}A_2^2(q^2)
    +\frac{\lambda_{D^*}(q^2-m_B^2+m_{D^*}^2)}{2m_{D^*}^2}A_1(q^2)A_2(q^2)\nonumber \\
    &\hspace{2cm}\left.\left.+\frac{2q^2 \lambda_{D^*}}{(m_B+m_{D^*})^2} V^2(q^2)\right\}
    +\frac{3}{2}\frac{m_l^2\lambda_{D^*}}{q^2} A_0^2(q^2) \left(1+\delta^{D^*l}_{H^+}(q^2)\right)
    \right],
\end{align}
where $\lambda_{D^*}=(m_B^2-m_{D^*}^2-q^2)^2-4m_{D^*}^2q^2$. Note that compared to the process
$\bar{B}\rightarrow D l^- \bar{\nu}$, various form factors contribute to $\bar{B}\rightarrow D^* l^- \bar{\nu}$.
However the charged Higgs boson only generates the corrections
proportional to $A_0^2(q^2)$, and the corrections $\delta^{D^*l}_{H^+}(q^2)$
are given by
\begin{align}
  \delta^{D^*l}_{H^+}(q^2)&=-\frac{q^2}{m_l(m_b+m_c)}\frac{{\rm Re}\left[
        \rho^{ll}_e V_{cb} (\rho_d^\dagger V^\dagger +V^\dagger \rho_u)_{bc}\right]}
        {\sqrt{2} G_{\rm F} m_{H^+}^2 |V_{cb}|^2} \nonumber \\
        &\hspace{0.5cm}+\frac{q^4}{m_l^2(m_b+m_c)^2}\frac{(\rho_e^\dagger \rho_e)^{ll}
          \left|(\rho_d^\dagger V^\dagger +V^\dagger \rho_u)_{bc}\right|^2}
        {8G_{\rm F}^2 m_{H^+}^4 |V_{cb}|^2}.
\label{delta_Dst}
\end{align}
Notice that similar to $\bar{B}\rightarrow Dl^- \bar{\nu}$ process, the charged Higgs boson contributions
to the decay rate is proportional to $\rho_e^{ll}$ or $(\rho_e^\dagger \rho_e)^{ll}$.
We also note that contrary to $\bar{B}\rightarrow Dl^- \bar{\nu}$ process, the charged Higgs contributions
are proportional to $(\rho_d^\dagger V^\dagger +V^\dagger \rho_u)_{bc}$ since the pseudo-scalar coupling
$\bar{c}\gamma_5 b H^+$ contributes to $\bar{B}\rightarrow D^* l^- \bar{\nu}$ process.

\section{$R(D^{(*)})$ in the general 2HDM}
$R(D^{(*)})$ are defined to measure a lepton flavor universality between $\tau$ and $l$ ($l=e,~\mu$) modes
in $\bar{B}\rightarrow D^{(*)} l^-\bar{\nu}$:
\begin{align}
  R(D^{(*)})=\frac{{\rm BR}(\bar{B}\rightarrow D^{(*)}\tau^- \bar{\nu})}
  {{\rm BR}(\bar{B}\rightarrow D^{(*)}l^- \bar{\nu})},
\end{align}
where $l=e$ or $\mu$.
So far, an apparent flavor non-universality between $\mu$ and $e$ modes has not been reported~\cite{Abdesselam2017kjf}. Therefore, we expect
that the deviation from the SM prediction of $R(D^{(*)})$ mainly comes from
$\bar{B}\rightarrow D^{(*)}\tau^- \bar{\nu}$ mode in the 2HDM.

As we discussed in the previous section, the charged Higgs boson contributions to $\bar{B}\rightarrow D^{(*)} l^-\bar{\nu}$
are proportional to $\rho_e^{ll}$ or $(\rho_e^\dagger \rho_e)^{ll}$ $(l=e,\mu,\tau)$. 
Therefore, $\rho_e^{\tau\tau}$, $\rho_e^{e\tau}$ and $\rho_e^{\mu\tau}$ induce the corrections to
$\bar{B}\rightarrow D^{(*)}\tau^- \bar{\nu}$, on the other hand, $\rho_e^{i\mu}$ and $\rho_e^{i e}$ $(i=e,\mu,\tau)$
break the lepton flavor universality between $\mu$ and $e$ modes, and hence here we assume
$\rho_e^{i\mu}$ and $\rho_e^{i e}$ are negligibly small.


The charged Higgs boson contributions in $\bar{B}\rightarrow D^{(*)} \tau^- \bar{\nu}$ are proportional
to $(\rho_d^\dagger V^\dagger \mp V^\dagger \rho_u)_{bc}$:
\begin{align}
  (\rho_d^\dagger V^\dagger \mp V^\dagger \rho_u)_{bc}&=\sum_i( \rho_d^{ib*} V_{ci}^* \mp V^*_{ib} \rho_u^{ic}),\\
  &=\rho_d^{sb*} V_{cs}^*+\rho_d^{db*} V^*_{cd}+\rho_d^{bb*}V^*_{cb}\mp
  (V^*_{tb} \rho_u^{tc}+ V^*_{cb}\rho_u^{cc}+V^*_{ub} \rho_u^{uc}).
\end{align}
Because of the hierarchical structure of the CKM matrix as
\begin{align}
|V_{cs}|\sim |V_{tb}|\sim 1 > |V_{cd}|\sim 0.2\gg |V_{cb}|\sim 0.04 \gg |V_{ub}|\sim 0.004,
\end{align}
effects of Yukawa couplings $\rho_d^{bb}$, $\rho_u^{cc}$ and $\rho_u^{uc}$ are small even if these
are of the order of one. Therefore, we consider the effects of $\rho_u^{tc}$, $\rho_d^{sb}$ and $\rho_d^{db}$.

In order to see how large deviations of $R(D^{(*)})$ from the SM prediction are possible in the 2HDM,
we consider the following three typical scenarios:
\begin{enumerate}
  \renewcommand{\labelenumi}{\arabic{enumi}).}
\item non-zero $\rho_u^{tc}$
  
\item non-zero $\rho_u^{tc}$ and $\rho_d^{sb}$ (and $\rho_d^{db}$)

\item non-zero $\rho_u^{tc}$ and $\rho_e^{\mu\tau}$ (and $\rho_e^{e\tau}$)

\end{enumerate}
in addition to the non-zero $\rho_e^{\tau\tau}$. We will also show the predictions in the Type II
2HDM as a comparison.

%
%
%
%
%

Before we study scenarios listed above, first we discuss the constraints from measurements of Higgs boson couplings,
a top decay $t\rightarrow hc$ and lepton flavor violating Higgs boson decays
on the relevant Yukawa couplings in these scenarios.
\subsection{Constraints from measurements of Higgs boson couplings and a top decay $t\rightarrow hc$}
\subsubsection{Constraints from measurements of Higgs boson couplings}
In the general 2HDM, even if the extra Yukawa couplings $\rho_f$ are negligible, the SM-like Higgs boson couplings with the SM particles
are modified and multiplied by the Higgs mixing parameter $s_{\beta \alpha}$.
Therefore, the measurements of the Higgs boson couplings constrain the Higgs mixing parameter.
The current limit~\cite{BeluscaMaito2016dqe} is\footnote{When the extra Yukawa couplings
  $\rho$ are all negligible, the SM-like Higgs boson couplings
  with the SM particles in our 2HDM are
  the same as those in the Type I 2HDM with a large $\tan \beta$. See, for example, Ref.~\cite{BeluscaMaito2016dqe}.}
\begin{align}
  \left| c_{\beta\alpha}\right|\le 0.4.
\end{align}

If $\rho_e^{\tau\tau}$ Yukawa coupling is not negligible, it contributes to the Higgs boson decay $h\rightarrow \tau^+\tau^-$.
The measurement of the signal strength $\mu_\tau$ constrains the $\rho_e^{\tau\tau} c_{\beta\alpha}$:
\begin{align}
  \mu_\tau\equiv \frac{{\rm BR}(h\rightarrow \tau^+\tau^-)}{{\rm BR}(h\rightarrow \tau^+\tau^-)_{\rm SM}}
  \simeq \left|s_{\beta\alpha}+\frac{\rho_e^{\tau\tau} c_{\beta\alpha} v}{\sqrt{2} m_\tau}\right|^2.
\end{align}
The current measurement of $\mu_\tau$ is given by~\cite{signalstrengthtautau}
\begin{align}
  \mu_\tau=1.11^{+0.24}_{-0.22}.
\end{align}
The result of the constraint on $\rho_e^{\tau\tau} c_{\beta\alpha}$ at the 95\% confidence level (C.L.) is given by
\begin{align}
  -2\times 10^{-3}\le \rho_e^{\tau\tau}c_{\beta\alpha}\le 3\times 10^{-3}.
\end{align}

\subsubsection{Constraint from the top decay $t\rightarrow h c$}  
Non-zero $\rho_u^{tc}$ as well as $\rho_u^{ct}$ generate an exotic top quark decay $t\rightarrow h c$.
The decay branching ratio is obtained by
\begin{align}
  {\rm BR}(t\rightarrow h c)&=\frac{c_{\beta\alpha}^2 (|\rho_u^{tc}|^2+|\rho_u^{ct}|^2)}{64\pi}\frac{m_t}{\Gamma_t}
  \left(1-\frac{m_h^2}{m_t^2}\right)^2, \\
  &\simeq 3\times 10^{-3}~\left(\frac{\rho_u^{tc} c_{\beta\alpha}}{0.15}\right)^2.
  \label{tDecay}
\end{align}
Here we have adopted $\Gamma_t=1.41~{\rm GeV}$ for the total decay rate of the top quark and we assumed that
$\rho_u^{ct}$ is negligible in the numerical estimate of Eq.~(\ref{tDecay}). The current experimental limit is set as
\begin{align}
  {\rm BR}(t\rightarrow h c)\le 4\times 10^{-3},
\end{align}
at the 95\% C.L.~\cite{Khachatryan2016atv}.

\subsubsection{Constraints from lepton flavor violating Higgs boson decays}
A search for the lepton flavor violating Higgs boson decays $h\rightarrow e\tau,~\mu\tau$ constrains
on the lepton flavor violating Yukawa couplings $\rho_e^{e\tau}$ and $\rho_e^{\mu\tau}$.
The latest data from the CMS collaboration at the LHC
with an integrated luminosity of 35.9 fb$^{-1}$ at $\sqrt{s}=13$ TeV is given by
\begin{align}
  {\rm BR}(h\rightarrow \mu\tau)&\le0.25~\%,\\
  {\rm BR}(h\rightarrow e \tau)&\le0.61~\%,
\end{align}
at 95\% C.L.~\cite{LFVmutau}. The branching ratio ${\rm BR}(h\rightarrow \mu\tau)$ in the general
2HDM is given by
\begin{align}
  {\rm BR}(h\rightarrow \mu\tau)&=\frac{c_{\beta\alpha}^2\left(|\rho_e^{\mu\tau}|^2+|\rho^{\tau\mu}_e|^2\right)m_h}
  {16\pi \Gamma_h},\\
  &=0.24\% \left(\frac{\rho_e^{\mu\tau}c_{\beta\alpha}}{2\times 10^{-3}}\right)^2,
\end{align}
where $\Gamma_h$ is the total decay rate of the Higgs boson and we have used $\Gamma_h=4.1$ MeV,
and we have neglected $\rho_e^{\tau\mu}$ for this numerical estimate. Similarly we get
\begin{align}
  {\rm BR}(h\rightarrow e\tau)&=0.62\% \left(\frac{\rho_e^{e\tau}c_{\beta\alpha}}{3.2\times 10^{-3}}\right)^2.
\end{align}

Note that if the Higgs mixing parameter $c_{\beta\alpha}$ is small $(|c_{\beta\alpha}|\le10^{-2}-10^{-3})$, these $\rho_f$
Yukawa couplings relevant to $R(D^{(*)})$ can be still of $O(0.1)-O(1)$.

\subsection{Scenario (1): non-zero $\rho_u^{tc}$ and $\rho_e^{\tau\tau}$}
This scenario has been suggested by Refs.~\cite{Crivellin2012ye,Tanaka2012nw} as an interesting solution to
explain the $R(D^{(*)})$ anomalies in the effective theory language. In this scenario,
the charged Higgs boson contributions $\delta_{H^+}^{D^{(*)}\tau}$ in Eqs.~(\ref{delta_D},~\ref{delta_Dst}) are expressed
by
\begin{align}
  1+\delta_{H^+}^{D^{(*)}\tau}(q^2)=\left|
  1\pm \frac{q^2\rho_e^{\tau\tau}\rho_u^{tc}V_{cb}V_{tb}^*}{2\sqrt{2}G_{\rm F} m_{H^+}^2  m_\tau(m_b\mp m_c)|V_{cb}|^2}
  \right|^2,
  \label{delta_scenairo1}
\end{align}
where double sign corresponds to a case for $D$ and $D^*$, respectively.
For the parameterization of the Cabbibo-Kobayashi-Maskawa (CKM) matrix, we use the Particle Data Group (PDG)
convention. We note that in the PDG convention, $V_{cb}$ and $V_{tb}$ are real and positive.
When $\rho_e^{\tau\tau}$ is real and negative, one can increase $R(D^*)$ by increasing $\rho_u^{tc}$ positively,
but $R(D)$ is decreased. However, as one increases $\rho_u^{tc}$ further, $R(D)$ also starts to increase so that
the current world average values of $R(D^{(*)})$ can be explained. As noticed in Ref.~\cite{Crivellin2012ye},
this scenario requires relatively large Yukawa coupling $\rho_u^{tc}$ in order to explain $R(D^{(*)})$.
However, the constraints from the other processes in this scenario would be important.
Therefore, we will clarify the allowed parameter space consistent with the experimental results.

\subsubsection{Constraint from $B_c^-\rightarrow \tau^-\bar{\nu}$}
In general, the effective operators for
$\bar{B}\rightarrow D^* \tau^- \bar{\nu}$ which are generated by the charged Higgs boson also
contribute to $B_c^-\rightarrow \tau^- \bar{\nu}$ process. Therefore, large extra contributions to
$\bar{B}\rightarrow D^* \tau^- \bar{\nu}$ process would be  strongly constrained
by this process~\cite{Alonso2016oyd,Akeroyd2017mhr}.\footnote{
See also Ref.\cite{Li:2016vvp}.}
The branching ratio for $B_c^-\rightarrow \tau^-\bar{\nu}$ is obtained by
\begin{align}
  {\rm BR}(B_c^-\rightarrow \tau^-\bar{\nu}) &=\tau_{B_c}\frac{m_{B_c}m_\tau^2 f_{B_c}^2 G_{\rm F}^2 |V_{cb}|^2}{8\pi}
  \left(1-\frac{m_\tau^2}{m_{B_c^2}}\right)^2\left[1+\delta_{H^+}^{D^*\tau}(m_{B_c}^2)\right],
  \label{BR_Bc}
\end{align}
where $\tau_{B_c}$ is the lifetime of $B_c^-$ meson and the charged Higgs boson contribution $\delta_{H^+}^{D^*\tau}$
is given in Eq.~(\ref{delta_Dst}). For non-zero $\rho_e^{\tau\tau}$ and $\rho_u^{tc}$, the numerical estimate shows
\begin{align}
  {\rm BR}(B_c^-\rightarrow \tau^- \bar{\nu}) &=\tau_{B_c}\frac{m_{B_c}m_\tau^2 f_{B_c}^2 G_{\rm F}^2 |V_{cb}|^2}{8\pi}
  \left(1-\frac{m_\tau^2}{m_{B_c^2}}\right)^2 \nonumber \\
&\hspace{2cm}\times  \left|1-\frac{m_{B_c}^2 \rho_e^{\tau\tau} \rho_u^{tc}V_{tb}^*}
  {2\sqrt{2}G_{\rm F}m_{H^+}^2 m_\tau (m_b+m_c)V_{cb}^*}\right|^2,\\
  &=(2\%)\left|
  1+3\left(\frac{500~{\rm GeV}}{m_{H^+}}\right)^2\left(\frac{0.04}{V_{cb}/V_{tb}}\right)^*
  \left(\frac{\rho_e^{\tau\tau}}{-0.5}\right)\left(\frac{\rho_u^{tc}}{0.5}\right)\right|^2.
\end{align}
As studied in Ref.~\cite{Alonso2016oyd}, the life time of the $B_c^-$ meson provides a constraint ${\rm BR}(B_c^-\rightarrow \tau^- \bar{\nu})\le30\%$.
Furthermore, the recent study suggested that the LEP data taken at the Z peak requires
${\rm BR}(B_c^-\rightarrow \tau^- \bar{\nu})\le10\%$~\cite{Akeroyd2017mhr}.
We simply adopt both constraints as references.\footnote{
  The constraint obtained in Ref.~\cite{Akeroyd2017mhr} relies on the theoretical value of
  ${\rm BR}(B_c^-\rightarrow J/\psi l\bar{\nu})$, which may be subject to debate. Therefore, we use
  both constraints ${\rm BR}(B_c^-\rightarrow \tau^- \bar{\nu})\le30\%$ and $10\%$ as references.}
In the Scenario (1), the constraint ${\rm BR}(B_c^-\rightarrow \tau^- \bar{\nu})\le30\%~(10\%)$ corresponds to,
for example, $-0.8\le\rho_u^{tc}\le0.5~(-0.5\le\rho_u^{tc}\le0.2)$ for $m_{H^+}=500$ GeV and $\rho_e^{\tau\tau}=-0.5$ if $\rho_u^{tc}$ is real.
In order to explain the current world average for $R(D^*)$ [$R(D^*)= 0.304$], large Yukawa coupling
$\rho_u^{tc}$ is required ($\rho_u^{tc}\sim +1$ for $m_{H^+}=500$ GeV and $\rho_e^{\tau\tau}=-0.5$)~\cite{Crivellin2012ye}.
Therefore, the constraint severely restricts the possibility to have a large deviation from the SM prediction
for $R(D^*)$.

In order to study the effects of the complex phases in the Yukawa couplings, we parameterize the Yukawa coupling
$\rho_e^{\tau\tau}$ as
\begin{align}
\rho_e^{\tau\tau}=\rho_{\tau\tau}e^{i\delta_{\tau\tau}},
\end{align}
where $\rho_{\tau\tau}$ is a real parameter and the phase $\delta_{\tau\tau}$ is assumed to be
$0\le \delta_{\tau\tau} \le \frac{\pi}{2}$. The phase of Yukawa coupling $\rho_u^{tc}$ is effectively absorbed
into the phase of $\delta_{\tau\tau}$ in $R(D^{(*)})$, and therefore we assume that $\rho_u^{tc}$ is a real 
(positive or negative) parameter.\footnote{As we show in Appendix C, the phase of $\rho_u^{tc}$ affects
  an imaginary part of $B_{d,s}-\bar{B}_{d,s}$ mixing. Therefore, the measurements of the imaginary part
  of $B_{d,s}-\bar{B}_{d,s}$ mixing have a potential to distinguish
the origin of the phase.}
The leading effect of $\delta_{\tau\tau}$ comes from the interference between
the SM and charged Higgs contributions, and it is proportional to ``$\rho_u^{tc}\rho_{\tau\tau}\cos\delta_{\tau\tau}$'' in
$R(D^{(*)})$. Therefore, the range of $\delta_{\tau\tau}$ ($0\le \delta_{\tau\tau}\le \frac{\pi}{2}$) is sufficient to see
the possible predictions for the $R(D^{(*)})$ when $\rho_u^{tc}\rho_{\tau\tau}$ is allowed to be positive or negative.
\begin{figure}[ht]
  \begin{center}
    \includegraphics[width=0.7\textwidth]{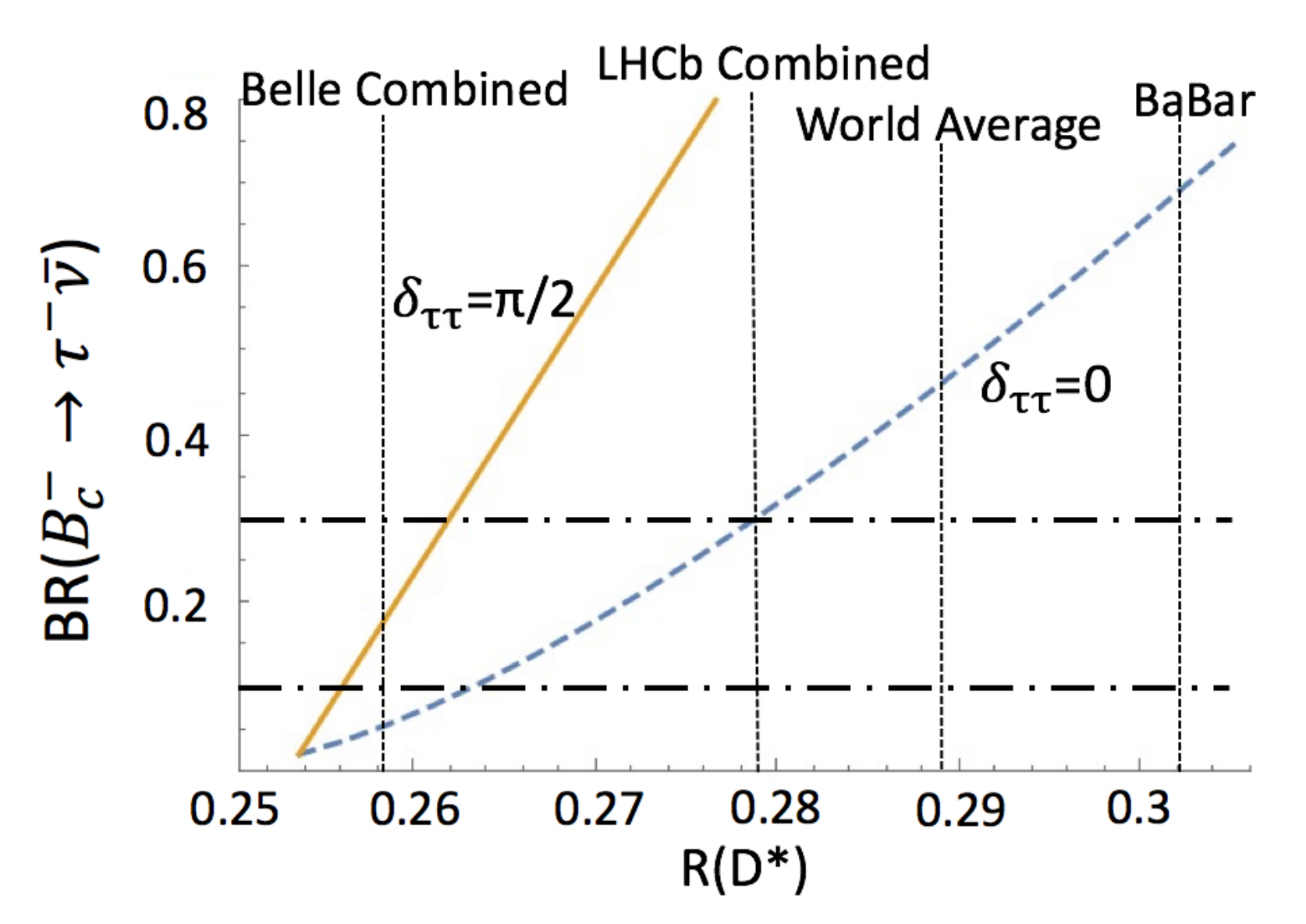}
    \caption{The correlations between $R(D^*)$ and ${\rm BR}(B_c^-\rightarrow \tau^-\bar{\nu})$ are shown
      in cases with $\delta_{\tau\tau}=0$ (dashed blue line) and $\frac{\pi}{2}$ (solid orange line). Here $\delta_{\tau\tau}$ is a phase of $\rho_e^{\tau\tau}$ which
      is parameterized by $\rho_e^{\tau\tau}=\rho_{\tau\tau}e^{i\delta_{\tau\tau}}$. The suggested constraints
      ${\rm BR}(B_c^-\rightarrow \tau^-\bar{\nu})\le30\%$ and $10\%$ are shown by horizontal dashed-dotted lines.
      The lower limits at 1$\sigma$
      for BaBar, Belle, LHCb and world average are also shown. }
    \label{Bclife}
  \end{center}
\end{figure}

In Figure~\ref{Bclife}, the correlations between $R(D^*)$ and ${\rm BR}(B_c^-\rightarrow \tau^-\bar{\nu})$ are shown
for $\delta_{\tau\tau}=0$ (dashed blue line) and $\frac{\pi}{2}$ (solid orange line). 
When the phase $\delta_{\tau\tau}$ is zero, the predicted value of $R(D^*)$ can not be larger
than $0.279~(0.263)$ because of constraints ${\rm BR}(B_c^-\rightarrow \tau^-\bar{\nu})\le30\%~(10\%)$~\cite{Alonso2016oyd,Akeroyd2017mhr}.
We found that when the phase $\delta_{\tau\tau}$ is non-zero, the constraint
on $B_c^-\rightarrow \tau^-\bar{\nu}$ becomes much stronger. For example, for $\delta_{\tau\tau}=\frac{\pi}{2}$, the upper limit
on $R(D^*)$ is about 0.262~(0.256) for ${\rm BR}(B_c^-\rightarrow \tau^-\bar{\nu})\le30\%~(10\%)$.
Note that the correlations do not depend on the detail of values of
model parameters other than the phase parameter $\delta_{\tau\tau}$ in the current scenario.
\begin{figure}[ht]
  \begin{center}
    \includegraphics[width=0.8\textwidth]{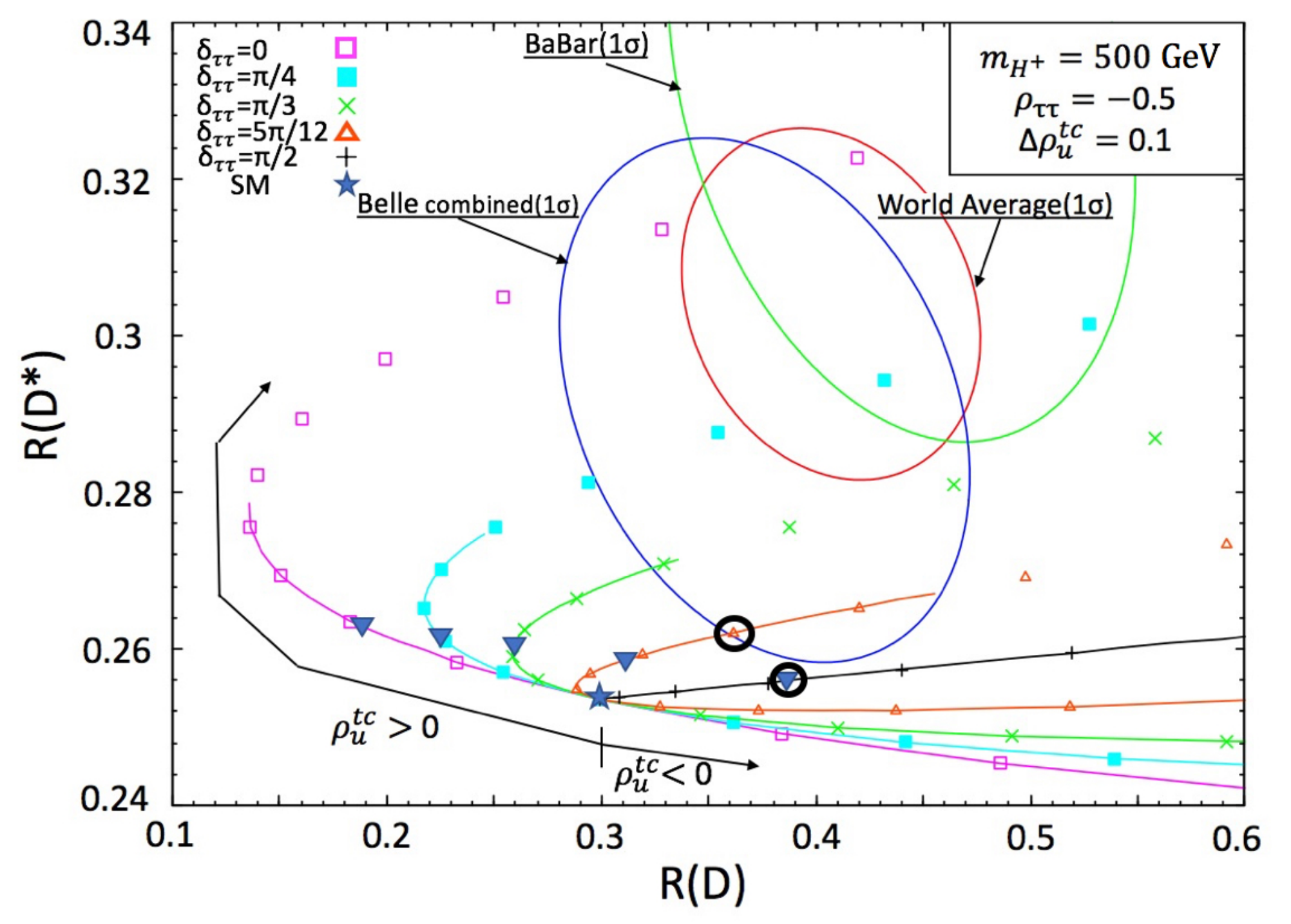}
    \caption{The predicted values for $R(D)$ and $R(D^*)$ are shown in the scenario (1). The current
      experimental limits $(1\sigma)$ for BaBar, Belle, and world average are also shown. Here we have set
      $m_{H^+}=500$ GeV and $\rho_{\tau\tau}=-0.5$ as a reference set of parameters. The Yukawa coupling $\rho_e^{\tau\tau}$
      is parameterized by $\rho_e^{\tau\tau}=\rho_{\tau\tau} e^{i\delta_{\tau\tau}}$. The value of $\rho_u^{tc}$ is
      varied from the SM point $(\rho_u^{tc}=0)$ (blue star-mark) by $0.1$ step positively ($\rho_u^{tc}>0$)
      or negatively ($\rho_u^{tc}<0$) with different phase parameters $\delta_{\tau\tau}=0$ (open pink  quadrangle),
      $\frac{\pi}{4}$ (filled cyan quadrangle), $\frac{\pi}{3}$ (green $\times$), $\frac{5\pi}{12}$ (orange triangle)
      and $\frac{\pi}{2}$ (black $+$).
      The solid line in each value of $\delta_{\tau\tau}$ shows the regions
      which are consistent with the constraint ${\rm BR}(B_c^-\rightarrow \tau\bar{\nu})\le30\%$.
      If we impose the constraint ${\rm BR}(B_c^-\rightarrow \tau^-\bar{\nu})\le10\%$, the regions above filled blue triangle
      in each value of $\delta_{\tau\tau}$ are excluded.
      The points indicated by a black circle are adopted as reference points for the LHC physics study discussed later. See the text in details.
    }
    \label{scenario1}
  \end{center}
\end{figure}

In Figure~\ref{scenario1}, the predicted values for $R(D)$ and $R(D^*)$ in the Scenario (1) are shown.
The current experimental $1\sigma$ limits for BaBar, Belle, and world average are also shown.
Here we have set $m_{H^+}=500$ GeV and $\rho_{\tau\tau}=-0.5$ as a reference set of parameters.
The value of $\rho_u^{tc}$ is varied from the SM point $(\rho_u^{tc}=0)$ (blue star-mark) by $0.1$ step positively ($\rho_u^{tc}>0$)
or negatively ($\rho_u^{tc}<0$) with different phase values $\delta_{\tau\tau}=0$ (open pink quadrangle),
$\frac{\pi}{4}$ (filled cyan quadrangle),
$\frac{\pi}{3}$ (green $\times$), $\frac{5\pi}{12}$ (orange triangle) and $\frac{\pi}{2}$ (black $+$) as shown in Figure~\ref{scenario1}. 
The solid line in each phase value $\delta_{\tau\tau}$
shows the regions which are consistent with the constraint ${\rm BR}(B_c^-\rightarrow \tau^-\bar{\nu})\le30\%$.
If we impose the constraint ${\rm BR}(B_c^-\rightarrow \tau^-\bar{\nu})\le10\%$, the regions above filled blue triangle
in each value of $\delta_{\tau\tau}$ are excluded.

Because of the constraints on ${\rm BR}(B_c^-\rightarrow \tau^-\bar{\nu})$, the predicted $R(D^*)$ are strongly restricted.
When $\rho_u^{tc}$ is positive and the phase parameter $\delta_{\tau\tau}$ is non-zero, the imaginary value of $\rho_e^{\tau\tau}\rho_u^{tc}$
can increase $R(D)$.
As one increases $\delta_{\tau\tau}$ and $\rho_u^{tc}$, $R(D)$ can get larger before the value of $R(D^*)$ reaches the constraint
from $B_c^-\rightarrow\tau^-\bar{\nu}$ as one can see in Figure~\ref{scenario1}.
The predicted $R(D)$ and $R(D^*)$ can not explain the current world average values. They can (can not) be
within the $1\sigma$ region of the combined Belle result if the constraint ${\rm BR}(B_c^-\rightarrow \tau^-\bar{\nu})\le30\%$
($10\%$) is imposed.
Here we have fixed a charged Higgs mass to be 500 GeV ($m_{H^+}=500$ GeV) and Yukawa coupling $\rho_{\tau\tau}$ to be $-0.5$
($\rho_e^{\tau\tau}=-0.5 e^{i\delta_{\tau\tau}}$).   The new physics
contributions to both $R(D^{(*)})$ and ${\rm BR}(B_c^-\rightarrow \tau^- \bar{\nu})$ scale like
\begin{align}
\left(\frac{\rho_e^{\tau\tau}}{m_{H^+}}\right)\left(\frac{\rho_u^{tc}}{m_{H^+}}\right).
\end{align}
Therefore, even when we change $m_{H^+}$ and $\rho_{\tau\tau}$, the predicted values of $R(D^{(*)})$
(and ${\rm BR}(B_c^-\rightarrow \tau^- \bar{\nu})$) are read from Figure~\ref{scenario1}.

For the LHC physics study discussed later, we choose two reference points shown by a black circle in Figure~\ref{scenario1}.
As a reference point 1, we select a point with $\rho_u^{tc}\rho_{\tau\tau}=-0.2$, $\delta_{\tau\tau}=\frac{5\pi}{12}$ and $m_{H^+}=500$ GeV, which
satisfies the constraint ${\rm BR}(B_c^-\rightarrow \tau^-\bar{\nu})\le30\%$ but not
${\rm BR}(B_c^-\rightarrow \tau^-\bar{\nu})\le10\%$. The reference point 1 predicts $(R(D),~R(D^*))=(0.361,0.262)$.
As a reference point 2, we choose a point with $\rho_u^{tc}\rho_{\tau\tau}=-0.156$, $\delta_{\tau\tau}=\frac{\pi}{2}$ and $m_{H^+}=500$ GeV, which
satisfies the constraint ${\rm BR}(B_c^-\rightarrow \tau^-\bar{\nu})\le10\%$. This point predicts $(R(D),~R(D^*))=(0.385,0.256)$.

\subsubsection{$q^2$ distribution in $d\Gamma(\bar{B}\rightarrow D\tau^-\bar{\nu})/dq^2$}
As stressed in Refs.~\cite{Tanaka2012nw,Celis2016azn}, the charged Higgs boson contribution significantly affects
the $q^2$ distribution in $d\Gamma(\bar{B}\rightarrow D\tau^-\bar{\nu})/dq^2$.
\begin{figure}[ht]
  \begin{center}
    \includegraphics[width=0.9\textwidth]{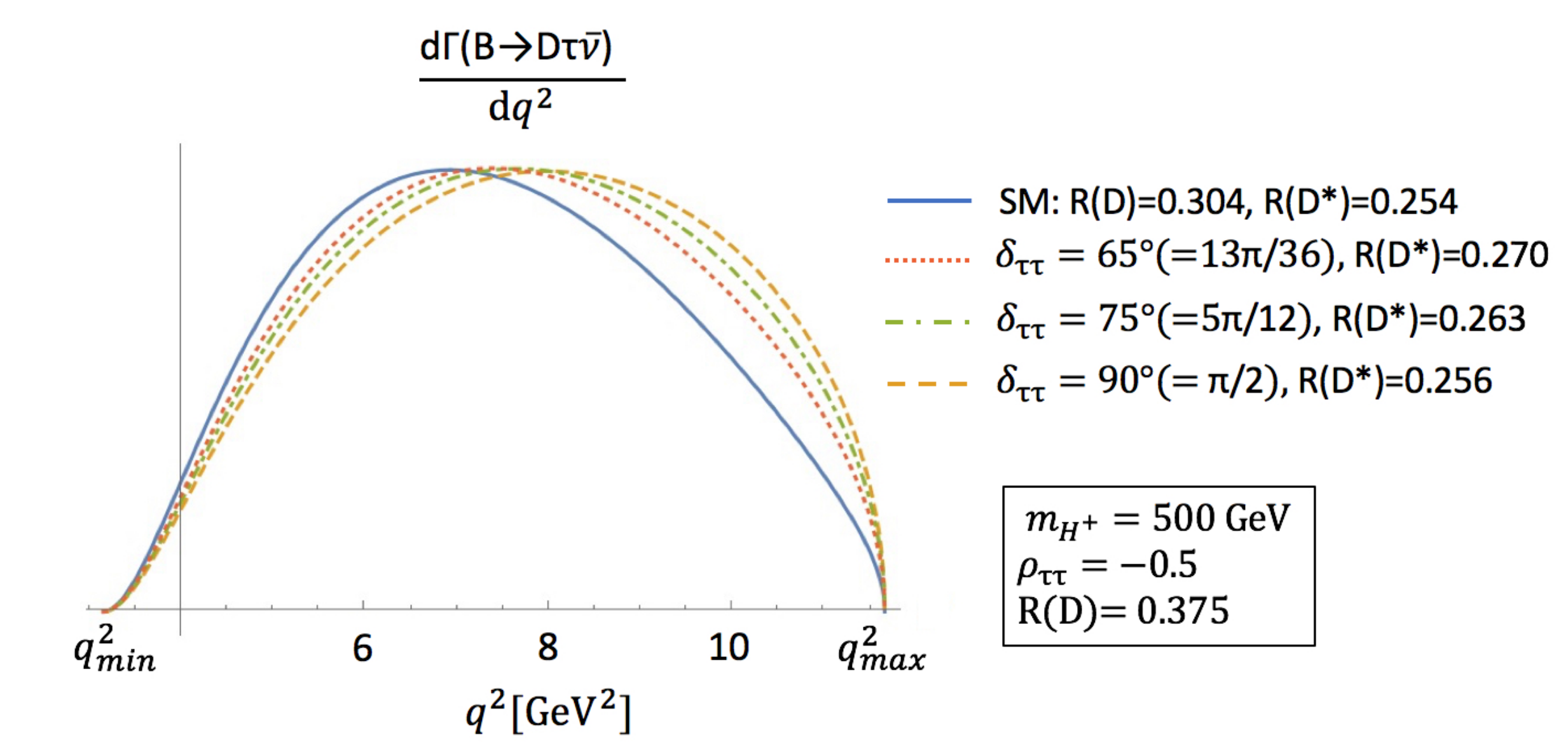}
    \caption{The $q^2$ distributions in $d\Gamma(\bar{B}\rightarrow D\tau^-\bar{\nu})/dq^2$
      are shown in cases with $(\delta_{\tau\tau},~R(D^*))=(65^\circ,~0.270)$,
      $(75^\circ,~0.263)$ and $(90^\circ,~0.256)$.
      Here we have set $m_{H^+}=500$ GeV and $\rho_{\tau\tau}=-0.5$. $R(D)$ is fixed to be $R(D)=0.375$
      by adjusting the value of $\rho_u^{tc}$ in each case. 
      To compare the shapes of the $q^2$ distributions, the heights of all plots are normalized to be the same.
      Here we also show the $q^2$ distribution in the SM case $(R(D)=0.304,~R(D^*)=0.254)$ as a comparison. 
      }
    \label{q2_dist}
  \end{center}
\end{figure}
In Figure~\ref{q2_dist}, the $q^2$ distributions in $d\Gamma(\bar{B}\rightarrow D\tau^-\bar{\nu})/dq^2$
are shown
in cases with $(\delta_{\tau\tau},~R(D^*))=(65^\circ,~0.270)$, $(75^\circ,~0.263)$ and $(90^\circ,~0.256)$.
Here we take $m_{H^+}=500$ GeV and $\rho_{\tau\tau}=-0.5$. The predicted value of $R(D)$ is fixed to be $R(D)=0.375$
by adjusting the value of $\rho_u^{tc}$ in each case. 
To compare the shapes of the $q^2$ distributions, the heights of all plots are normalized to be the same.
Here we also show the $q^2$ distribution in the SM case $(R(D)=0.304,~R(D^*)=0.254)$ as a comparison.
As one can see from Figure~\ref{q2_dist}, the difference from the SM distribution increases in large $q^2$
regions as the phase of $\rho_e^{\tau\tau}$ ($\delta_{\tau\tau}$) becomes larger.
The BaBar~\cite{Lees2013uzd} and Belle~\cite{Huschle2015rga} have provided results of the $q^2$ distribution. However, the general
2HDM can not explain the BaBar's results of $R(D^{(*)})$, and the Belle's result for the $q^2$ distribution
still has large uncertainty. Furthermore, their analyses for the $q^2$ distribution measurement rely
on the theoretical models. Therefore, here we do not explicitly impose a constraint from
the $q^2$ distributions in $d\Gamma(\bar{B}\rightarrow D\tau^-\bar{\nu})/dq^2$. However we would like to
stress that the precise measurement of the $q^2$ distributions in $d\Gamma(\bar{B}\rightarrow D\tau^-\bar{\nu})/dq^2$
would have a significant impact on this scenario.

\subsubsection{Effects on $b\rightarrow s(d)$ transition: $B_{d,s}-\bar{B}_{d,s}$ mixing, $b\rightarrow s\gamma$ and $b\rightarrow s l^+l^-$}
In the existence of $\rho_u^{tc}$, a charged Higgs boson induces new contributions to $b\rightarrow s(d)$ transition processes such as
$B_{d,s}-\bar{B}_{d,s}$ mixing, $b\rightarrow s\gamma$ and $b\rightarrow s l^+l^-$ at one-loop level.

For $B_{d,s}-\bar{B}_{d,s}$ mixing,
the detail expressions of the 2HDM contributions are shown in Appendix C.
The charged Higgs boson correction to $B_{d,s}$ meson mass difference is approximately expressed by
\begin{align}
  \Delta m_{B_{d_i}}^{\rm 2HDM}  &=  \Delta m_{B_{d_i}}^{\rm SM}+\delta(\Delta m_{B_{d_i}})^{\rm 2HDM},\nonumber \\
  \delta(\Delta m_{B_{d_i}})^{\rm 2HDM}
  &=-\frac{m_{B_{d_i}}F_{B_{d_i}}^2 B_{B_{d_i}}}{192 \pi^2 m_{H^+}^2} {\rm Re}\left[(V_{tb} V_{t d_i}^*)^2\right]
  \left|\rho_u^{tc}\right|^4 G_1(x_c,x_c),
\end{align}
where a function $G_1$ is defined in Eq.~(\ref{G1_func}) in Appendix C and $G_1(x_c,x_c)\simeq 1$ for
$x_c=m_c^2/m_{H^+}^2\ll 1$, and
$m_{B_{d_i}}$, $F_{B_{d_i}}$ and $B_{B_{d_I}}$ are a mass, a decay constant and a bag parameter
of $B_{d_i}$ meson, respectively and their values for our numerical analysis are listed in
Appendix D. The numerical estimates of the charged Higgs boson contribution are obtained by
\begin{align}
  \delta (\Delta m_{B_d})^{\rm 2HDM}
  &=-0.05\times \left(\frac{500~{\rm GeV}}{m_{H^+}}\right)^2\left(\frac{|\rho_u^{tc}|}{1.0}\right)^4~[{\rm ps}^{-1}],
 \label{Bd}  \\  
 \delta(\Delta m_{B_s})^{\rm 2HDM}
  &=-2\times \left(\frac{500~{\rm GeV}}{m_{H^+}}\right)^2\left(\frac{|\rho_u^{tc}|}{1.0}\right)^4~[{\rm ps}^{-1}].
  \label{Bs}
\end{align}
As one sees, the charged Higgs boson contributions with non-zero $\rho_u^{tc}$ prefer the negative values.
The measured values $\Delta m_{B_{d,s}}^{\rm exp}$~\cite{Amhis2016xyh} are
\begin{align}
\Delta m_{B_d}^{\rm exp}=0.5064\pm 0.0019~[{\rm ps}^{-1}],~~~  \Delta m_{B_s}^{\rm exp} =17.757\pm 0.021~[{\rm ps}^{-1}],
\end{align}
and the SM predictions $\Delta m_{B_{d,s}}^{\rm SM}$ (95 \% C.L. region)~\cite{Bobeth:2016llm} are given by
\begin{align}
0.45~[{\rm ps}^{-1}] \le \Delta m_{B_d}^{\rm SM} \le 0.78~[{\rm ps}^{-1}],~~~  16.2~[{\rm ps}^{-1}]\le \Delta m_{B_s}^{\rm SM}\le21.9~[{\rm ps}^{-1}].
\end{align}
Since the uncertainty of the measured values is very small in contrast to the SM predictions, we take into
account the uncertainty of the SM predictions, and we apply the 95\% C.L. allowed region to the
2HDM contribution as follows:
\begin{align}
  &-0.27~[{\rm ps}^{-1}]\le\delta(\Delta m_{B_d})^{\rm 2HDM}\le0.06~[{\rm ps}^{-1}],\nonumber \\
  &-4.1~[{\rm ps}^{-1}]\le\delta(\Delta m_{B_s})^{\rm 2HDM}\le1.6~[{\rm ps}^{-1}].
  \label{BBbar_constraints}
\end{align}
Compared with Eqs.~(\ref{Bd},~\ref{Bs}), interesting parameter regions
($\rho_u^{tc}\sim 0.2-0.5$ for $m_{H^+}=500$ GeV) in Scenario (1) are consistent with the current limit on
$\Delta m_{B_{d,s}}$. 

The effective operators relevant to $b\rightarrow s \gamma$ and $b\rightarrow s l^+l^-$ are given by
\begin{align}  
{\cal L_{\rm eff}}= &\frac{4G_{\rm F}}{\sqrt{2}}V_{tb} V^*_{ts}\left[\frac{e}{16\pi^2}
    C_7 m_b (\bar{s}\sigma_{\mu\nu}P_R b) F^{\mu\nu}
    +\frac{g_3}{16\pi^2} C_8 m_b(\bar{s}\sigma_{\mu\nu} T^a P_R b) G^{a,\mu\nu} \right.\nonumber \\
    &+ \left.\frac{e^2}{16\pi^2}C_{9(l)} (\bar{s}\gamma^\mu P_L b) (\bar{l}\gamma_\mu l)
    +\frac{e^2}{16\pi^2}C_{10(l)} (\bar{s}\gamma^\mu P_L b) (\bar{l}\gamma_\mu\gamma_5 l)
    \right],
\end{align}
where $l$ is a charged lepton $l=e,~\mu$ or $\tau$. We summarize the detail expressions induced by the charged Higgs boson
via $\rho_u$ Yukawa couplings for $C_{7,8,9,10}$ in Appendix C.
The charged Higgs boson contributions with non-zero $\rho_u^{tc}$ are expressed by
\begin{align}
  C_7^{\rm 2HDM} &=\frac{1}{4\sqrt{2}G_{\rm F} m_{H^+}^2}|\rho_u^{tc}|^2 G_\sigma(x_c),\\
  C_8^{\rm 2HDM} &=\frac{1}{4\sqrt{2}G_{\rm F} m_{H^+}^2}|\rho_u^{tc}|^2 G_{\sigma 2}(x_c),\\
  C_{9(l)}^{\rm 2HDM} &=-\frac{1}{2\sqrt{2} G_{\rm F}m_{H^+}^2}|\rho_u^{tc}|^2 G_\gamma(x_c)
  +\frac{-1+4 s_W^2}{8\pi\alpha}|\rho_u^{tc}|^2 G_Z(x_c),\\
  C_{10(l)}^{\rm 2HDM} &= \frac{1}{8\pi\alpha}|\rho_u^{tc}|^2 G_Z(x_c),
\end{align}
where $x_c=m_c^2/m_{H^+}^2 \ll 1$ for $m_{H^+}=500$ GeV and various functions are defined in Appendix C and the
approximate expressions are obtained as
\begin{align}
  G_\sigma(x_c)&=Q_u G_{\sigma 1}(x_c)+Q_{H^+} G_{\sigma 2}(x_c)\simeq -\frac{7}{36},\\
  G_{\sigma 2} (x_c) & \simeq -\frac{1}{6},\\
  G_\gamma(x_c)&=Q_u G_{\gamma 1}(x_c)+Q_{H^+} G_{\gamma 2}(x_c)\simeq -\frac{19+12\log x_c}{54},\\
  G_Z(x_c) &\simeq 0,
\end{align}
for $x_c\ll 1$. Here $Q_u$ and $Q_{H^+}$ are charges of up-type quarks ($+2/3$) and a charged Higgs boson
($+1$).
We note that a term proportional to $G_\gamma(x_c)$ in $C_{9(l)}^{\rm 2HDM}$ originates from
$\gamma$ penguin contribution, on the other hand, terms proportional to $G_Z(x_c)$
in $C_{9,10(l)}^{\rm 2HDM}$ come from $Z$ penguin contribution. Therefore, $C_{9,10(l)}^{\rm 2HDM}$
are universal to all lepton flavor $l$. We also note that in $G_\gamma(x)$ there is a log-enhancement
when $x$ is small.

For $b\rightarrow s\gamma$,
Wilson coefficients $C_{7,8}^{\rm 2HDM}$ are estimated as
\begin{align}
  C_7^{\rm 2HDM} &=-0.012 \left(\frac{500~{\rm GeV}}{m_{H^+}}\right)^2\left(
  \frac{|\rho_u^{tc}|}{1.0}\right)^2,\\
  C_8^{\rm 2HDM} &=-0.010 \left(\frac{500~{\rm GeV}}{m_{H^+}}\right)^2\left(
  \frac{|\rho_u^{tc}|}{1.0}\right)^2.
\end{align}
The low-energy Wilson coefficient $C_{7}(\mu_b)$ is
evaluated by using the renormalization group equation and the charged Higgs boson contribution
$C_7^{\rm 2HDM}(\mu_b)$ is expressed by
\begin{align}
   C_7^{\rm 2HDM}(\mu_b)&=\eta^{\frac{16}{23}}  C_7^{\rm 2HDM}+\frac{8}{3}\left(\eta^{\frac{14}{23}}
   -\eta^{\frac{16}{23}}\right) C_8^{\rm 2HDM},\nonumber \\
   &=-0.009 \left(\frac{500~{\rm GeV}}{m_{H^+}}\right)^2\left(
  \frac{|\rho_u^{tc}|}{1.0}\right)^2,
\end{align}
where $\eta=\alpha_s(\mu)/\alpha_s(\mu_b)$ and we have taken $\mu=m_W$, $\mu_b=5$ GeV and $\alpha_s(m_Z)=0.118$ in our numerical
estimate.
As discussed in Ref.~\cite{C7C7p}, for example, the allowed regions of the new physics contributions
$ C_7^{\rm NP}(\mu_b)$ from a global fit analysis of $b\rightarrow s$ transition observables are constrained as
\begin{align}
  -0.04&\le C_7^{\rm NP}(\mu_b) \le 0.0
\end{align}
at $1\sigma$ level. Therefore, the interesting regions in Scenario (1) ($\rho_u^{tc}\sim 0.2-0.5$ for $m_{H^+}=500$ GeV)
are consistent with this constraint.

For $b\rightarrow s l^+l^-$ process, the contributions to $C_{9,10}$ play an important role.
The numerical estimates for the charged Higgs boson contributions with non-zero $\rho_u^{tc}$ are
given by
\begin{align}
  C_{9(l)}^{\rm 2HDM} &=-0.27 \left(\frac{500~{\rm GeV}}{m_{H^+}}\right)^2\left(\frac{|\rho_u^{tc}|}{1.0}\right)^2,
  \label{C9}\\
  C_{10(l)}^{\rm 2HDM} &\simeq 0,
  \label{C10}
\end{align}
where the $Z$ penguin contribution is suppressed, and hence only $C_{9(l)}^{\rm 2HDM}$ receives the non-negligible
correction from the $\gamma$ penguin contribution, which has a ``$\log x_c$'' enhancement.

As we discussed in the introduction, the current experimental situation on $b\rightarrow s l^+l^-$ processes is involved.
In the angular observable of $B\rightarrow K^* \mu^+\mu^-$, the discrepancies between the SM prediction and the measured value
have been reported. (See Refs.~\cite{P5pBelle,P5pATLAS,P5pCMS,P5pBaBar,P5pLHCb}.) The LHCb results
have also indicated deviations from the SM predictions in lepton flavor universality measurements
$R_K^{(*)}={\rm BR}(B\rightarrow K^{(*)} \mu^+\mu^-)/{\rm BR}(B\rightarrow K^{(*)} e^+ e^-)$.
In Ref.~\cite{Altmannshofer2017fio}, for example,
a global fit analysis of angular observables in $B\rightarrow K^* \mu^+\mu^-$ as well as branching ratios of
$B\rightarrow K^{(*)}\mu^+\mu^-$ and $B_s\rightarrow \phi\mu^+\mu^-$ suggests that the best fit values of
new physics contributions to $C_{9,10}$ are preferred to be away from zero,
$(C_{9(\mu)}^{\rm NP},C_{10(\mu)}^{\rm NP})=(-1.15,+0.26)$. 
As shown in Eqs.~(\ref{C9},~\ref{C10}),
the interesting regions in Scenario (1) ($\rho_u^{tc}\sim 0.2-0.5$ for $m_{H^+}=500$ GeV) can not explain these discrepancies and
the fit may be only slightly improved, compared to the SM predictions.\footnote{The Scenario (1) can not explain $R_K^{(*)}$ anomalies
  because the charged Higgs boson effects with non-zero $\rho_u^{tc}$ are universal to all charged leptons $l$
  in $b\rightarrow s l^+l^-$.} Since the anomalies in $b\rightarrow s l^+l^-$ are still subject to discuss, we do not impose
the constraints from this process in our analysis.

In addition to these $b\rightarrow s(d)$ transition processes, in the presence of $\rho_u^{tc}$ and $\rho_e^{\tau\tau}$,
$B_s\rightarrow \tau^+\tau^-$ process is induced. However, we have checked that the constraint from this process
is still very weak, and hence we only list the expression of $B_s\rightarrow \tau^+\tau^-$ process in Appendix C.

\subsection{Scenario (2): non-zero $\rho_u^{tc}$, $\rho_e^{\tau\tau}$ and $\rho_d^{sb}$ ($\rho_d^{db}$)}
In addition to the non-zero $\rho_e^{\tau\tau}$ and $\rho_u^{tc}$, we introduce other non-zero Yukawa couplings
$\rho_d^{sb}$ (and/or $\rho_d^{db}$) in this scenario.
This scenario has been suggested by, for example, 
Refs.~\cite{Lees2013uzd,Freytsis2015qca,Celis2016azn} in the effective theory description.

To study the typical parameter space which is consistent with the other experimental data, we parameterize the Yukawa couplings
as follows:
\begin{align}
  r_s&=\frac{\rho_d^{sb}}{\rho_u^{tc*}},~~
  r_d=\frac{\rho_d^{db}}{\rho_u^{tc*}},~~a=r_s\frac{V_{cs}}{V_{tb}}+r_d\frac{V_{cd}}{V_{tb}}.
\end{align}
Note that
\begin{align}
(\rho_d^\dagger V^\dagger\pm V^\dagger \rho_u)_{bc}=V_{tb}^*\rho_u^{tc}(a^* \pm 1).
\end{align}
Therefore for non-zero $r_s$~(and/or $r_d$), it is possible to realize
Re$[\rho_e^{\tau\tau}(\rho_d^\dagger V^\dagger\pm V^\dagger \rho_u)_{bc}]<0$, so that
one may increase both $R(D)$ and $R(D^*)$
without having too large Yukawa couplings.
\begin{figure}[ht]
  \begin{center}
   \includegraphics[width=0.9\textwidth]{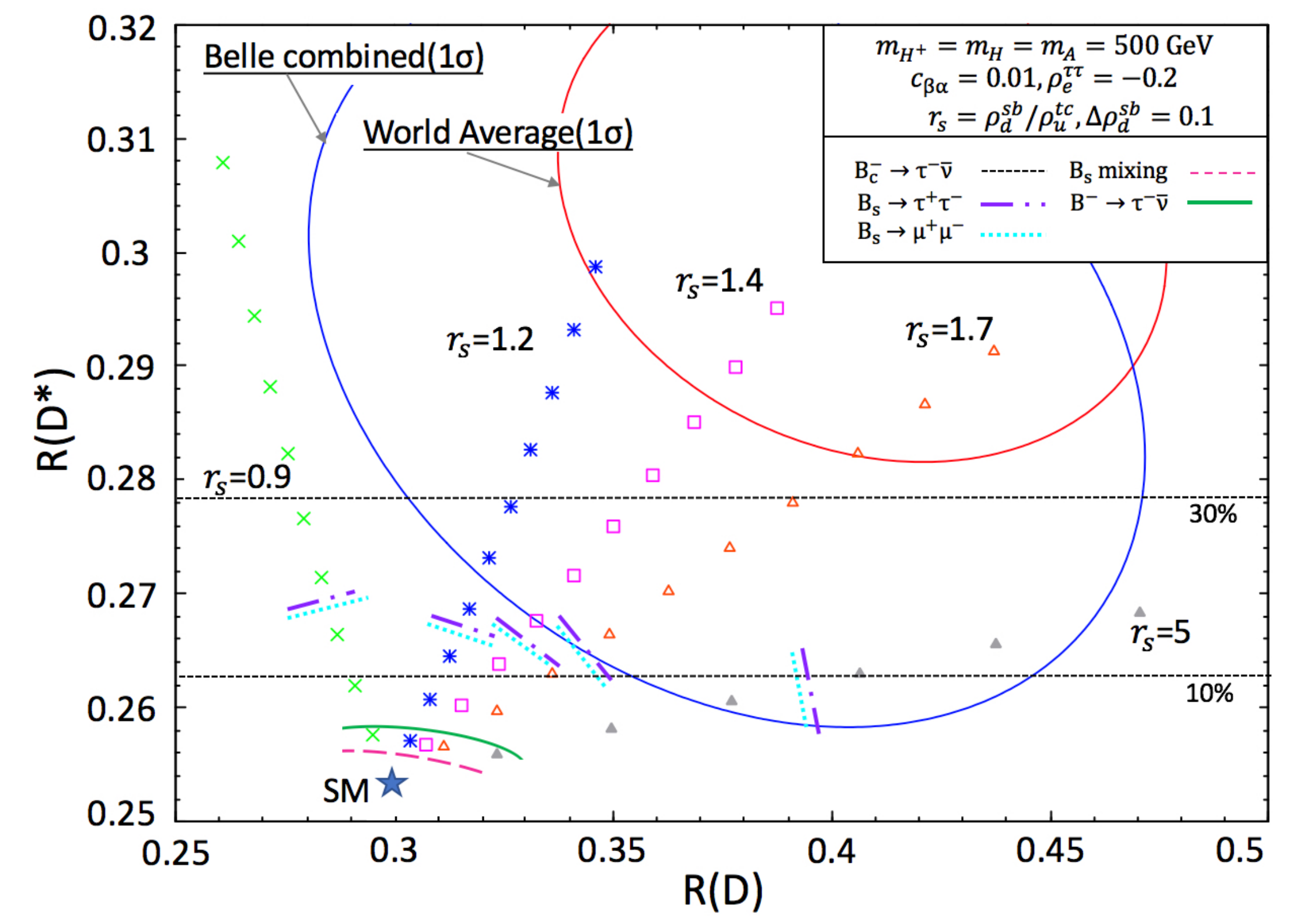}
    \caption{The predicted values for $R(D)$ and $R(D^*)$ are shown in the Scenario (2). The current
      experimental limits $(1\sigma)$ for Belle and world average are also shown. Here we have assumed that
      the relevant Yukawa couplings are all real, and we have
      set $m_{H^+}=m_{H/A}=500$ GeV, $\rho_e^{\tau\tau}=-0.2$ and $c_{\beta\alpha}=0.01$.
      The value of $\rho_d^{sb}$ is varied from the SM point $(\rho_d^{sb}=0)$ (blue star-mark) by
      $0.1$ step positively with different values of $r_s(=\rho_d^{sb}/\rho_u^{tc})$, $r_s=0.9$ (green $\times$),
      1.2 (blue $*$), 1.4 (open pink quadrangle),
      1.7 (open orange triangle) and 5 (filled grey triangle). We also show the constraints from $B_c^-\rightarrow \tau^- \bar{\nu}$,
      $B_s-\bar{B}_s$ mixing, $B_s\rightarrow \tau^+\tau^-$ and $B^-\rightarrow \tau^- \bar{\nu}$. See the text
      in details.
    }
    \label{scenario2}
  \end{center}
\end{figure}

In Figure~\ref{scenario2}, the predicted values for $R(D)$ and $R(D^*)$ are shown in the Scenario (2). The current
experimental limits $(1\sigma)$ for Belle and world average are also shown. Here we have assumed that the
Yukawa coupling $\rho_e^{\tau\tau}$, $\rho_u^{tc}$ and $\rho_d^{sb}$ are all real, and we have
set $m_{H^+}=m_{H/A}=500$ GeV, $\rho_e^{\tau\tau}=-0.2$, and $c_{\beta\alpha}=0.01$.
The value of $\rho_d^{sb}$ is varied from the SM point $(\rho_d^{sb}=0)$ (blue star-mark) by
$0.1$ step positively with different values of $r_s(=\rho_d^{sb}/\rho_u^{tc})$, $r_s=0.9$ (green $\times$), 1.2 (blue $*$), 1.4 (open pink quadrangle),
1.7 (open orange triangle) and 5 (filled grey triangle).
As one can see from Figure~\ref{scenario2}, for $r_s\sim 1-2$ and $\rho_d^{sb}\sim 1$, 
the predicted values of $R(D)$ and $R(D^*)$ can be within the $1\sigma$ of the world average.

However, as we will see below, the strong constraints come from not only $B_c^-\rightarrow \tau^- \bar{\nu}$
but also $B_{d,s}-\bar{B}_{d,s}$ mixing, $B_{d,s}\rightarrow\mu^+\mu^-$ and $\tau^+\tau^-$,
and $B^-\rightarrow \tau^- \bar{\nu}$ processes.

\subsubsection{Constraint from $B_c^-\rightarrow \tau^-\bar{\nu}$}
For non-zero $\rho_d^{sb (db)}$ as well as non-zero $\rho_e^{\tau\tau}$ and $\rho_u^{tc}$,
the branching ratio for $B_c^-\rightarrow \tau^- \bar{\nu}$ is given by the same expression shown in
Eq.~(\ref{BR_Bc}):
\begin{align}
  {\rm BR}(B_c^-\rightarrow \tau^- \bar{\nu}) &=\tau_{B_c}\frac{m_{B_c}m_\tau^2 f_{B_c}^2 G_{\rm F}^2 |V_{cb}|^2}{8\pi}
  \left(1-\frac{m_\tau^2}{m_{B_c^2}}\right)^2 \nonumber \\
&\hspace{2cm}\times  \left|1-\frac{m_{B_c}^2V_{tb}^* \rho_e^{\tau\tau} \rho_u^{tc}(1+a^*)}
       {2\sqrt{2}m_\tau (m_b+m_c)V_{cb}^* G_{\rm F}m_{H^+}^2}\right|^2.
\end{align}
From the constraint on ${\rm BR}(B_c^-\rightarrow \tau^-\bar{\nu})$, the upper limit on $\rho_u^{tc}$ is modified
by about $1/|1+a^*|$, compared with the Scenario (1).
In Figure~\ref{scenario2}, the constraint from $B_c^-\rightarrow \tau^- \bar{\nu}$ is shown
by horizontal dashed black lines ($30\%$ and $10\%$). As discussed in Scenario (1),
we should note that $R(D^*)$ can not be larger than 0.279 (0.263)
due to the constraint ${\rm BR}(B_c^-\rightarrow \tau^- \bar{\nu})\le30\%$ ($10\%$) as shown in Figure~\ref{scenario2},
since $B_c^-\rightarrow \tau^- \bar{\nu}$ and $\bar{B}\rightarrow D^* \tau^- \bar{\nu}$ are correlated.

\subsubsection{Constraint from $B_{d,s}-\bar{B}_{d,s}$ mixing}
In the presence of $\rho_d^{sb,db}$, the flavor changing neutral current appears at the tree level. It would be
strongly constrained by $B_{d,s}-\bar{B}_{d,s}$ mixing ($B_{d,s}$ mass differences, $\Delta m_{B_{d,s}}$).
For $B_s$ mass difference, the contributions induced by the neutral Higgs boson mediations at the tree level are
given by
%
\begin{align}
 \delta( \Delta m_{B_s})^{\rm 2HDM}&=-\frac{5R}{48}m_{B_s}B_{B_s}F_{B_s}^2{\rm Re}[(\rho_d^{sb})^2]
  \left(\frac{c_{\beta\alpha}^2}{m_h^2}+\frac{s_{\beta\alpha}^2}{m_H^2}-\frac{1}{m_A^2}\right),\\
  &\simeq 4~\left(\frac{\rho_d^{sb}}{0.1}\right)^2 
 \left[\left(\frac{\Delta^2/m_H^2}{10^{-3}}\right)\left(\frac{500~{\rm GeV}}{m_H}\right)^2
    -1.5\left(\frac{c_{\beta\alpha}}{0.01}\right)^2\right] ~~[{\rm ps}^{-1}].
\end{align}
The detail expression is shown in Appendix C, and numerical values of the parameters are listed in Appendix D,
and here we have assumed that $\Delta^2=m_H^2-m_A^2\ll m_H^2,~m_A^2$.

To satisfy the constraint shown in Eq.~(\ref{BBbar_constraints}), the degeneracy between $H$ and $A$ (that is, small $\Delta^2$)
has to be realized\footnote{Note that $\Delta^2\simeq \lambda_5 v^2$. Therefore, the small
  $\Delta^2$ corresponds to the small Higgs quartic coupling $\lambda_5$.}.
For example, $\Delta^2/m_H^2$ has to be much smaller than
$10^{-3}$ for $m_H\simeq 500$ GeV and $\rho_d^{sb}\sim 0.1$. Therefore, we simply assume
$\Delta^2=0$ in our studies below. To satisfy the constraint, the coupling $\rho_d^{sb} c_{\beta\alpha}$
has to be small. For example, for $c_{\beta\alpha}=0.01$, $\rho_d^{sb}\le0.08$.
In Figure~\ref{scenario2}, the constraints from $\Delta m_{B_s}$ are shown by long dashed pink line.
The regions above the long dashed pink line are excluded.

For $B_d$ mass difference, one can get the expression from one for $B_s$ by replacing
its flavor index $s$ with $d$. To satisfy the constraint on $\Delta m_{B_d}$ shown
in Eq.~(\ref{BBbar_constraints}), the coupling $\rho_d^{db} c_{\beta\alpha}$ also
has to be small. For example, for $c_{\beta\alpha}=0.01$, $\rho_d^{db}\le0.02$.
Since the constraint is very severe, $\rho_d^{db}$ can not generate the significant contributions
to $R(D^{(*)})$. Therefore, we do not include the effects of $\rho_d^{db}$ in our studies.

\subsubsection{Constraint from $B_s\rightarrow \mu^+\mu^-$ and $\tau^+\tau^-$}
The measured value~\cite{Bsmumuexp,Aaij:2017vad} of branching ratio for $B_s\rightarrow \mu^+\mu^-$ is
\begin{align}
  {\rm BR}(B_s\rightarrow \mu^+\mu^-)=2.9^{+0.7}_{-0.6}\times 10^{-9}.
\end{align}
The SM prediction~\cite{BsmumuSM} is given by
\begin{align}
{\rm BR}(B_s\rightarrow \mu^+\mu^-)_{\rm SM}=(3.25\pm 0.17)\times 10^{-9},
\end{align}
which agrees with the current measured value, and hence the new physics contributions would be
strongly constrained.

The 2HDM contributions mediated by the neutral Higgs bosons ($h,~H,~A$) at the tree level and 
the decay rate for $B_s\rightarrow \mu^+\mu^-$ are presented in Appendix C. If $\rho_d^{bs}$ and $\rho_e^{\mu\mu}$ Yukawa couplings
are negligible, non-zero Wilson coefficient is obtained by
\begin{align}
  \frac{G_{\rm F}^2 m_W^2}{\pi^2} (C_{S(\mu)}^{bs})^{\rm 2HDM}&=\frac{\rho_d^{sb*}m_\mu s_{\beta\alpha} c_{\beta\alpha}}{\sqrt{2}v}
  \left(\frac{1}{m_h^2}-\frac{1}{m_H^2}\right),
\end{align}
The branching ratio for $B_s\rightarrow \mu^+\mu^-$ in the current scenario is given by
\begin{align}
  {\rm BR}(B_s\rightarrow \mu^+\mu^-)_{\rm 2HDM} &={\rm BR}(B_s\rightarrow \mu^+\mu^-)_{\rm SM}+
  \Delta {\rm BR}(B_s\rightarrow \mu^+\mu^-),\\
         \Delta {\rm BR}(B_s\rightarrow \mu^+\mu^-) &=\frac{f_{B_s}^2m_{B_s}^5\tau_{B_s}}{32 \pi (m_b+m_s)^2}\left(1-\frac{4m_\mu^2}{m_{B_s}^2}
  \right)^{\frac{3}{2}}\left|\frac{G_{\rm F}^2 m_W^2}{\pi^2}(C_{S(\mu)}^{bs})^{\rm 2HDM}\right|^2,\nonumber \\
  &=1.1\times 10^{-9}\left(\frac{\rho_d^{sb}}{0.33}\right)^2\left(\frac{c_{\beta\alpha}}{0.01}\right)^2,
\end{align}
Here note that the small Higgs mixing parameter $c_{\beta\alpha}$ can suppress the 2HDM contribution.
In Figure~\ref{scenario2}, the constraints from $B_s\rightarrow \mu^+\mu^-$ are shown by dotted cyan lines in each
$r_s$ case. The regions above the line are excluded.

In addition to $B_s\rightarrow \mu^+\mu^-$ process, $B_s\rightarrow \tau^+\tau^-$ process could also
put a strong constraint on this scenario. In the presence of non-zero $\rho_d^{sb}$ and $\rho_e^{\tau\tau}$,
the 2HDM contributions to Wilson coefficients are given by
\begin{align}
  \frac{G_{\rm F}^2 m_W^2}{\pi^2} (C_{S(\tau)}^{bs})^{\rm 2HDM}&=\frac{\rho_d^{sb}c_{\beta\alpha}}{2m_h^2}
  \left(\rho_e^{\tau\tau}c_{\beta\alpha}+\frac{\sqrt{2}m_\tau s_{\beta\alpha}}{v}\right)
  \nonumber \\
  &\hspace{1cm}+\frac{\rho_d^{sb} s_{\beta\alpha}}{2m_H^2}\left(\rho_e^{\tau\tau} s_{\beta\alpha}-
  \frac{\sqrt{2}m_\tau c_{\beta\alpha}}{v}\right),\\
  \frac{G_{\rm F}^2 m_W^2}{\pi^2} (C_{P(\tau)}^{bs})^{\rm 2HDM}&=\frac{\rho_d^{sb}\rho_e^{\tau\tau}}{2m_A^2},  
\end{align}
where we have assumed that $\rho_d^{sb}$ and $\rho_e^{\tau\tau}$ are real.

The current experimental bound on this process is given by the LHCb collaboration~\cite{Aaij2017xqt}:
\begin{align}
{\rm BR}(B_s\rightarrow \tau^+\tau^-)\le 6.8\times 10^{-3}
\end{align}
at 95\% C.L. Since the SM prediction is $(7.73\pm 0.49)\times 10^{-7}$~\cite{Bobeth2013uxa},
we neglect it in our analysis below. The branching ratio is given by
\begin{align}
  {\rm BR}(B_s\rightarrow \tau^+\tau^-)_{\rm 2HDM}&=\frac{\tau_{B_s} f_{B_s}^2 m_{B_s}^5}{128\pi(m_b+m_s)^2}
  \sqrt{1-\frac{4m_\tau^2}{m_{B_s}^2}} (\rho_d^{sb}\rho_e^{\tau\tau})^2 \nonumber \\
  &\hspace{1cm}\times\left[
    \frac{ s_{\beta\alpha}^4}{m_H^4}(1+\Delta_{\tau})^2(1-\frac{4m_\tau^2}{m_{B_S}^2})
    +\frac{1}{m_A^4}\right],
\end{align}
where
\begin{align}
  \Delta_\tau=c_{\beta\alpha}\left\{
  \frac{c_{\beta\alpha} m_H^2}{s_{\beta\alpha} m_h^2}+\frac{\sqrt{2}m_\tau}{\rho_e^{\tau\tau}v}
  \left(\frac{m_H^2}{m_h^2}-1\right)\right\}.
\end{align}
Note that $\Delta_\tau$ is small when the Higgs mixing parameter $c_{\beta\alpha}$ is small ($c_{\beta\alpha}\sim 0.01$)
and $\rho_e^{\tau\tau}\sim O(0.1)$. The numerical estimate shows
\begin{align}
  {\rm BR}(B_s\rightarrow \tau^+\tau^-)_{\rm 2HDM}\simeq 6.6\times 10^{-3}\left(\frac{\rho_d^{sb}}{0.35}\right)^2
  \left(\frac{\rho_e^{\tau\tau}}{-0.2}\right)^2\left(\frac{500~{\rm GeV}}{m_H}\right)^4,
\end{align}
where we have assumed $m_A=m_H$.
In Figure~\ref{scenario2}, the constraints from $B_s\rightarrow \tau^+\tau^-$ are shown by dashed-dotted purple lines in each $r_s$ case.
The regions above the line are excluded.

\subsubsection{Constraint from $B^-\rightarrow \tau^- \bar{\nu}$}
The charged Higgs boson can also contribute to $B^-\rightarrow \tau^- \bar{\nu}$ process.~\footnote{
An importance of the correlation between $B\to D^{(*)}\tau\bar{\nu}$ and $B \to \tau \bar{\nu}$ in some models
has been discussed in Refs.~\cite{Itoh:2004ye,Nierste:2008qe,Nandi:2016wlp}.}
The current experimental value~\cite{Huschle2015rga,Btotaunu} is given by 
\begin{align}
{\rm BR}(B^-\rightarrow \tau^- \bar{\nu})=(1.09\pm0.24)\times10^{-4}.
\end{align}
The decay branching ratio for $B^-\rightarrow \tau^- \bar{\nu}$ in the 2HDM can be obtained
from one for $B_c^-\rightarrow \tau^- \bar{\nu}$ by replacing its flavor index $c$  with $u$.
Especially, if non-zero $\rho_d^{sb,db}$ are assumed (and $\rho_u^{iu}=0$ for $i=u,c,t$), the branching ratio in the 2HDM is given by
\begin{align}
  {\rm BR}(B^-\rightarrow \tau^-\bar{\nu})_{\rm 2HDM} &={\rm BR}(B^-\rightarrow \tau^-\bar{\nu})_{\rm SM}
  \left|
  1-\frac{m_B^2\rho_e^{\tau\tau}Y_{ub}^{*}}{2\sqrt{2}m_\tau m_bG_{\rm F}m_{H^+}^2}\right|^2,
\end{align}
where the SM ratio and $Y_{ub}$ are expressed by
\begin{align}
    {\rm BR}(B^-\rightarrow \tau^-\bar{\nu})_{\rm SM} &=
  \tau_{B}\frac{m_{B}m_\tau^2 f_{B}^2 G_{\rm F}^2 |V_{ub}|^2}{8\pi}
  \left(1-\frac{m_\tau^2}{m_B^2}\right)^2=0.948\times10^{-4},\\
  Y_{ub}&=\rho_d^{sb}\frac{V_{us}}{V_{ub}}+\rho_d^{db}\frac{V_{ud}}{V_{ub}}.
\end{align}
Here, to calculate the SM prediction, we have used the best fitted values of the CKM matrix elements ($|V_{ub}|=0.409\times10^{-2}$).
Since $|V_{us}/V_{ub}|\simeq 6\times 10$ and $|V_{ud}/V_{ub}|\simeq 3\times 10^2$, the flavor violating couplings $\rho_d^{sb,db}$
are strongly constrained. 
The numerical estimate of the correction to the branching ratio $\Delta {\rm BR}~(={\rm BR}_{\rm 2HDM}-{\rm BR}_{\rm SM})$
is given by
\begin{align}
  \frac{\Delta {\rm BR}}{{\rm BR}_{\rm SM}} &\simeq 0.37\left(\frac{\rho_e^{\tau\tau}}{-0.2}\right)
  \left(\frac{\rho_d^{sb}}{0.1}+\frac{\rho_d^{db}}{0.02}\right)
  \left(\frac{500~{\rm GeV}}{m_{H^+}}\right)^2 \nonumber \\
  &+0.32 \left(\frac{\rho_e^{\tau\tau}}{-0.2}\right)^2
  \left(\frac{\rho_d^{sb}}{0.1}+\frac{\rho_d^{db}}{0.02}\right)^2  
  \left(\frac{500~{\rm GeV}}{m_{H^+}}\right)^4.
\end{align}

For example, for $m_{H^+}$=500 GeV and $\rho_e^{\tau\tau}=-0.2$, which corresponds
to the same parameter set as in Figure~\ref{scenario2}, the constraint from $B^-\rightarrow \tau^- \bar{\nu}$
is given by
\begin{align}
\rho_d^{sb}\le 0.1,~~~~~\rho_d^{db}\le 0.02,
\end{align} 
at $95\%$ C.L.
In Figure~\ref{scenario2}, the constraint from $B^-\rightarrow \tau^-\bar{\nu}$
is shown by the solid green line. The regions above the line are excluded.

Since the experimental constraints are severe, the effects on $R(D^{(*)})$ from the Yukawa couplings
$\rho_d^{sb,db}$ in the Scenario (2) are very limited. Therefore, we will not study the Scenario (2) furthermore.


\subsection{Scenario (3): non-zero $\rho_u^{tc}$, $\rho_e^{\tau\tau}$ and $\rho_e^{\mu\tau}$ (and/or $\rho_e^{e\tau}$)}
The lepton flavor violating Yukawa couplings $\rho_e^{\mu\tau}$ and $\rho_e^{e\tau}$ can also affect
$\bar{B}\rightarrow D^{(*)} \tau^- \bar{\nu}$. Therefore we study their effects.
The charged Higgs contributions are given by
\begin{align}
  1+\delta_{H^+}^{D^{(*)}\tau}(q^2)&=\left|
  1\pm \frac{q^2}{m_\tau(m_b\mp m_c)}\frac{V_{cb}V^*_{tb}\rho_e^{\tau\tau}\rho_u^{tc}}{2\sqrt{2}G_{\rm F} m_{H^+}^2 |V_{cb}|^2}
    \right|^2 \nonumber\\
    &+\frac{q^4}{m_\tau^2 (m_b\mp m_c)^2}\frac{|V_{tb}|^2|\rho_u^{tc}|^2(|\rho_e^{e\tau}|^2+|\rho_e^{\mu\tau}|^2)}
           {8G_{\rm F}^2m_{H^+}^4 |V_{cb}|^2},
           \label{cont_scenario3}
\end{align}
where double sign corresponds to $R(D)$ and $R(D^*)$, respectively.
As discussed in the Scenario (1), negative $\rho_e^{\tau\tau}$ and positive $\rho_u^{tc}$ increase $R(D^*)$, but decrease
$R(D)$. However, since the flavor violating contributions
do not interfere with the SM contributions, the effects behave like the imaginary part of $\rho_e^{\tau\tau}$ discussed in
the Scenario (1), and always increase both $R(D)$ and $R(D^*)$. 
Here, to study the effects of $\rho_e^{\mu\tau}$, we take $m_{H^+}=500$ GeV and $\rho_e^{\tau\tau}=-0.15$ as a reference
set of parameters. We parameterize the flavor violating coupling $\rho_e^{\mu\tau}$ as
\begin{align}
  r_\tau=\left|\frac{\rho_e^{\mu\tau}}{\rho_e^{\tau\tau}}\right|.
\end{align}
\begin{figure}[ht]
  \begin{center}
  \includegraphics[width=0.9\textwidth]{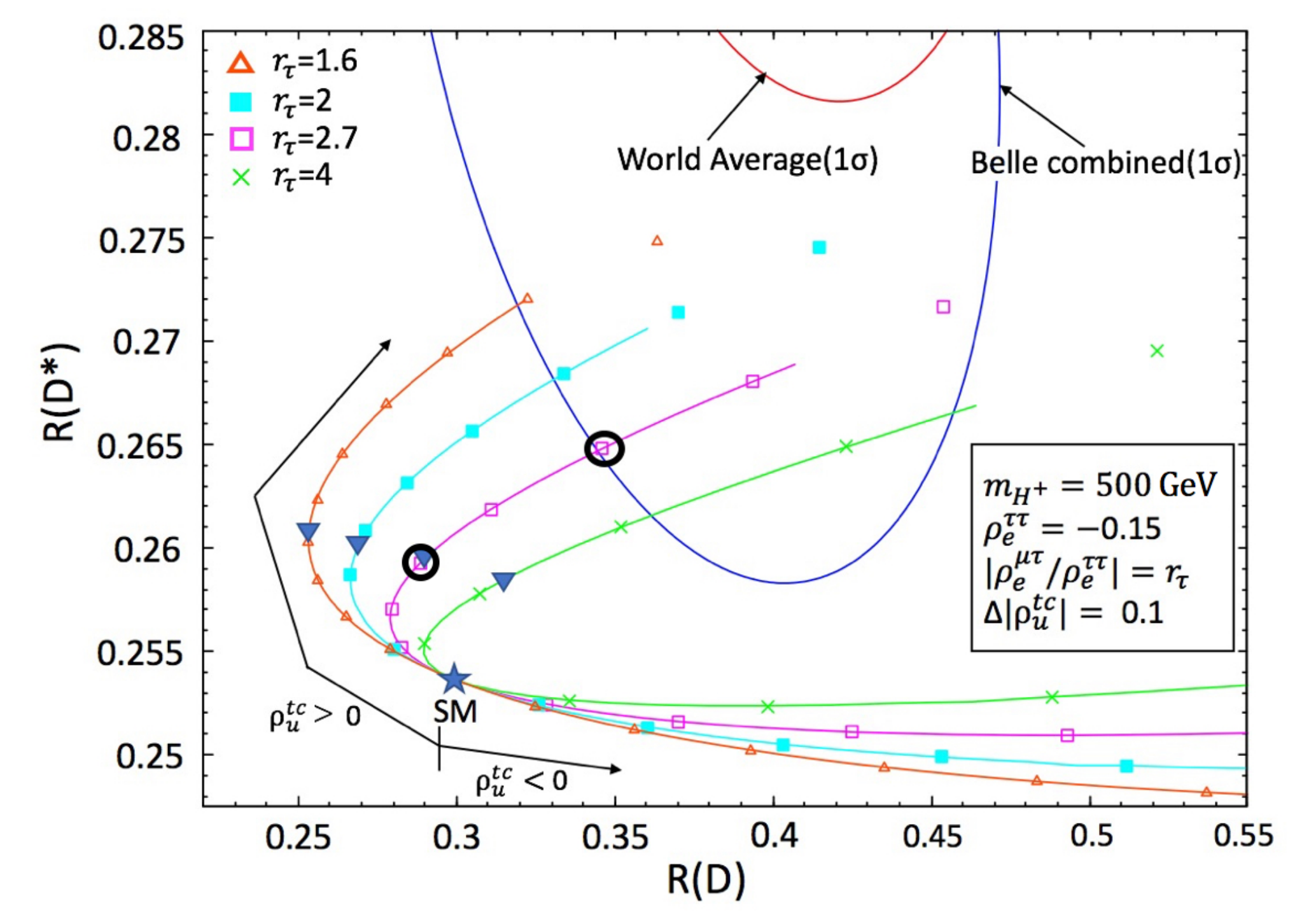}
    \caption{The predicted values for $R(D)$ and $R(D^*)$ are shown in the scenario (3). The currthrt
      experimental limits $(1\sigma)$ for Belle, and world average are also shown. Here we have
      set $m_{H^+}=500$ GeV, $\rho_e^{\tau\tau}=-0.15$.
      The value of $\rho_u^{tc}$ is assumed to be real and is varied from the SM point $(\rho_d^{sb}=0)$ (blue star-mark) by
      $0.1$ step positively $(\rho_u^{tc}>0)$ and negatively $(\rho_u^{tc}<0)$ with different values of
      $r_\tau(=|\rho_e^{\mu\tau}/\rho_e^{\tau\tau}|)$, $r_\tau=1.6$ (orange triangle), 2 (filled cyan quadrangle), 2.7 (open pink quadrangle),
      and 4 (green $\times$). The regions with the solid line in each $r_\tau$
      are consistent with the constraint ${\rm BR}(B_c^-\rightarrow \tau^- \bar{\nu})\le30\%$. The regions above filled blue triangle
      in each value of $r_\tau$ are excluded if ${\rm BR}(B_c^-\rightarrow \tau^- \bar{\nu})\le10\%$ is adopted. 
        The points indicated by a black circle are reference points for the LHC physics study discussed later.
    }
    \label{scenario3}
  \end{center}
\end{figure}

In Figure~\ref{scenario3}, the predicted values of $R(D)$ and $R(D^*)$ are shown in the Scenario (3). The current experimental limit
($1\sigma$) for Belle and world average are also shown. The value of $\rho_u^{tc}$ is assumed to be real and is varied from $\rho_u^{tc}=0$
(SM point, star-mark) by 0.1 step positively ($\rho_u^{tc}>0$) and negatively ($\rho_u^{tc}<0$) with different values of
$r_\tau(=|\rho_e^{\mu\tau}/\rho_e^{\tau\tau}|)$,
$r_\tau=1.6$ (orange triangle), 2 (filled cyan quadrangle), 2.7 (open pink quadrangle), and 4 (green $\times$). The regions with the solid line
in each $r_\tau$ are consistent with
the constraint ${\rm BR}(B_c^-\rightarrow \tau^- \bar{\nu})\le30\%$. The regions above filled blue triangle in each $r_\tau$ are
excluded if ${\rm BR}(B_c^-\rightarrow \tau^- \bar{\nu})\le10\%$ is imposed. 
Similar to the effect of imaginary part of the $\rho_u^{tc}$ in the Scenario (1),
the constraint
on $R(D^*)$ get stronger when $\rho_e^{\mu\tau}$ becomes larger.
We see that the predicted values of $R(D^{(*)})$ can not explain the current world average. On the other hand, they can
(can not) be within 1$\sigma$ region of Belle's result as shown in Figure~\ref{scenario3} if
${\rm BR}(B_c^-\rightarrow \tau^- \bar{\nu})\le30\%$ ($10\%$) is adopted.
From Eq.~(\ref{cont_scenario3}), the effect of non-zero $\rho_e^{e\tau}$ on $R(D^{(*)})$ is expected to be the same
as one of $\rho_e^{\mu\tau}$.

So far we have fixed parameters $m_{H^+}=500$ GeV and $\rho_e^{\tau\tau}=-0.15$.
The predicted values of $R(D^{(*)})$ depend on parameter sets $\rho_e^{\tau\tau}\rho_u^{tc}/m_{H^+}^2$ and
$|\rho_u^{tc}\rho_e^{i\tau}/m_{H^+}^2|^2$ where $i=e,~\mu$. Therefore, even if we change $m_{H^+}$ and $\rho_e^{\tau\tau}$
but these parameter sets are fixed, the predicted values do not change.
For the LHC physics study discussed later, we choose two reference points in the Scenario (3).
As a reference point 1, we take a point with $\rho_u^{tc}\rho_e^{\tau\tau}=-0.075$, $r_\tau=2.7$ and $m_{H^+}=500$ GeV,
which satisfies the constraint
${\rm BR}(B_c^-\rightarrow \tau^- \bar{\nu})\le30\%$, but not ${\rm BR}(B_c^-\rightarrow \tau^- \bar{\nu})\le10\%$.
As a reference point 2 in Scenario (3), we choose a point with
$\rho_u^{tc}\rho_e^{\tau\tau}=-0.045$, $r_\tau=2.7$ and $m_{H^+}=500$ GeV, which satisfies the constraint
${\rm BR}(B_c^-\rightarrow \tau^- \bar{\nu})\le10\%$.

Below, we have several comments.
The $q^2$ distribution in $d\Gamma(\bar{B}\rightarrow D\tau^-\bar{\nu})/dq^2$ is similar to the one
in Scenario (1).  The difference from the SM distribution increases in large $q^2$ regions as $\rho_e^{\mu\tau}$
becomes larger.

The effects in the $b\rightarrow s$ transition processes are also the same as one in Scenario (1).

A new interesting process is $\tau\rightarrow \mu\gamma$ ($\tau\rightarrow e\gamma$)
when both $\rho_e^{\tau\tau}$ and $\rho_e^{\mu\tau}$ ($\rho_e^{e\tau}$) Yukawa couplings are non-zero.
We will discuss the $\tau\rightarrow \mu\gamma$ process in the next section.

\subsubsection{$\tau\rightarrow \mu\gamma$}
In the presence of non-zero $\rho_e^{\tau\tau}$ and $\rho_e^{\mu\tau}$, $\tau\rightarrow \mu\gamma$ process is
generated. As discussed in Ref.~\cite{Omura2015xcg}, not only 1-loop contribution but also Barr-Zee type
2-loop contribution are important. The 1-loop contribution tends to be suppressed when $H$ and $A$ are degenerate.
On the other hand, the Barr-Zee type 2-loop contribution is affected by $\rho_u^{tt}$, which does not have an impact on
$R(D^{(*)})$.
\begin{figure}[ht]
  \begin{center}
    \includegraphics[width=0.6\textwidth]{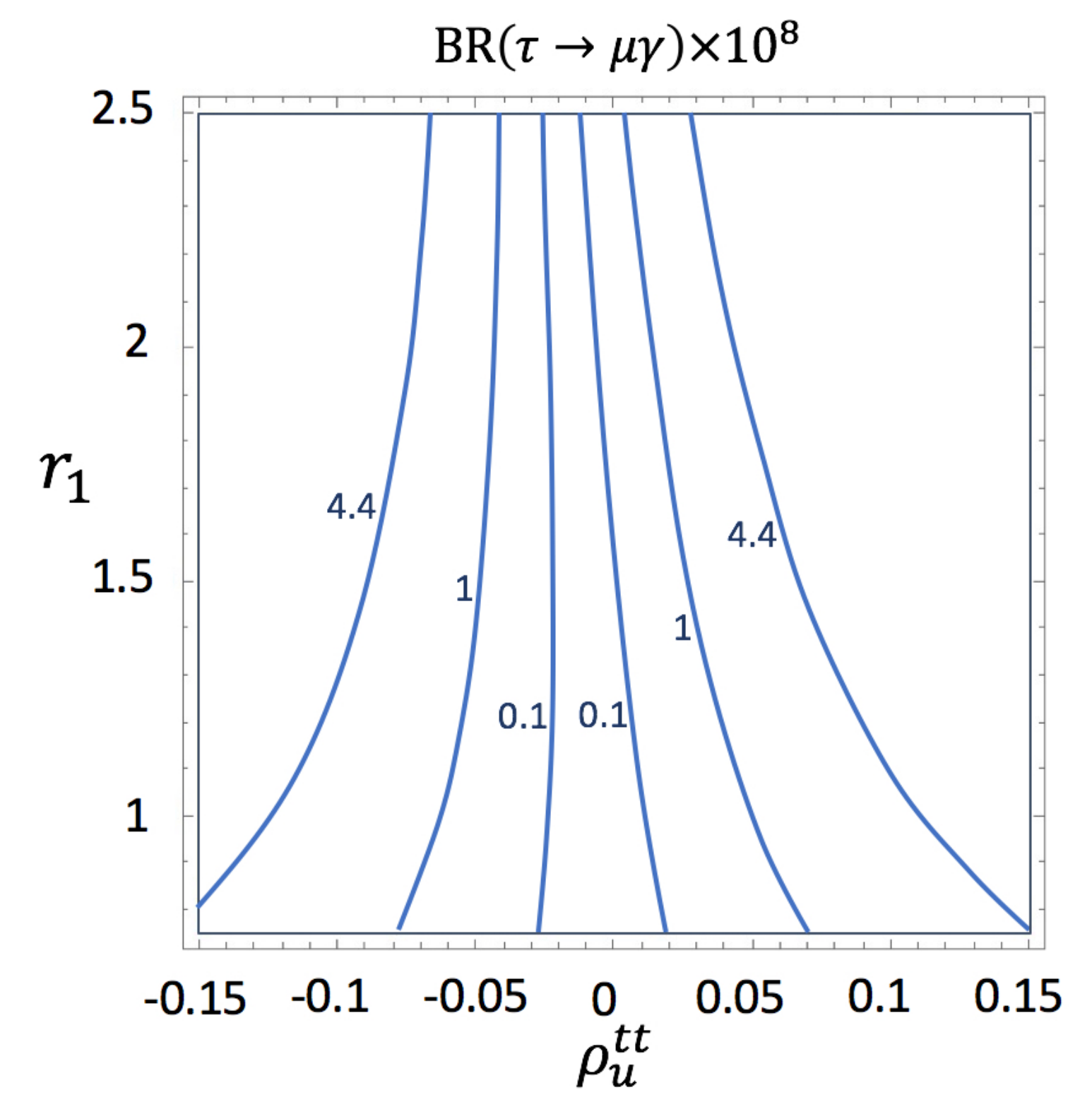}
\caption{${\rm BR}(\tau \rightarrow \mu\gamma)$ is shown as a function of $\rho_u^{tt}$ and
      $r_1~(=\frac{\rho_e^{\mu\tau}}{-0.27}=\frac{\rho_e^{\tau\tau}}{-0.1})$. 
      Here we have assumed that $c_{\beta\alpha}=0.001$, $m_{H/A}=m_{H^+}=500$ GeV
      and $r_\tau=|\rho_e^{\mu\tau}/\rho_e^{\tau\tau}|=2.7$. Note that the current experimental upper limit on
    ${\rm BR}(\tau \rightarrow \mu\gamma)$ is $4.4\times 10^{-8}$.}    
    \label{taumugamma_scenario3}
  \end{center}
\end{figure}
In Figure~\ref{taumugamma_scenario3}, the predicted branching ratio ${\rm BR}(\tau \rightarrow \mu\gamma)$
is shown as a function of $\rho_u^{tt}$ and $r_1~(=\frac{\rho_e^{\mu\tau}}{-0.27}=\frac{\rho_e^{\tau\tau}}{-0.1})$.
Here we have set $c_{\beta\alpha}=0.001$, $m_{H/A}=m_{H^+}=500$ GeV and 
$r_\tau=|\rho_e^{\mu\tau}/\rho_e^{\tau\tau}|=2.7$. We note that this parameter set is consistent with
the constraint from $h\rightarrow \mu\tau$. As seen in Figure~\ref{taumugamma_scenario3},
the interesting parameter region for $R(D^{(*)})$ is consistent with the current experimental upper limit
on ${\rm BR}(\tau \rightarrow \mu\gamma)$ [${\rm BR}(\tau \rightarrow \mu\gamma)\le4.4\times 10^{-8}$]
unless the $|\rho_u^{tt}|$ is large ($|\rho_u^{tt}|>0.1$).\footnote{When $\rho_u^{tt}$ is not zero, the charged Higgs boson
  contributions to $b\rightarrow s$ transition processes are modified. When $\rho_u^{tt}$ is small ($|\rho_u^{tt}|\le0.1$),
  the effects are limited. In Appendix C, we present the effects with non-zero
  $\rho_u^{tt}$ on the $b\rightarrow s$ transition processes.}

\subsection{Type II two Higgs doublet model}
\begin{figure}[ht]
  \begin{center}
    \includegraphics[width=0.8\textwidth]{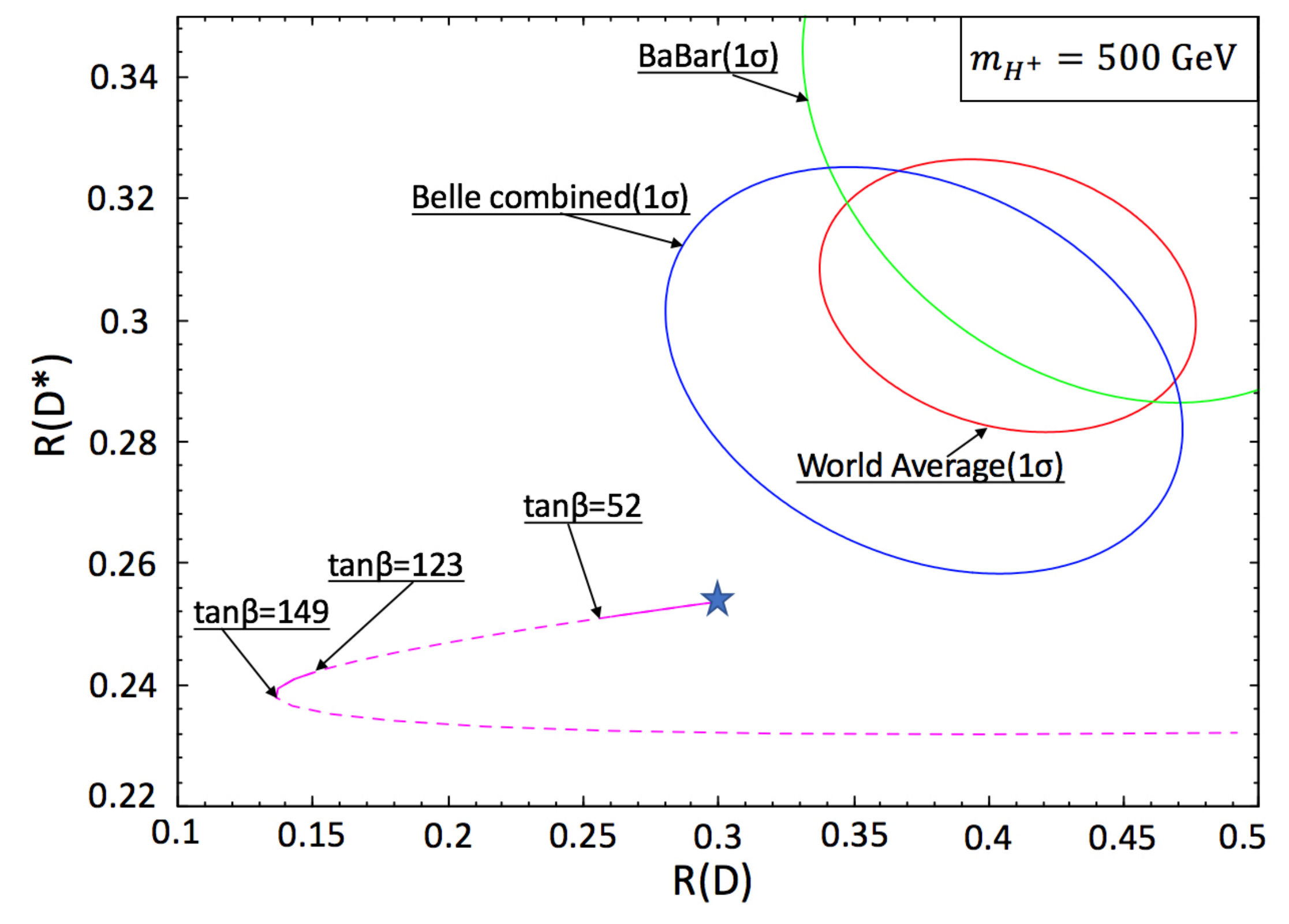}
    \caption{The predicted values for $R(D)$ and $R(D^*)$ are shown in the type II 2HDM. The current
      experimental limits $(1\sigma)$ for Belle, and world average are also shown. Here we have
      set $m_{H^+}=500$ GeV, and $\tan\beta$ is varied from zero (SM point, star-mark).
      The constraint from $B^-\rightarrow \tau^- \bar{\nu}$ is imposed and the regions with solid line are
      allowed, on the other hand, the regions with the dashed line are excluded.
    }
    \label{type2}
  \end{center}
\end{figure}
So far, we have discussed $R(D^{(*)})$ in the general 2HDM. As a comparison, here we discuss
the typical prediction of $R(D^{(*)})$ in type II 2HDM. The Yukawa interactions in the type II 2HDM
are restricted by the fermion masses and $\tan\beta$, which is a ratio between vacuum expectation values
of two neutral Higgs components, and hence the predicted values of $R(D^{(*)})$ are
very limited in contrast to those in the general 2HDM.
In Figure~\ref{type2}, the predicted values for $R(D)$ and $R(D^*)$ are shown in the type II 2HDM.
The current experimental limits $(1\sigma)$ for BaBar, Belle and world average are also shown. Here
we assume that $m_{H^+}=500$ GeV, and $\tan\beta$ is varied from zero (SM point, star-mark). The constraint from
$B^-\rightarrow \tau^- \bar{\nu}$ is imposed, and the regions with solid line are
allowed, on the other hand, the regions with dashed line are excluded by the $B^-\rightarrow \tau^- \bar{\nu}$ constraint.
As can be seen from Figure~\ref{type2}, the predicted values can not reach the region within the $1\sigma$ of BaBar,
Belle nor the world average. We would like to stress that the predictions of the general 2HDM would be
very different from those in the type II 2HDM.

\section{Constraints and implications at the LHC}
So far, we have discussed the possible effects on $R(D^{(*)})$ and various constraints from the flavor
physics in each scenario. In this section, we study productions and decays of heavy Higgs bosons at the LHC
and discuss possible constraints from the LHC results
and implications at the LHC.

\subsection{Productions for heavy neutral Higgs bosons}
\subsubsection{Scenario (1)}
Since the Yukawa coupling $\rho_e^{\tau\tau}$ plays a crucial role to induce the important contribution to
$R(D^{(*)})$, the heavy neutral Higgs boson search in $H/A\rightarrow \tau^+\tau^-$ mode
at the LHC experiment would be important~\cite{highpt}.
When the Higgs mixing parameter $c_{\beta\alpha}$ is small, the gluon-gluon fusion production for heavy neutral
Higgs bosons $H$ and $A$ relies on the interaction via $\rho_u^{tt}$, which does not affect $R(D^{(*)})$.
Here we use HIGLU~\cite{Spira1995mt} to calculate the production cross section for $H$ and $A$ via the gluon-gluon
fusion process at the NNLO and they are given by
\begin{align}
\sigma (pp\rightarrow H)=1.4\left|\rho_u^{tt}\right|^2~[{\rm pb}],~~\sigma(pp\rightarrow A)=2.3\left|\rho_u^{tt}\right|^2~[{\rm pb}],
\end{align}
at $\sqrt{s}=8$ TeV for $m_{H/A}=500$ GeV, and
\begin{align}
\sigma (pp\rightarrow H)=4.9\left|\rho_u^{tt}\right|^2~[{\rm pb}],~~\sigma(pp\rightarrow A)=8.3\left|\rho_u^{tt}\right|^2~[{\rm pb}],
\end{align}
at $\sqrt{s}=13$ TeV for $m_{H/A}=500$ GeV.

\begin{figure}[ht]
  \begin{center}
    \includegraphics[width=0.7\textwidth]{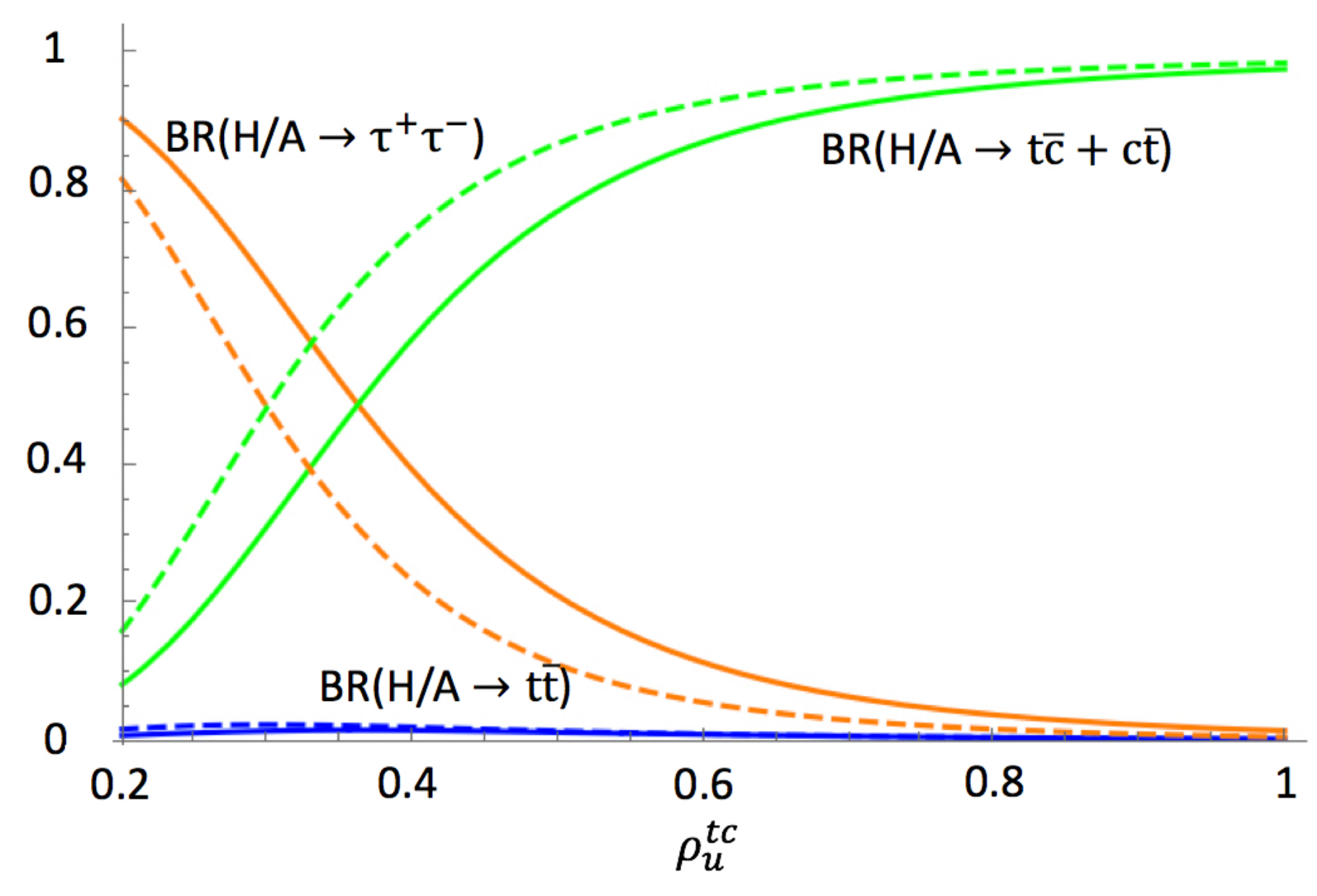}
    \caption{The decay branching ratios of $H$ and $A$ are shown as a function of $\rho_u^{tc}$. 
      For solid lines,  we take a reference point 1 ($\rho_u^{tc}\rho_{\tau\tau}=-0.2$, $\delta_{\tau\tau}=\frac{5\pi}{12}$ and $m_{H^+}=500$ GeV).
      For dashed lines, a reference point 2 ($\rho_u^{tc}\rho_{\tau\tau}=-0.156$, $\delta_{\tau\tau}=\frac{\pi}{2}$ and $m_{H^+}=500$ GeV)
      is taken.      
      We have also assumed that $m_{H/A}=m_{H^+}$ and $c_{\beta\alpha}=0.001$ as a reference parameter set for Scenario (1).
      Here $\rho_u^{tt}$ is also fixed to be $0.1$ ($\rho_u^{tt}=0.1$).}
\label{neutral_decayBR}
\end{center}
\end{figure}
For the LHC physics study in Scenario (1), we consider reference point 1 and 2, shown by black circle points
in Figure~\ref{scenario1}. As discussed in Scenario (1), the reference point 1 is defined by a point with
$\rho_u^{tc}\rho_{\tau\tau}=-0.2$, $\delta_{\tau\tau}=\frac{5\pi}{12}$ and $m_{H^+}=500$ GeV, and
the reference point 2 is
$\rho_u^{tc}\rho_{\tau\tau}=-0.156$, $\delta_{\tau\tau}=\frac{\pi}{2}$ and $m_{H^+}=500$ GeV.
The decay branching ratios of $H$ and $A$ are shown as a function of $\rho_u^{tc}$ in Figure~\ref{neutral_decayBR}.\footnote{
  When $H$, $A$ and $H^+$ are not degenerate, two body decays such as $H\rightarrow AZ$ and $H\rightarrow H^\pm W^\mp$ may
  be allowed. However, we note that for the heavy Higgs boson like $m_H\simeq 500$ GeV, non-degeneracy for such two body decays
  requires large Higgs quartic couplings $\lambda_{4,5}$.}
For solid lines (dashed lines), we have taken the reference point 1 (reference point 2). Here we assume
that $m_{H/A}=m_{H^+}$ and $c_{\beta\alpha}=0.001$ as a reference parameter set for Scenario (1).
We also fix $\rho_u^{tt}=0.1$.\footnote{We note that the phase $\delta_{\tau\tau}$ does not affect the decay
  $H/A\rightarrow \tau^+\tau^-$.}

When we fix $m_{H^+}$ ($m_{H^+}=500$ GeV) and a product of the Yukawa couplings $\rho_u^{tc}\rho_e^{\tau\tau}$,
the predicted values of $R(D^{(*)})$ are also fixed. In the reference parameter set, on the other hand, the decay branching ratios
for $H/A$ depend on $\rho_u^{tc}$ (fixing a value of $\rho_u^{tc}\rho_e^{\tau\tau}$) as shown in Figure~\ref{neutral_decayBR}.
When $\rho_u^{tc}$ is small (large), the dominant decay mode is $H/A\rightarrow \tau^+\tau^-$
($H/A\rightarrow t\bar{c}+c\bar{t} $) because $\rho_{\tau\tau}$ $(\rho_u^{tc})$ is large. For $\rho_u^{tt}=0.1$,
a decay branching ratio for $H/A\rightarrow t\bar{t}$ is always small.
For a fixed value of $\rho_u^{tc}$, an absolute value of $\rho_e^{\tau\tau}$ in the reference point 1 ($|\rho_e^{\tau\tau}|=0.2/\rho_u^{tc}$)
is larger than one in the reference point 2 ($|\rho_e^{\tau\tau}|=0.156/\rho_u^{tc}$), and hence the branching ratio
of $H/A\rightarrow \tau^+\tau^-$ ($H/A\rightarrow t\bar{c}+c\bar{t}$) in the reference point 1 is larger (smaller)
than one in the reference point 2, as seen in Figure~\ref{neutral_decayBR}.

\begin{figure}[ht]
  \begin{center}
    \includegraphics[width=0.8\textwidth]{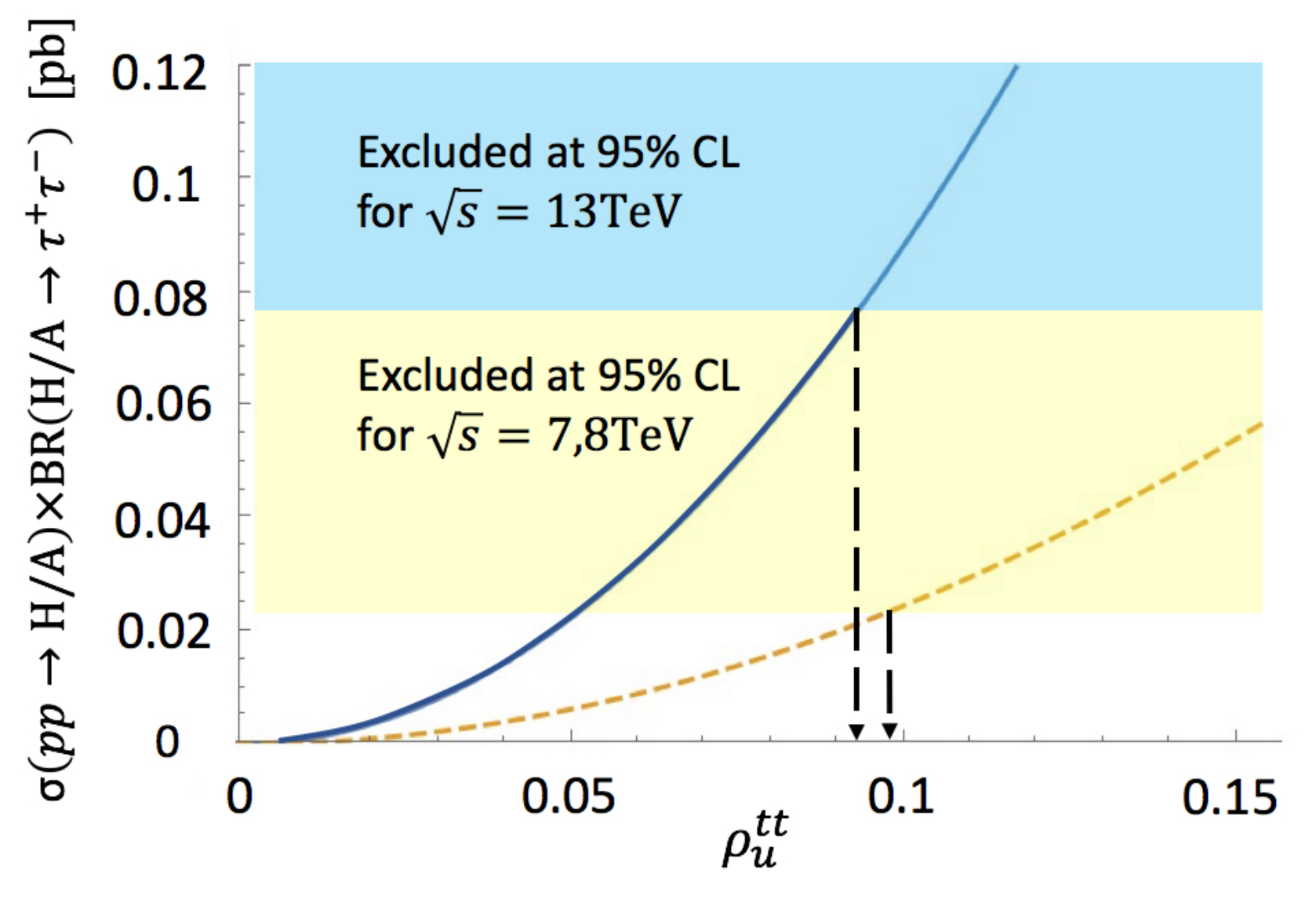}
    \caption{The cross section $\sigma(pp\rightarrow H/A) \times {\rm BR}(H/A\rightarrow \tau^+\tau^-)$ [pb]
      via the gluon-gluon fusion production at $\sqrt{s}=8$ TeV (dashed orange line)
      and $\sqrt{s}=13$ TeV (solid blue line) are shown as a function of $\rho_u^{tt}$. 
      Here we show the sum of the $H$ and $A$ contributions. We have taken
      $\rho_u^{tc}=0.3$ in the reference parameter set with the reference point 1, mentioned in Figure~\ref{neutral_decayBR}.
      We also show the regions which are excluded by the LHC experiment (95\% C.L.) at $\sqrt{s}=7$
      and 8 TeV~\cite{ATLAStautau7,CMSStautau7}
      (lemon yellow shaded region (lower)) and at $\sqrt{s}=13$ TeV~\cite{ATLAStautau13,CMStautau13} (sky blue shaded region (upper)), respectively.
      }
\label{Run1vsRun2}
\end{center}
\end{figure}
In Figure~\ref{Run1vsRun2}, the cross sections $\sigma(pp\rightarrow H/A)\times {\rm BR}(H/A\rightarrow \tau^+\tau^-)$ [pb]
via the gluon-gluon fusion production at $\sqrt{s}=8$ TeV  (dashed orange line)
and $\sqrt{s}=13$ TeV (solid blue line) are shown as a function of $\rho_u^{tt}$. Here the sum of
$H$ and $A$ contributions are taken into account. We have taken $\rho_u^{tc}=0.3$ in the reference parameter set with the reference point 1,
mentioned in Figure~\ref{neutral_decayBR}.
We also show the regions which are excluded (95 \% C.L.) by the LHC experiment at $\sqrt{s}=7$
and 8 TeV~\cite{ATLAStautau7,CMSStautau7}
(sky blue shaded region (upper)) and at $\sqrt{s}=13$ TeV~\cite{ATLAStautau13,CMStautau13} (lemon yellow shaded region (lower)), respectively. As can be seen
from Figure~\ref{Run1vsRun2}, the Run 2 results
already put a stronger constraint on $\rho_u^{tt}$.

\begin{figure}[ht]
  \begin{center}
    \includegraphics[width=0.7\textwidth]{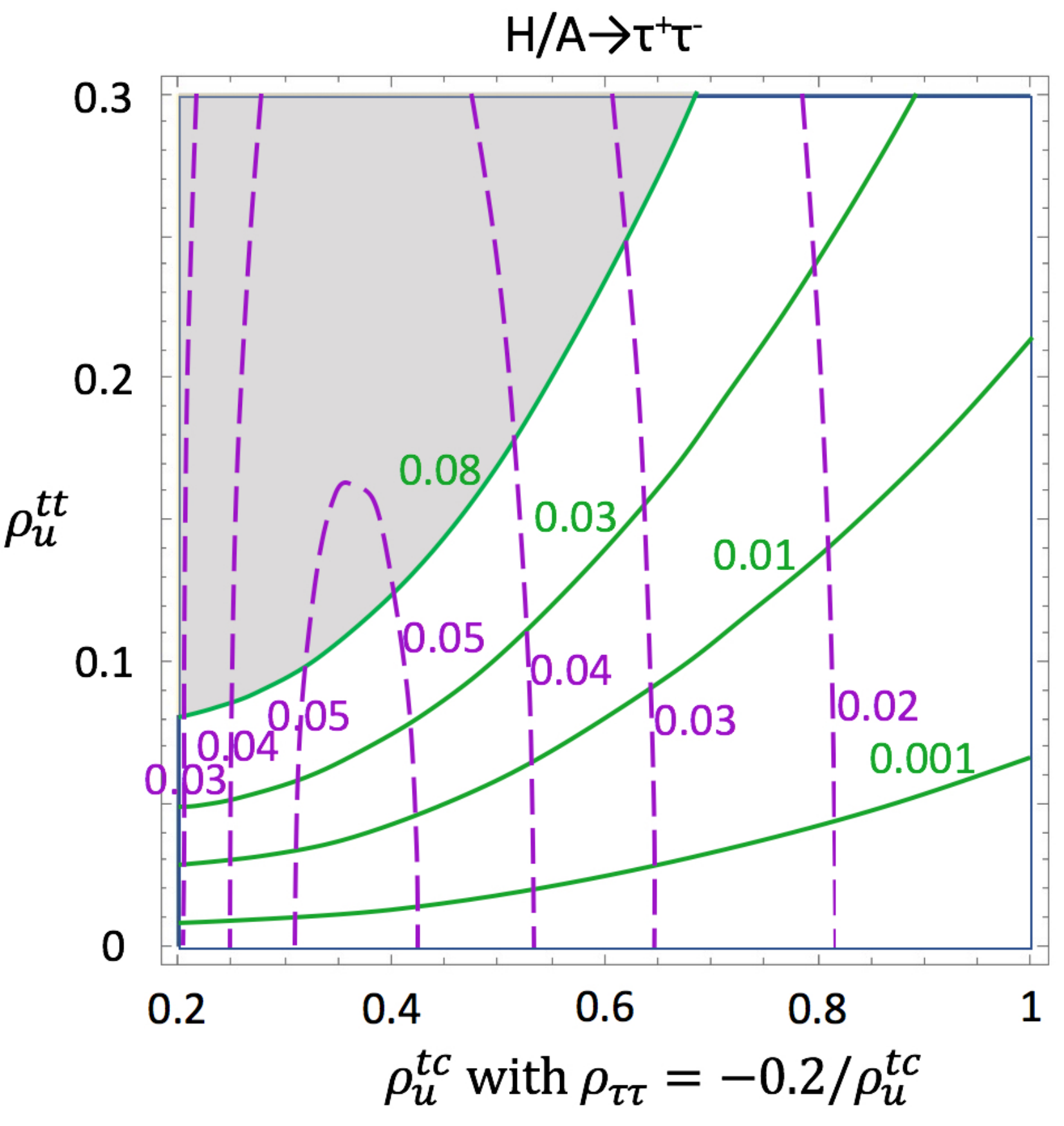}
    \caption{The cross section $\sigma(pp\rightarrow H/A) \times {\rm BR}(H/A\rightarrow \tau^+\tau^-)$ [pb]
      via the gluon-gluon fusion production (solid green lines) and the cross section
      $\sigma(pp \rightarrow t+ H/A) \times {\rm BR}(H/A\rightarrow \tau^+\tau^-)$ [pb] via $c g\rightarrow t+ H/A$ process
      (dashed purple lines) at $\sqrt{s}=13$ TeV
      are shown as a function of $\rho_u^{tc}$ and $\rho_u^{tt}$. Here the sum of $H$ and $A$ contributions
      are shown. Here we have taken the reference parameter set with the reference point 1, mentioned in Figure~\ref{neutral_decayBR}.
      The shaded region is excluded at 95 \% C.L. by the search for
      $\sigma(pp\rightarrow H/A) \times {\rm BR}(H/A\rightarrow \tau^+\tau^-)$ via the gluon-gluon fusion production
      at the LHC experiments~\cite{ATLAStautau13,CMStautau13}.}
\label{cross_tautau}
\end{center}
\end{figure}
In Figure~\ref{cross_tautau}, the cross section $\sigma(pp\rightarrow H/A) \times {\rm BR}(H/A\rightarrow \tau^+\tau^-)$
[pb] via the gluon-gluon fusion production (solid green lines) at $\sqrt{s}=13$ TeV
is shown as a function of $\rho_u^{tc}$ and $\rho_u^{tt}$.
Here the sum of $H$ and $A$ contributions are shown.
We have taken the reference parameter set with the reference point 1, mentioned in Figure~\ref{neutral_decayBR}, and the shaded region is
excluded at 95 \% C.L. by the search for $\sigma(pp\rightarrow H/A) \times {\rm BR}(H/A\rightarrow \tau^+\tau^-)$
at the LHC experiment~\cite{ATLAStautau13,CMStautau13}.

\begin{figure}[ht]
  \begin{center}
    \includegraphics[width=0.4\textwidth]{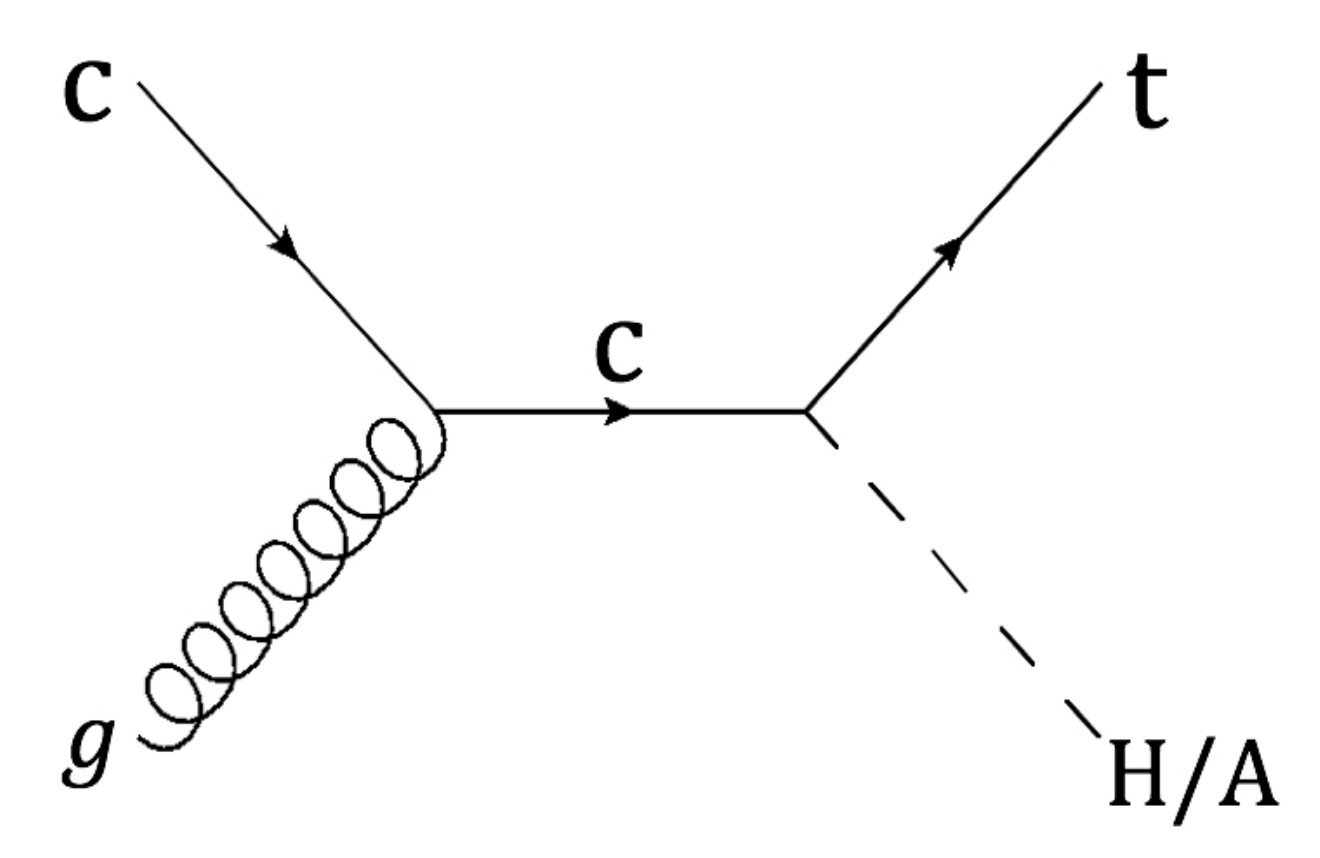}
    \caption{A Feynman diagram for neutral Higgs boson production via $gc\rightarrow t+H/A$ process.}
\label{Feyn_gcth}
\end{center}
\end{figure}
In addition to the gluon-gluon fusion production for the neutral Higgs bosons ($H$ and $A$),
in the existence of the sizable $\rho_u^{tc}$ Yukawa coupling,
the production via $\rho_u^{tc}$ ($gc\rightarrow t +H/A)$) shown in Figure~\ref{Feyn_gcth} would be important.
We calculate $\sigma(pp\rightarrow t+H/A)$
via $gc\rightarrow t+H/A$ process by using the calchep~\cite{Belyaev2012qa} with CTEQ leading order PDF.
The result is given by
\begin{align}
\sigma(pp\rightarrow t+H/A)=0.8~|\rho_u^{tc}|^2~[{\rm pb}],
\end{align}
at $\sqrt{s}=13$ TeV for $m_{H/A}=500$ GeV.
In Figure~\ref{cross_tautau}, we also
show the results of $\sigma(pp\rightarrow t+H/A)\times {\rm BR}(H/A\rightarrow \tau^+\tau^-)$ [pb] (dashed purple lines)
as a function of $\rho_u^{tc}$ and $\rho_u^{tt}$.
As one can see, the cross sections are sizable. For the reference point 2 with the fixed $\rho_u^{tc,tt}$, the cross sections
$\sigma(pp\rightarrow H/A) \times {\rm BR}(H/A\rightarrow \tau^+\tau^-)$ and
$\sigma(pp\rightarrow t+H/A) \times {\rm BR}(H/A\rightarrow \tau^+\tau^-)$
are slightly smaller than those for the reference point 1 because of smaller ${\rm BR}(H/A\rightarrow \tau^+\tau^-)$
as shown in Figure~\ref{neutral_decayBR}.

\begin{figure}[ht]
  \begin{center}
    \includegraphics[width=0.49\textwidth]{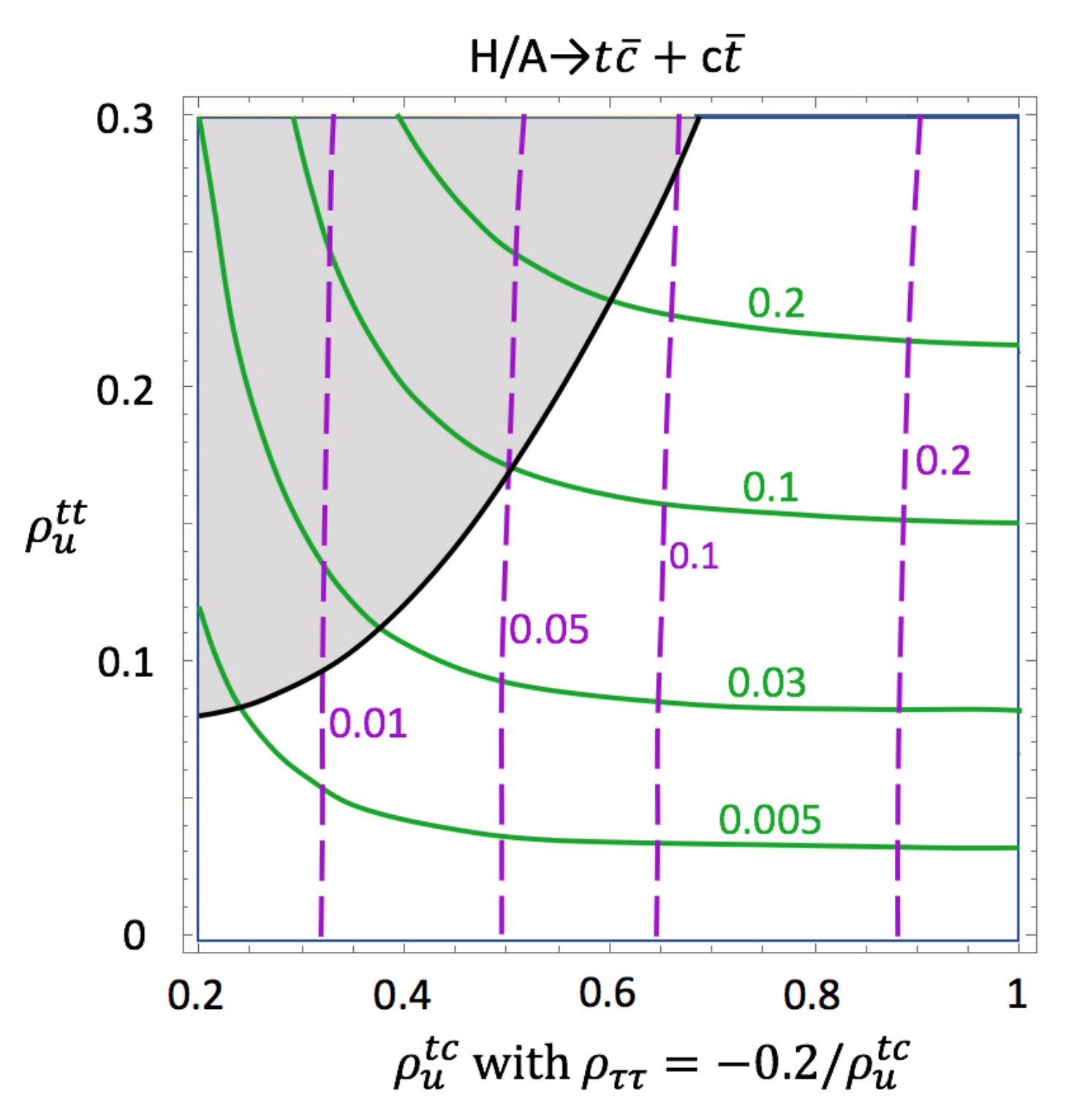}
    \includegraphics[width=0.49\textwidth]{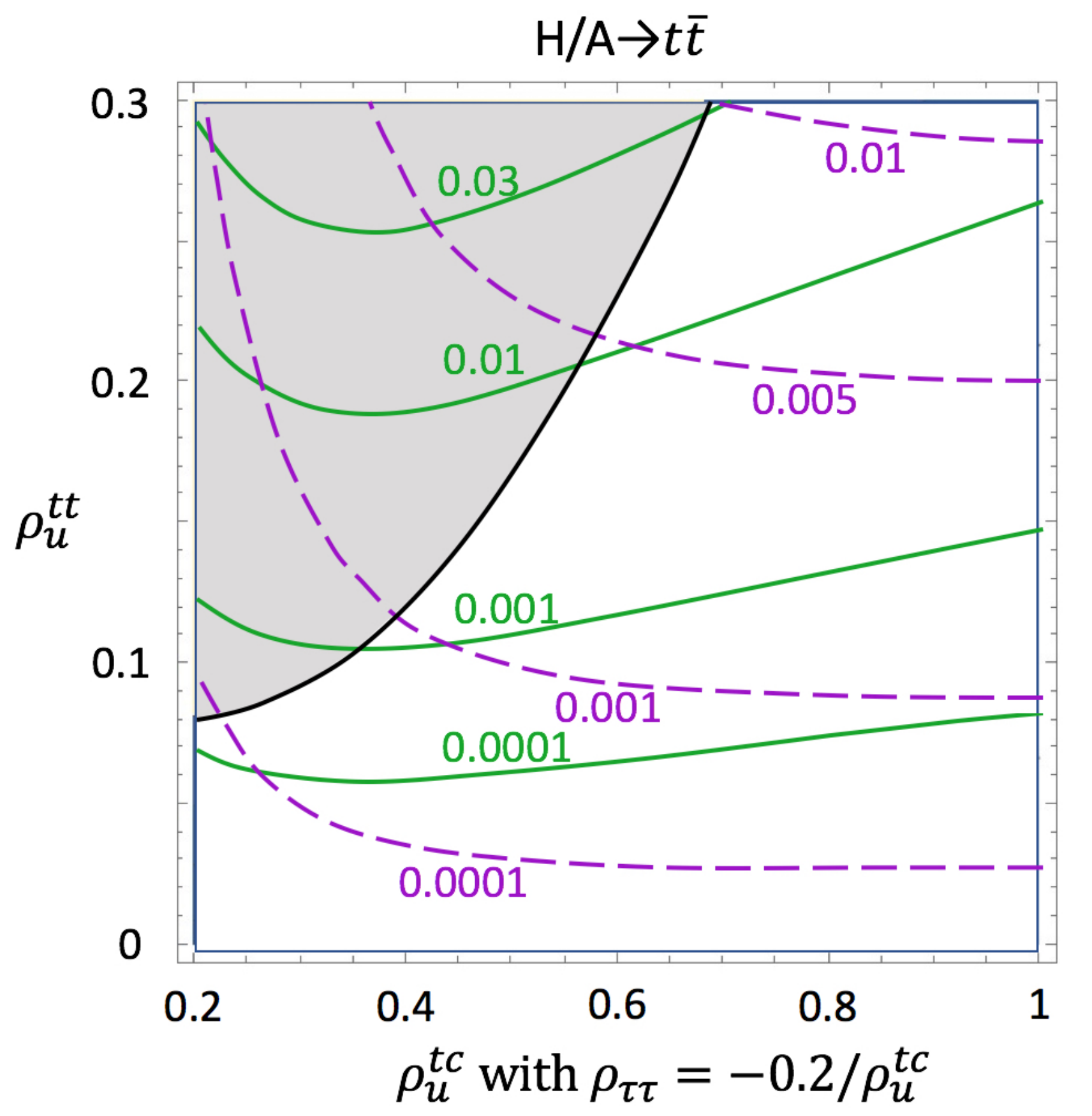}    
    \caption{[{\bf Left}] The cross section $\sigma(pp\rightarrow H/A) \times {\rm BR}(H/A\rightarrow t\bar{c}+c\bar{t})$ [pb]
      via the gluon-gluon fusion production (solid green lines) and the cross section
      $\sigma(pp \rightarrow t+ H/A) \times {\rm BR}(H/A\rightarrow t\bar{c}+c\bar{t})$ [pb] via $c g\rightarrow t+ H/A$ process
      (dashed purple lines) and [{\bf Right}] the cross section
      $\sigma(pp\rightarrow H/A) \times {\rm BR}(H/A\rightarrow t\bar{t})$ [pb]
      via the gluon-gluon fusion production (solid green lines) and the cross section
      $\sigma(pp \rightarrow t+ H/A) \times {\rm BR}(H/A\rightarrow t\bar{t})$ [pb] via $c g\rightarrow t+ H/A$ process
      (dashed purple lines)
      are shown as a function of $\rho_u^{tc}$ and $\rho_u^{tt}$ at $\sqrt{s}=13$ TeV.
      The same parameter set is taken as in Figure~\ref{cross_tautau}.
      The shaded region is excluded at 95 \% C.L. by the search for
      $\sigma(pp\rightarrow H/A) \times {\rm BR}(H/A\rightarrow \tau^+\tau^-)$ via the gluon-gluon fusion production
      at the LHC experiment~\cite{ATLAStautau13,CMStautau13}.}
\label{cross_other}
\end{center}
\end{figure}
Since the heavy neutral Higgs bosons significantly decay to top and charm quarks as shown in
Figure~\ref{neutral_decayBR}, the exotic decay mode would be important.
In a left figure of Figure~\ref{cross_other}, we present the cross sections
$\sigma(pp\rightarrow H/A)\times {\rm BR}(H/A\rightarrow t\bar{c}+c\bar{t})$ [pb]
via the gluon-gluon fusion process (solid green lines) and 
$\sigma(pp\rightarrow t+H/A)\times {\rm BR}(H/A\rightarrow t\bar{c}+c\bar{t})$ [pb]
via $c g\rightarrow t+ H/A$ process (dashed purple lines)
as a function of $\rho_u^{tc}$ and $\rho_u^{tt}$ at $\sqrt{s}=13$ TeV.
The same parameter set is taken as in Figure~\ref{cross_tautau}.
As $\rho_u^{tc}$ gets larger, the production cross section via $cg\rightarrow t+H/A$ becomes
significantly larger. For the reference point 2 with the fixed values of $\rho_u^{tc,tt}$, the cross sections
$\sigma(pp\rightarrow H/A)\times {\rm BR}(H/A\rightarrow t\bar{c}+c\bar{t})$ and
$\sigma(pp\rightarrow t+H/A)\times {\rm BR}(H/A\rightarrow t\bar{c}+c\bar{t})$ are even larger than
those for the reference point 1 because of larger ${\rm BR}(H/A\rightarrow t\bar{c}+c\bar{t})$, as seen in
Figure~\ref{neutral_decayBR}.

On the other hand, the production cross sections
$\sigma(pp\rightarrow H/A)\times {\rm BR}(H/A\rightarrow t\bar{t})$ [pb]
via the gluon-gluon fusion process and 
$\sigma(pp\rightarrow t+H/A)\times {\rm BR}(H/A\rightarrow t\bar{t})$ [pb]
via $c g\rightarrow t+ H/A$ process tend to be smaller than other modes because of the smaller
branching ratio. In a right figure of Figure~\ref{cross_other}, the cross sections
$\sigma(pp\rightarrow H/A)\times {\rm BR}(H/A\rightarrow t\bar{t})$ [pb]
via the gluon-gluon fusion process (solid green lines) and 
$\sigma(pp\rightarrow t+H/A)\times {\rm BR}(H/A\rightarrow t\bar{t})$ [pb]
via $c g\rightarrow t+ H/A$ process (dashed purple lines)
are shown as a function of $\rho^{tc}_u$ and $\rho_u^{tt}$ at $\sqrt{s}=13$ TeV.

\begin{figure}[ht]
  \begin{center}
    \includegraphics[width=0.49\textwidth]{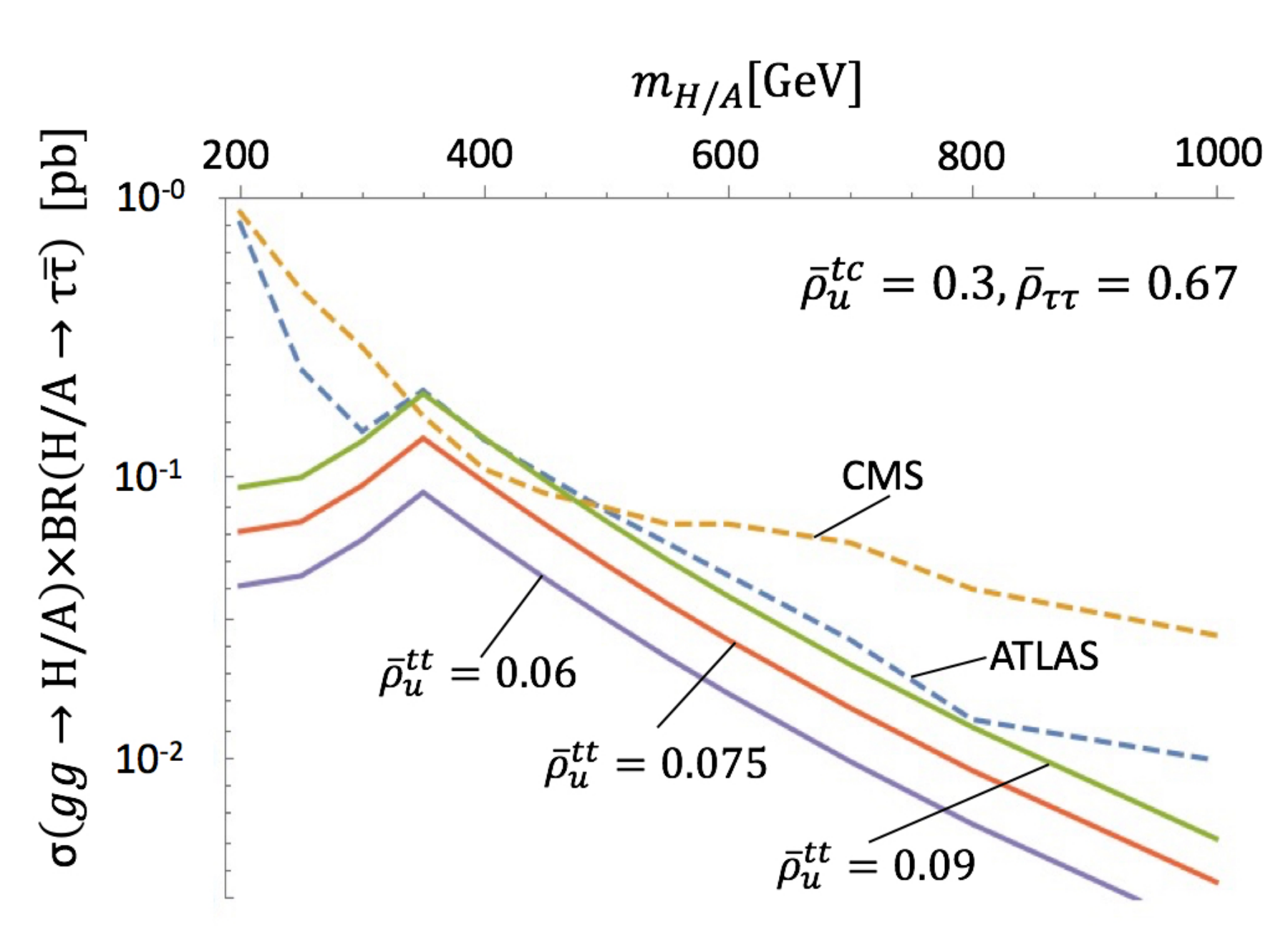}
    \includegraphics[width=0.49\textwidth]{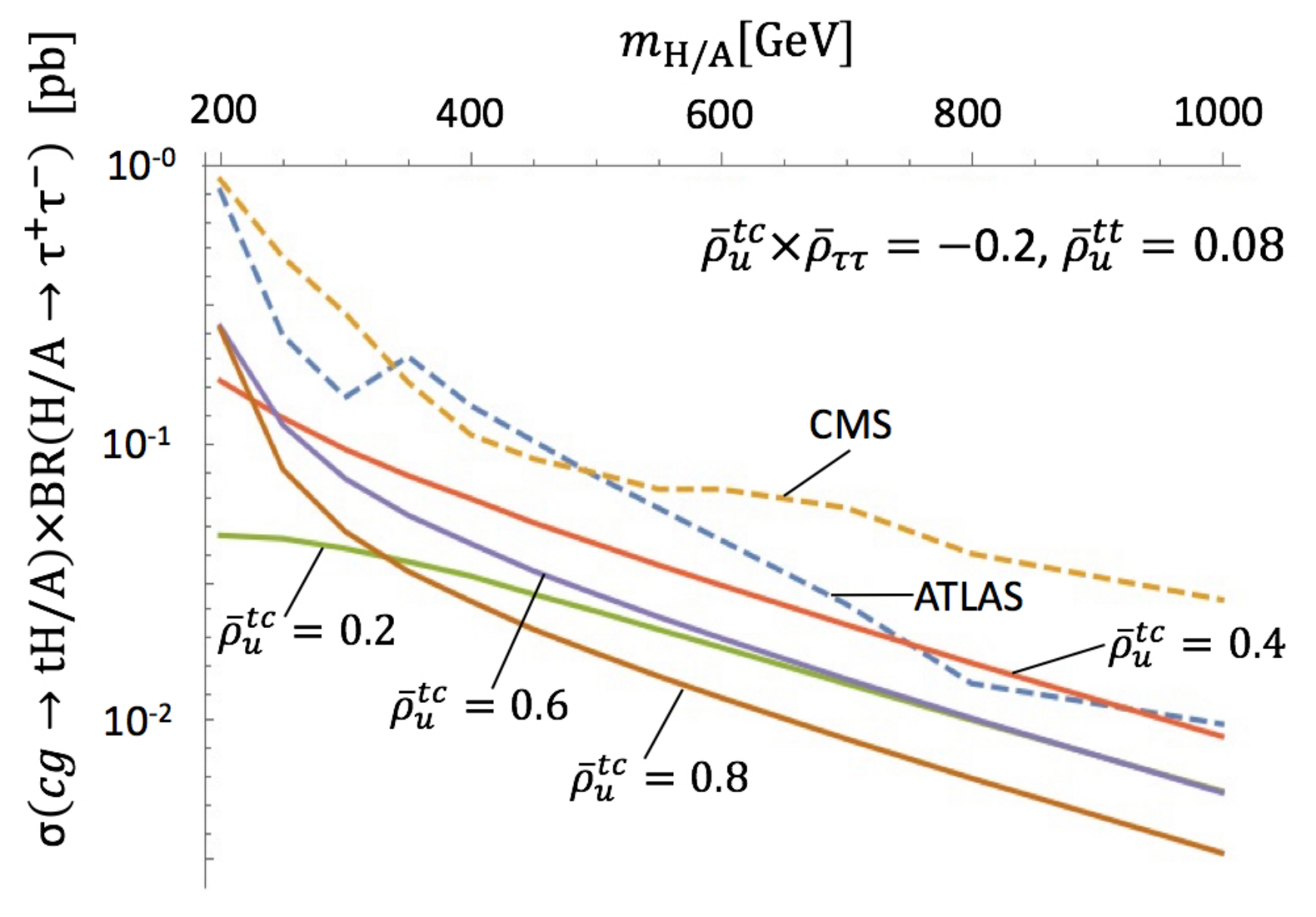}    
    \caption{[{\bf Left}] The cross sections $\sigma(pp\rightarrow H/A) \times {\rm BR}(H/A\rightarrow \tau^+\tau^-)$ [pb]
      via the gluon-gluon fusion production at $\sqrt{s}=13$ TeV are shown as a function of the heavy Higgs boson mass
      $m_{H/A}$ [GeV]. In order to fix the predicted values of $R(D^{(*)})$, the relevant Yukawa couplings are also
      scaled as $\rho_f^{ij}=\bar{\rho}_f^{ij}\left(\frac{m_{H/A}}{500~{\rm GeV}}\right)$. Here $\bar{\rho}_f^{ij}$ are
      the reference values for $m_{H/A}=m_{H^+}=500$ GeV, and we have set $\bar{\rho}_u^{tc}=0.3$ and
      $\bar{\rho}_{\tau\tau}=0.67$ (the reference point 1).
      The predicted cross sections with $\bar{\rho}_u^{tt}=0.06$, $\bar{\rho}_u^{tt}=0.075$
      and $\bar{\rho}_u^{tt}=0.09$ as well as the ATLAS (blue dotted line) and
      CMS (dotted orange line) limits (95 \% C.L.) are shown.
      [{\bf Right}] The cross sections
      $\sigma(pp \rightarrow t+ H/A) \times {\rm BR}(H/A\rightarrow \tau^+\tau^-)$ [pb] via $c g\rightarrow t+ H/A$ process
      at $\sqrt{s}=13$ TeV are shown as a function of $m_{H/A}$ [GeV]. We have taken $\bar{\rho}_u^{tt}=0.08$ and
      $\bar{\rho}_u^{tc}\bar{\rho}_{\tau\tau}=-0.2$ (the reference point 1).
      The predicted cross sections with $\bar{\rho}_u^{tc}=0.2,~0.4,~0.6$ and $0.8$ are shown.
      As a comparison, ATLAS and CMS limits (95 \% C.L.) on
      $\sigma(pp\rightarrow H/A) \times {\rm BR}(H/A\rightarrow \tau^+\tau^-)$ [pb] via
      the gluon-gluon fusion production at $\sqrt{s}=13$ TeV are also shown.
      }
\label{cross_scale}
\end{center}
\end{figure}
So far, we have fixed the heavy neutral Higgs boson masses ($m_{H/A}=500$ GeV).
In a left figure of Figure~\ref{cross_scale}, the cross sections
$\sigma(pp\rightarrow H/A) \times {\rm BR}(H/A\rightarrow \tau^+\tau^-)$ [pb]
via the gluon-gluon fusion process at $\sqrt{s}=13$ TeV are shown as a function of the heavy Higgs boson mass $m_{H/A}$ [GeV].
In order to fix the predicted values of $R(D^{(*)})$, we also scale the relevant Yukawa couplings as
\begin{align}
  \rho_f^{ij}=\bar{\rho}_f^{ij}\left(\frac{m_{H/A}}{500~{\rm GeV}}\right),
\end{align}
where $\bar{\rho}_f^{ij}$ are the reference values for $m_{H/A}=500$ GeV. Here we have assumed $m_{H/A}=m_{H^+}$.
We have set $\bar{\rho}_u^{tc}=0.3$ and $\bar{\rho}_{\tau\tau}=0.67$ (the reference point 1).
The predicted cross sections with $\bar{\rho}_u^{tt}=0.06$, $\bar{\rho}_u^{tt}=0.075$
and $\bar{\rho}_u^{tt}=0.09$ as well as the ATLAS and
CMS limits (95 \% C.L.) are shown. As can be seen from the figure, the constraints around
$m_{H/A}=400$ GeV are somewhat stronger than one at $m_{H/A}=500$ GeV because of the stronger CMS limit. In the small
($m_{H/A}<300$ GeV) and large ($m_{H/A}>800$ GeV) mass regions, the constraints become weaker.
In a right figure of Figure~\ref{cross_scale}, the cross sections
$\sigma(pp \rightarrow t+ H/A) \times {\rm BR}(H/A\rightarrow \tau^+\tau^-)$ [pb] via $c g\rightarrow t+ H/A$ process
at $\sqrt{s}=13$ TeV are shown as a function of $m_{H/A}$ [GeV]. We have taken $\bar{\rho}_u^{tt}=0.08$ and
$\bar{\rho}_u^{tc}\bar{\rho}_{\tau\tau}=-0.2$ (the reference point 1).
The predicted cross sections with $\bar{\rho}_u^{tc}=0.2,~0.4,~0.6$ and $0.8$ are shown.
As a comparison, ATLAS and CMS  limits (95 \% C.L.) on
$\sigma(pp\rightarrow H/A) \times {\rm BR}(H/A\rightarrow \tau^+\tau^-)$ [pb] via
the gluon-gluon fusion production at $\sqrt{s}=13$ TeV are also shown.


\subsubsection{Scenario (3)}
\begin{figure}[ht]
  \begin{center}
    \includegraphics[width=0.65\textwidth]{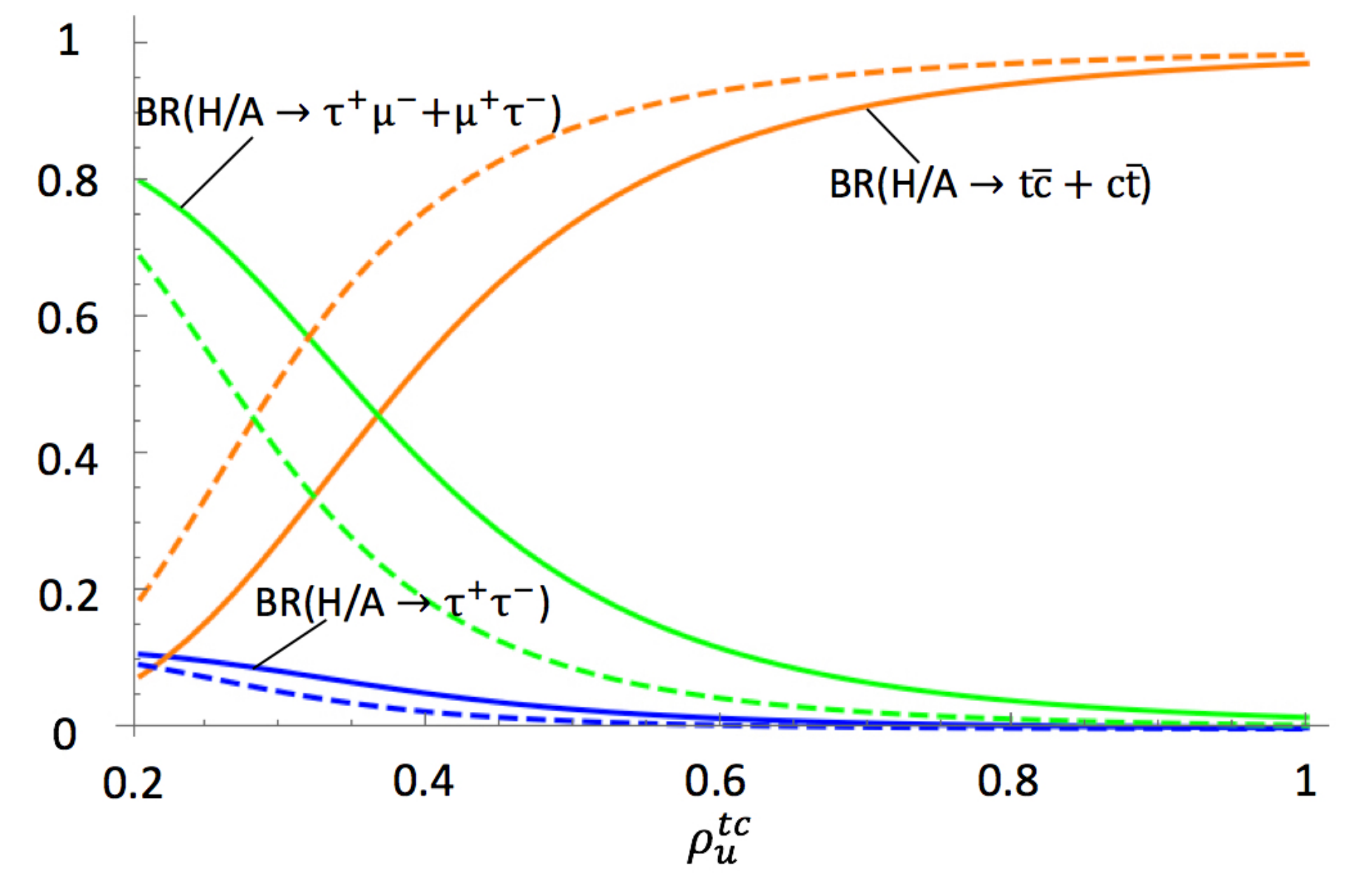}
    \caption{The decay branching ratios of $H/A$ are shown as a function of $\rho_u^{tc}$.
      For solid lines, the reference point 1 ($\rho_u^{tc}\rho_e^{\tau\tau}=-0.075$, $r_\tau(=|\rho_e^{\mu\tau}/\rho_e^{\tau\tau}|)=2.7$
      and $m_{H^+}=500$ GeV) is taken. For dashed lines, the reference point 2
      ($\rho_u^{tc}\rho_e^{\tau\tau}=-0.045$, $r_\tau(=|\rho_e^{\mu\tau}/\rho_e^{\tau\tau}|)=2.7$ and
      $m_{H^+}=500$ GeV) is set. Here we also assume that $m_{H/A}=m_{H^+}=500$ GeV and
      $c_{\beta\alpha}=0.001$ as a reference parameter set. Here a value of $\rho_u^{tt}$ is fixed to be $-0.1$.
      For this $\rho_u^{tt}$, BR$(H/A\rightarrow t\bar{t})$ is small (not shown in the figure).
      }
    \label{BR_H_scenario3}
  \end{center}
\end{figure}

A main difference between Scenario (1) and (3) is the existence of non-zero $\rho_e^{\mu\tau}$ coupling.
Because of non-zero $\rho_e^{\mu\tau}$, the size of $\rho_e^{\tau\tau}$ coupling can be smaller than one
for the Scenario (1) in the interesting regions for $R(D^{(*)})$.
For the LHC study in Scenario (3), we consider two reference points,
which are denoted by a black circle in Figure \ref{scenario3} for Scenario (3). 
As a reference point 1, parameters with $\rho_u^{tc}\rho_e^{\tau\tau}=-0.075$, $r_\tau(=|\rho_e^{\mu\tau}/\rho_e^{\tau\tau}|)=2.7$
and $m_{H^+}=500$ GeV are set, and as a reference point 2, those with
$\rho_u^{tc}\rho_e^{\tau\tau}=-0.045$, $r_\tau(=|\rho_e^{\mu\tau}/\rho_e^{\tau\tau}|)=2.7$ are taken.
As a reference parameter set for Scenario (3), we also assume that $m_{H/A}=m_{H^+}=500$ GeV and $c_{\beta\alpha}=0.001$.
In Figure~\ref{BR_H_scenario3}, the decay branching ratios of $H/A$ are shown as a function of $\rho_u^{tc}$.
Here we have taken the reference parameter set with the reference point 1 (solid lines)
and the reference point 2 (dashed lines) as mentioned above. A value of $\rho_u^{tt}$ is fixed to be $-0.1$.
For this $\rho_u^{tt}$, BR$(H/A\rightarrow t\bar{t})$ is small (not shown in the figure).
As we can see from the figure, BR$(H/A\rightarrow \tau^+\tau^-)$ can be much smaller than one in the Scenario (1), on the other
hand, BR$(H/A\rightarrow \mu^\pm\tau^\mp)$ can be of order one.
Since absolute values of $\rho_e^{\tau\tau}$ and $\rho_e^{\mu\tau}$ in the reference point 1 ($|\rho_e^{\tau\tau}|=0.075/\rho_u^{tc}$)
are larger than those in the reference point 2 ($|\rho_e^{\tau\tau}|=0.045/\rho_u^{tc}$) with the fixed value of $\rho_u^{tc}$,
BR$(H/A\rightarrow \tau^+\tau^-)$ and BR$(H/A\rightarrow \mu^\pm\tau^\mp)$ (BR$(H/A\rightarrow t\bar{c}+c\bar{t})$) in the reference point 1
are larger (smaller) than those in the reference point 2.

\begin{figure}[h]
  \begin{center}
    \includegraphics[width=0.49\textwidth]{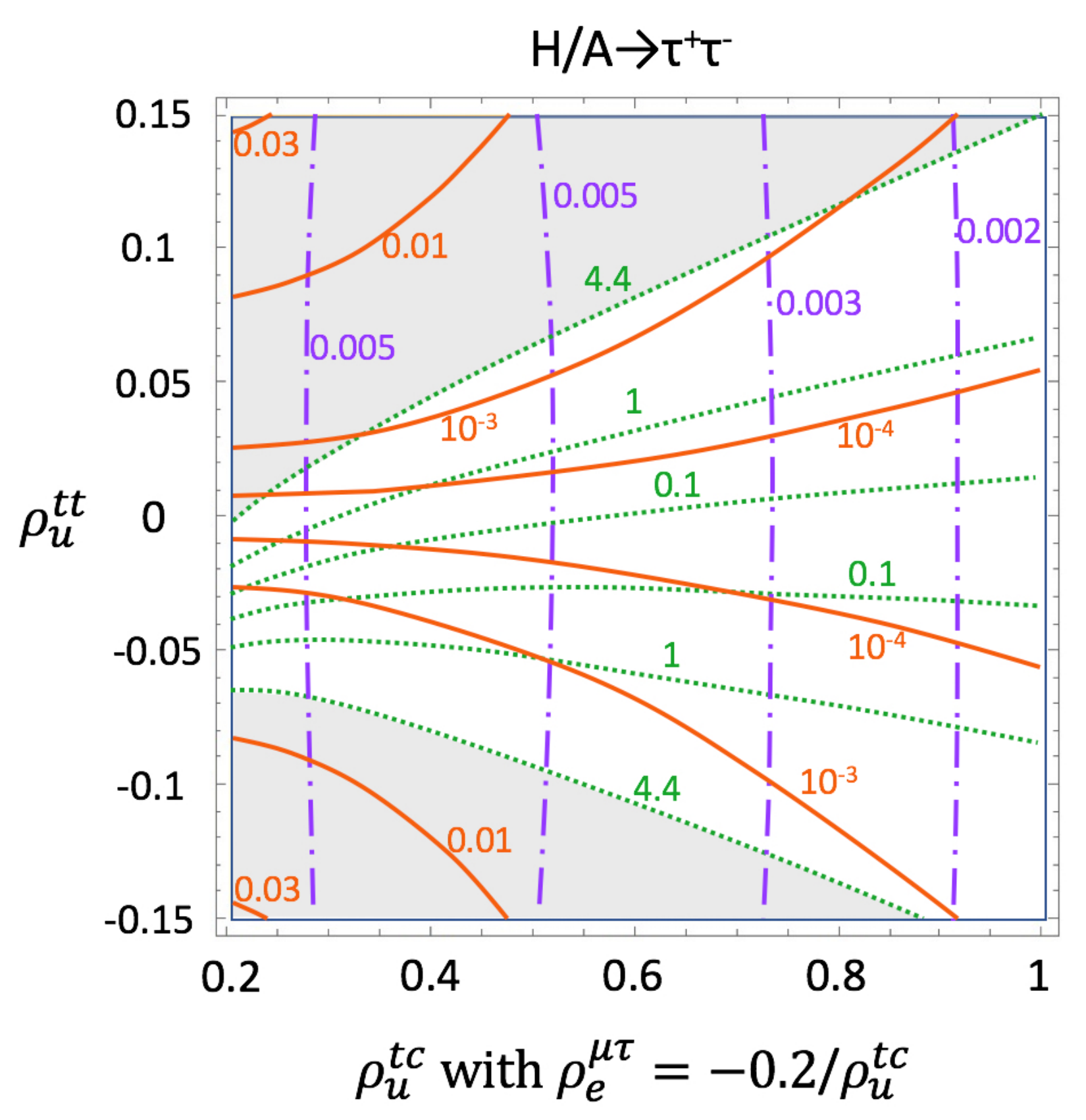}
    \includegraphics[width=0.49\textwidth]{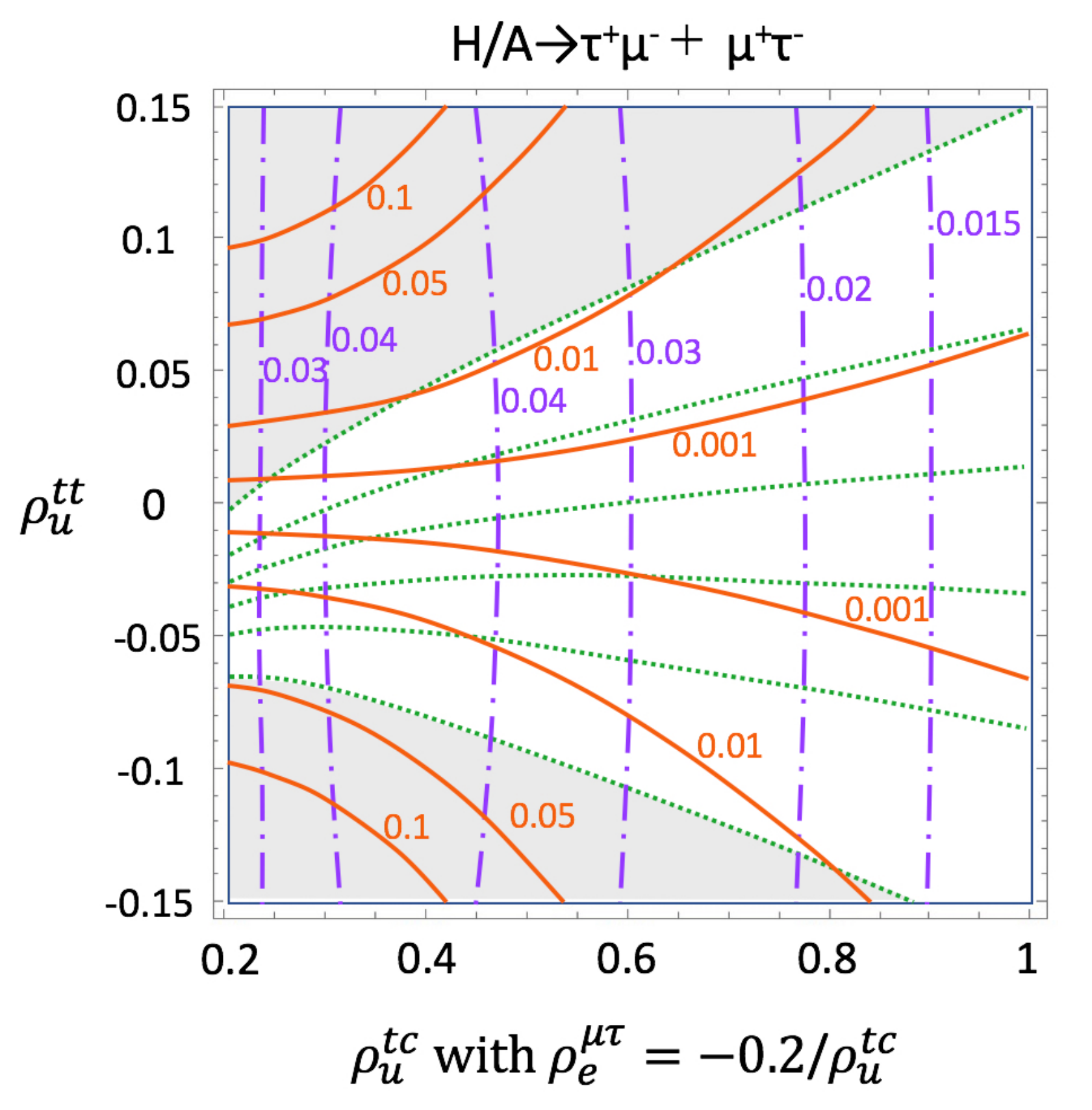}
    \includegraphics[width=0.49\textwidth]{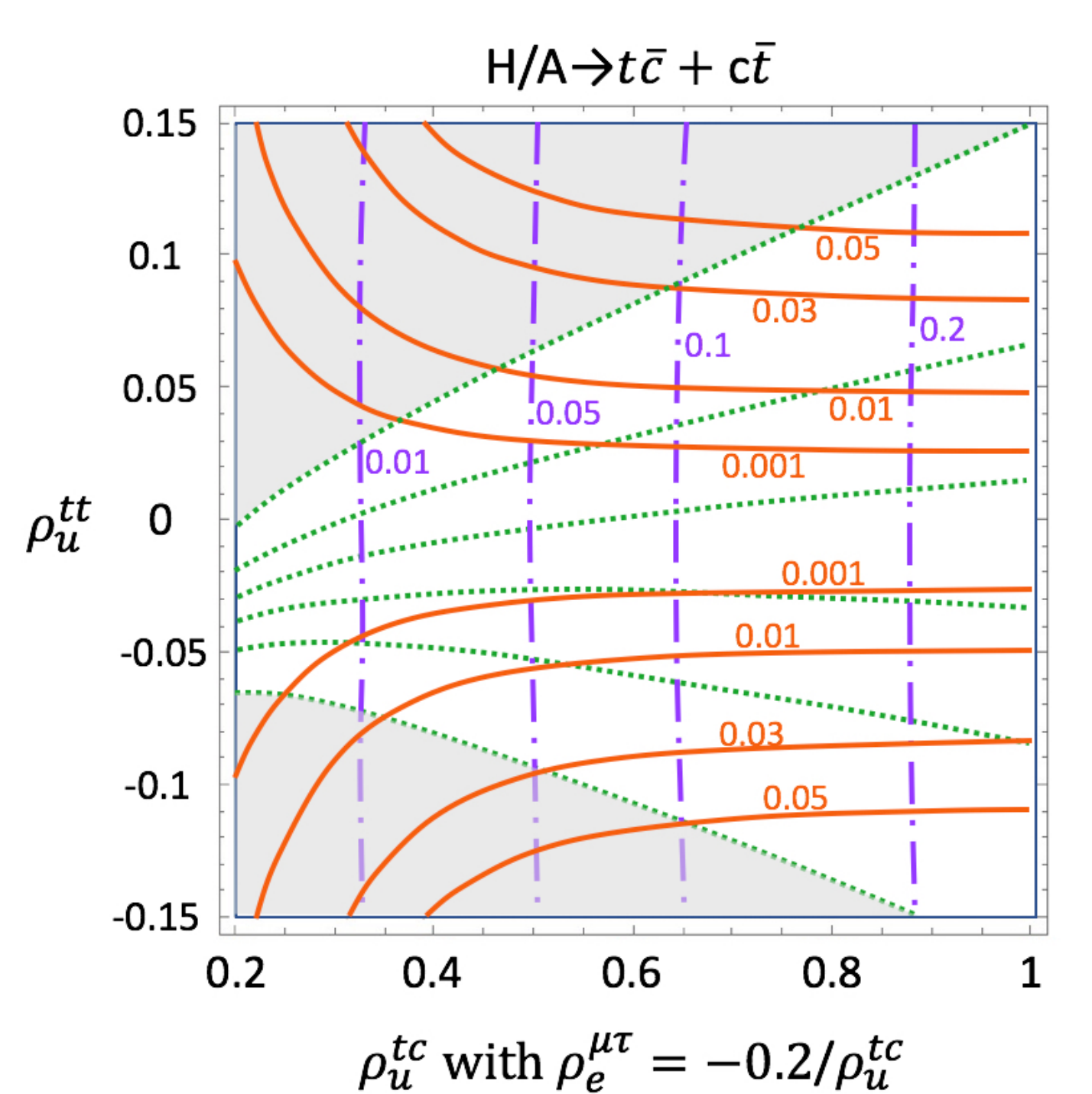}        
    \caption{The cross section $\sigma(pp\rightarrow H/A) \times {\rm BR}(H/A)$ [pb]
      via the gluon-gluon fusion production (solid orange lines) and the cross section
      $\sigma(pp \rightarrow t+ H/A) \times {\rm BR}(H/A)$ [pb] via $c g\rightarrow t+ H/A$ process
      (dashed-dotted purple lines) at $\sqrt{s}=13$ TeV
      are shown as a function of $\rho_u^{tc}$ and $\rho_u^{tt}$
      in cases with ${\rm BR}(H/A)={\rm BR}(H/A\rightarrow \tau^+\tau^-)$ [{\bf upper left figure}],
      ${\rm BR}(H/A)={\rm BR}(H/A\rightarrow \mu^\pm\tau^\mp)$ [{\bf upper right figure}],
      and ${\rm BR}(H/A)={\rm BR}(H/A\rightarrow t\bar{c}+c\bar{t})$ [{\bf lower figure}].      
      Here we have taken the reference parameter set with the reference point 1, mentioned in
      Figure~\ref{BR_H_scenario3}. The dotted green lines show the branching ratio of $\tau\rightarrow\mu \gamma$
      (in a unit of $10^{-8}$).
      The shaded regions are excluded by the current limit of $\tau\rightarrow \mu\gamma$
      [BR$(\tau\rightarrow \mu\gamma)\le4.4\times 10^{-8})$].
    }
    \label{H_pro_scenario3}
  \end{center}
\end{figure}

In Figure~\ref{H_pro_scenario3}, the cross section $\sigma(pp\rightarrow H/A) \times {\rm BR}(H/A)$ [pb]
via the gluon-gluon fusion production (solid orange lines) and the cross section
$\sigma(pp \rightarrow t+ H/A) \times {\rm BR}(H/A)$ [pb] via $c g\rightarrow t+ H/A$ process
(dashed-dotted purple lines) at $\sqrt{s}=13$ TeV
are shown as a function of $\rho_u^{tc}$ and $\rho_u^{tt}$
in cases with ${\rm BR}(H/A)={\rm BR}(H/A\rightarrow \tau^+\tau^-)$ (upper left figure),
${\rm BR}(H/A)={\rm BR}(H/A\rightarrow \mu^\pm\tau^\mp)$ (upper right figure),
and ${\rm BR}(H/A)={\rm BR}(H/A\rightarrow t\bar{c}+c\bar{t})$ (lower figure).
Here we have taken the reference parameter set with the reference point 1, mentioned in
Figure~\ref{BR_H_scenario3}.
The dotted green lines show the branching ratio of $\tau\rightarrow\mu \gamma$
(in a unit of $10^{-8}$).
The shaded regions are excluded by the current limit of $\tau\rightarrow \mu\gamma$
[BR$(\tau\rightarrow \mu\gamma)\le4.4\times 10^{-8}$].
In $H/A\rightarrow \tau^+\tau^-$ mode (upper left figure of Figure~\ref{H_pro_scenario3}), the
cross sections are much smaller compared to those in Scenario (1) because of the smaller
decay branching ratio. In $H/A\rightarrow \mu^\pm \tau^\mp$ mode (upper right figure of
Figure~\ref{H_pro_scenario3}), the cross section can be as large as $O(0.01)$ pb
especially in $c g\rightarrow t+ H/A$ production mode. On the other hand,
the cross sections of ${\rm BR}(H/A\rightarrow t\bar{c}+c\bar{t})$ mode can be as large
as those in a case of Scenario (1). For the reference point 2 with the fixed values of
$\rho_u^{tc,tt}$, the cross sections
$\sigma(pp\rightarrow H/A) \times {\rm BR}(H/A\rightarrow \mu^\pm\tau^\mp)$ and
$\sigma(pp \rightarrow t+ H/A) \times {\rm BR}(H/A \rightarrow \mu^\pm\tau^\mp)$
are smaller, on the other hand,
the cross sections $\sigma(pp\rightarrow H/A) \times {\rm BR}(H/A\rightarrow t\bar{c}+c\bar{t})$ and
$\sigma(pp \rightarrow t+ H/A) \times {\rm BR}(H/A \rightarrow t\bar{c}+c\bar{t})$
are larger than those for the reference point 1, because of ${\rm BR}(H/A)$ shown in
Figure~\ref{BR_H_scenario3}.

Therefore, again, the searches for the exotic productions and exotic decay modes of the heavy neutral
Higgs bosons would be important to probe the interesting parameter regions for $R(D^{(*)})$.

\clearpage

\subsection{Productions for charged Higgs boson}
\begin{figure}[ht]
  \begin{center}
    \includegraphics[width=0.5\textwidth]{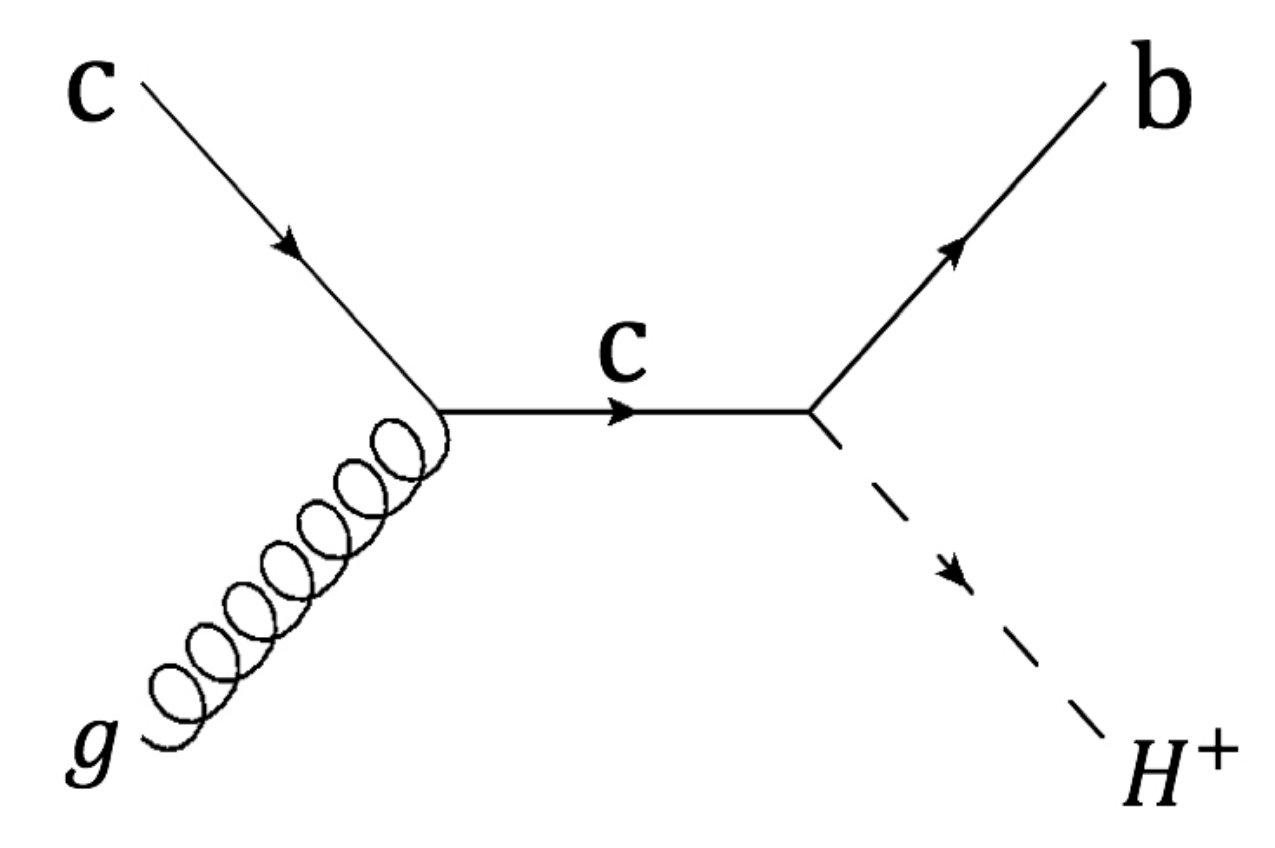}
    \caption{A Feynman diagram for charged Higgs boson production via $cg\rightarrow b+H^+$ process.}
    \label{Feynman_ch_production}
  \end{center}
\end{figure}

In the presence of non-zero $\rho_u^{tc}$ and $\rho_e^{\tau\tau}$, the charged Higgs boson production via
$cg\rightarrow b H^+$ shown in Figure~\ref{Feynman_ch_production}, decaying into $\tau^+\nu$ $(H^+\rightarrow \tau^+\nu)$
would be very important. As pointed out in Ref.~\cite{Altmannshofer2017poe}, this process is directly related to
$\bar{B}\rightarrow D^{(*)}\tau^-\bar{\nu}$ process, and hence the search for this charged Higgs boson production and decay mode
would be a crucial probe for the anomaly of $R(D^{(*)})$.
The production cross section via $cg\rightarrow b H^+$ is calculated by the calchep~\cite{Belyaev2012qa} with CTEQ leading order PDF,
\begin{align}
\sigma(pp\rightarrow bH^++\bar{b}H^-)=19~|\rho_u^{tc}|^2~[{\rm pb}].
\end{align}
Therefore the production cross section would be significantly large.

\begin{figure}[ht]
  \begin{center}
    \includegraphics[width=0.7\textwidth]{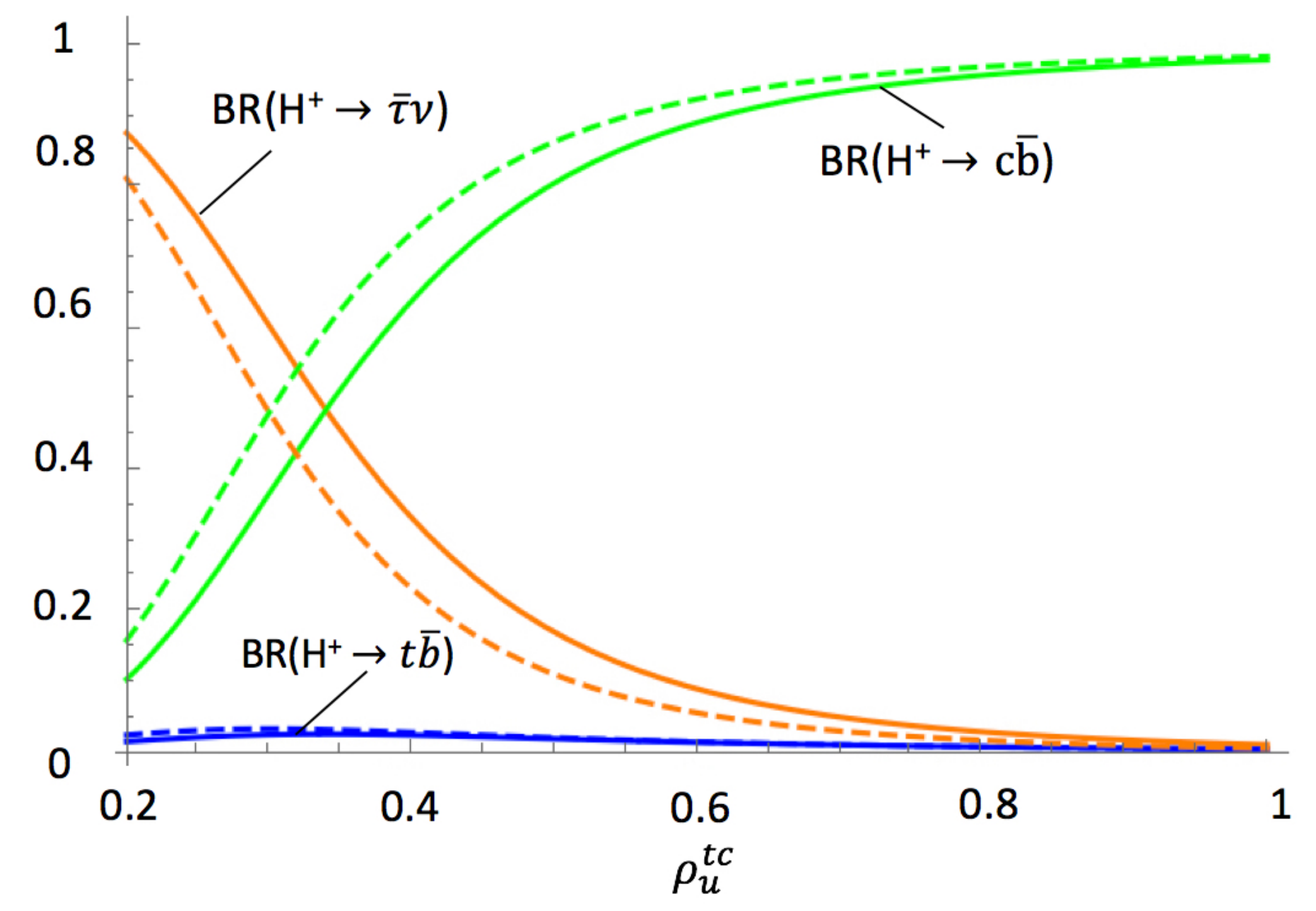}
    \caption{Decay branching ratios for charged Higgs boson $H^+$ are shown as a function of
      $\rho_u^{tc}$. Here we have taken a reference parameter set with the reference point 1 (solid lines)
      and the reference point 2 (dashed lines) for Scenario (1) as shown in
      Figure~\ref{neutral_decayBR}. We also assumed $\rho_u^{tt}=0.1$.}
    \label{decay_chargedH}
  \end{center}
\end{figure}

In Figure~\ref{decay_chargedH}, decay branching ratios for charged Higgs boson $H^+$ are shown
as a function of $\rho_u^{tc}$. Here we have assumed a reference parameter set with the reference point 1 (solid lines)
and the reference point 2 (dashed lines) for Scenario (1) as mentioned
in Figure~\ref{neutral_decayBR}. Here the coupling $\rho_u^{tt}$ is also fixed to be $0.1$.
We note that since the $\rho_e^{\tau\tau}$ coupling in the reference point 1 for Scenario (1) is larger
than one for the reference point 2, BR$(H^+\rightarrow \tau^+\bar{\nu})$ (BR$(H^+\rightarrow c\bar{b},t\bar{b})$)
for the reference point 1 is larger (smaller) than one for the reference point 2.

We also note that in Scenario (3), in the presence of $\rho_e^{\mu\tau}$ coupling, the charged Higgs boson
can decay via the $\rho_e^{\mu\tau}$ coupling. The decay product, however, is $\tau$ and $\nu_\mu$, where $\nu_\mu$
can not be observed. Therefore, the decay branching ratio for BR$(H^+\rightarrow \tau^+\nu)$ (where a summation
of neutrino flavor is taken into account) does not significantly change, compared to one in Scenario (1),
and hence the result for the charged Higgs boson decay branching ratios in Scenario (1) is almost hold in Scenario (3).

\begin{figure}[ht]
  \begin{center}
    \includegraphics[width=0.6\textwidth]{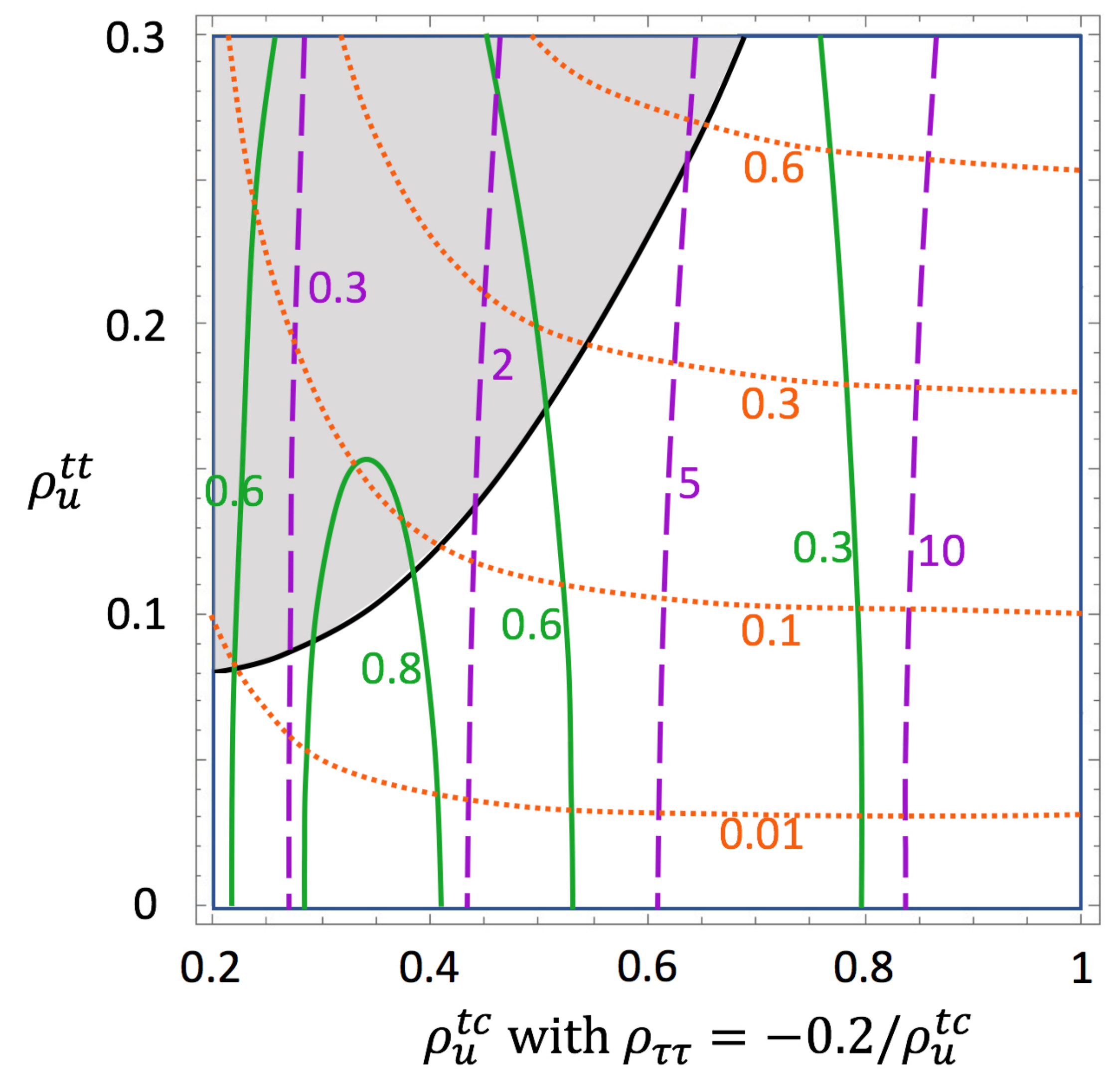}
    \caption{The cross sections $\sigma(pp\rightarrow b H^++\bar{b}H^-)\times {\rm BR}(H^\pm)$
      (via $cg \rightarrow b H^+$ process) at $\sqrt{s}=13$ TeV
      are shown as a function of $\rho_u^{tc}$ and $\rho_u^{tt}$. The size of cross sections $\sigma\times {\rm BR}$
      with ${\rm BR}(H^\pm)={\rm BR}(H^\pm \rightarrow \tau^\pm \nu),~{\rm BR}(H^\pm \rightarrow c\bar{b},~b\bar{c})$
      and ${\rm BR}(H^\pm \rightarrow t\bar{b},~b\bar{t})$ are shown in solid green lines, 
      dashed purple lines and dotted orange lines, respectively.
      Here the same parameter set is taken as in Figure~\ref{cross_tautau}.}
    \label{production_taunu_chargedH}
  \end{center}
\end{figure}

We show the cross sections $\sigma(pp\rightarrow b H^++\bar{b} H^-)\times {\rm BR}(H^\pm)$ [pb] at $\sqrt{s}=13$ TeV
as a function of $\rho_u^{tc}$ and $\rho_u^{tt}$ in Figure~\ref{production_taunu_chargedH}.
In Figure~\ref{production_taunu_chargedH}, the size of cross sections $\sigma\times {\rm BR}$
with ${\rm BR}(H^\pm)={\rm BR}(H^\pm \rightarrow \tau^\pm \nu)$,~${\rm BR}(H^\pm \rightarrow c\bar{b},~b\bar{c})$
and ${\rm BR}(H^\pm \rightarrow t\bar{b},~b\bar{t})$ are shown in solid green lines, dashed purple lines and
dotted orange lines, respectively.
Here the same parameter set is taken as in Figure~\ref{cross_tautau}.
For the reference point 2 in Scenario (1), as seen in Figure~\ref{decay_chargedH}, the cross sections
$\sigma(pp\rightarrow b H^++\bar{b} H^-)\times {\rm BR}(H^\pm)$ with ${\rm BR}(H^\pm)={\rm BR}(H^\pm \rightarrow \tau^\pm \nu)$
are slightly smaller, on the other hand,
$\sigma(pp\rightarrow b H^++\bar{b} H^-)\times {\rm BR}(H^\pm)$ with ${\rm BR}(H^\pm)={\rm BR}(H^\pm \rightarrow c\bar{b},~b\bar{c})$
and ${\rm BR}(H^\pm \rightarrow t\bar{b},~b\bar{t})$ are slightly larger than those for the reference point 1 with fixed values of
$\rho_u^{tc,tt}$.
As one can see from the figure, the cross sections are significantly large in the interesting regions for $R(D^{(*)})$.

As suggested in Ref.~\cite{Altmannshofer2017poe}, for $pp\rightarrow b H^+ \rightarrow b\tau^+ \nu$ process,
the main SM background comes from $pp\rightarrow j W^+ \rightarrow j\tau^+\nu$ where $j$ stands for a light quark
or a gluon jet misidentified as a $b$-quark jet, and another background process
is $pp\rightarrow b W^+\rightarrow b\tau^+\nu$. Although the cross sections of these background processes
are also large, the $p_T$ distribution of $\tau$ for the signal process would be significantly different
from those for the background events as discussed in Ref.~\cite{Altmannshofer2017poe}.
\begin{figure}[hb]
  \begin{center}
    \includegraphics[width=0.8\textwidth]{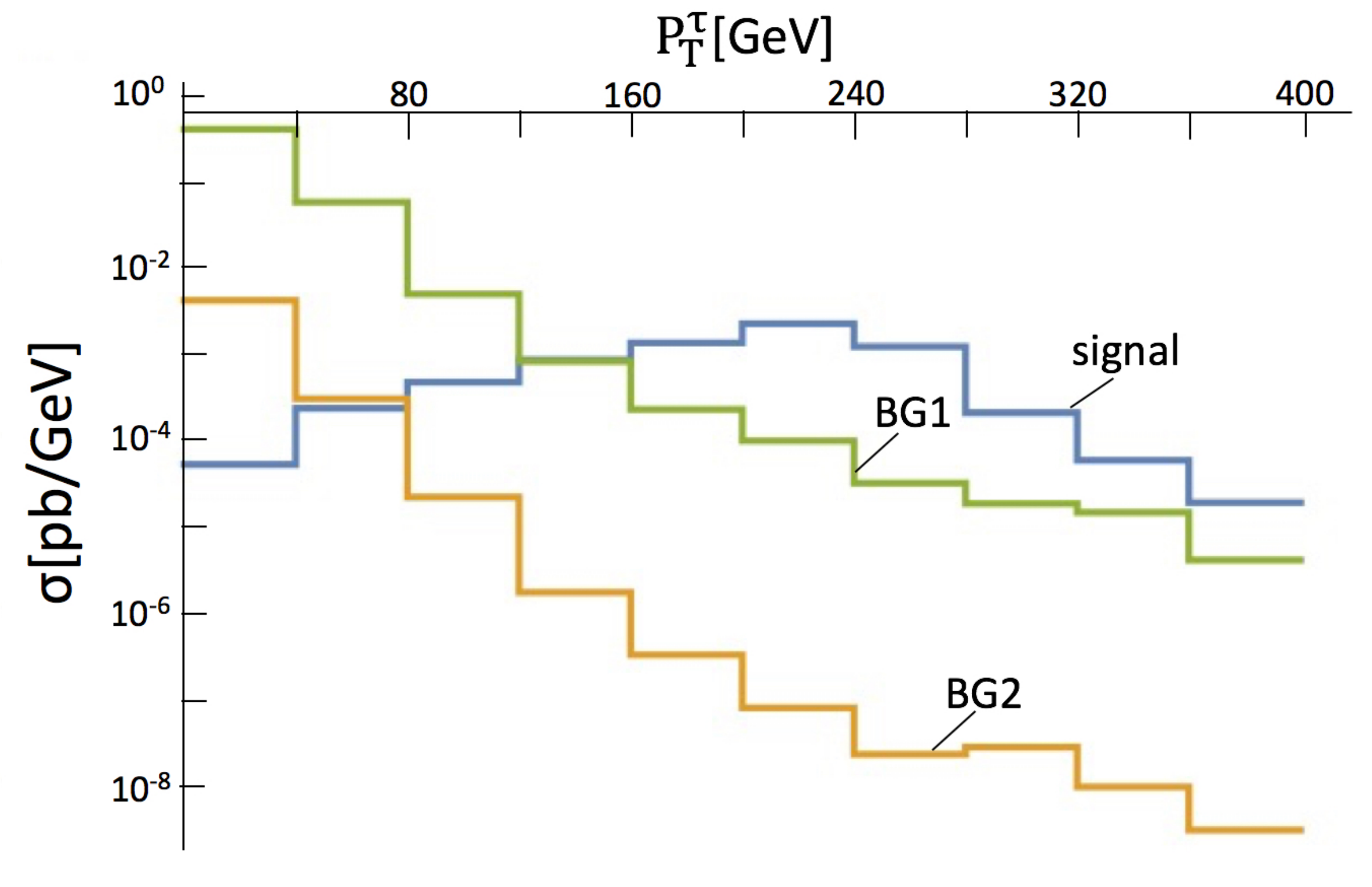}
    \caption{$p_T$ distributions of $\tau$ for the signal process $pp\rightarrow bH^+\rightarrow b\tau^+\nu$ (signal)
      and a background process $pp\rightarrow j W^+\rightarrow j \tau^+\nu$ where $j$ stands for a light quark (or a gluon)
      jet misidentified as a b-quark jet (BG1) and another background process $pp\rightarrow b W^+\rightarrow b\tau^+\nu$
      (BG2). Here we have assumed the light-parton misidentification probability is $1.5\%$ as studied in
      Ref.~\cite{Altmannshofer2017poe}, and we have imposed $p_T>20$ GeV for all $j$, $b$ and $\tau$.
      We have taken $m_{H^+}=500$ GeV, $\rho_u^{tc}=0.3$ and $\rho_{\tau\tau}=-0.67$ (the reference point 1 for Scenario (1)).}
    \label{pt_dist}
  \end{center}
\end{figure}
In Figure~\ref{pt_dist}, we show $p_T$ distributions of $\tau$ for the signal process $pp\rightarrow bH^+\rightarrow b\tau^+\nu$ (signal)
and a background process $pp\rightarrow j W^+\rightarrow j \tau^+\nu$ (BG1) and
another background process $pp\rightarrow b W^+\rightarrow b\tau^+\nu$ (BG2).
For BG1, we have shown numbers of events multiplied by $0.015$ because it is motivated by
the fact that the light-parton misidentification probability is $1.5\%$ as studied in
Ref.~\cite{Altmannshofer2017poe}. We have imposed $p_T>20$ GeV for all $j$, $b$ and $\tau$ as motivated by 
experimental studies.
We have taken $m_{H^+}=500$ GeV, $\rho_u^{tc}=0.3$ and $\rho_{\tau\tau}=-0.67$ (the reference point 1 for Scenario (1)).
As we can see from the figure, the $p_T$ distribution
for the signal events would be significantly different from those for the background events. Therefore, we expect
that even current data of the LHC has a potential to discriminate the signal events from the backgrounds.

\begin{figure}[ht]
  \begin{center}
    \includegraphics[width=0.49\textwidth]{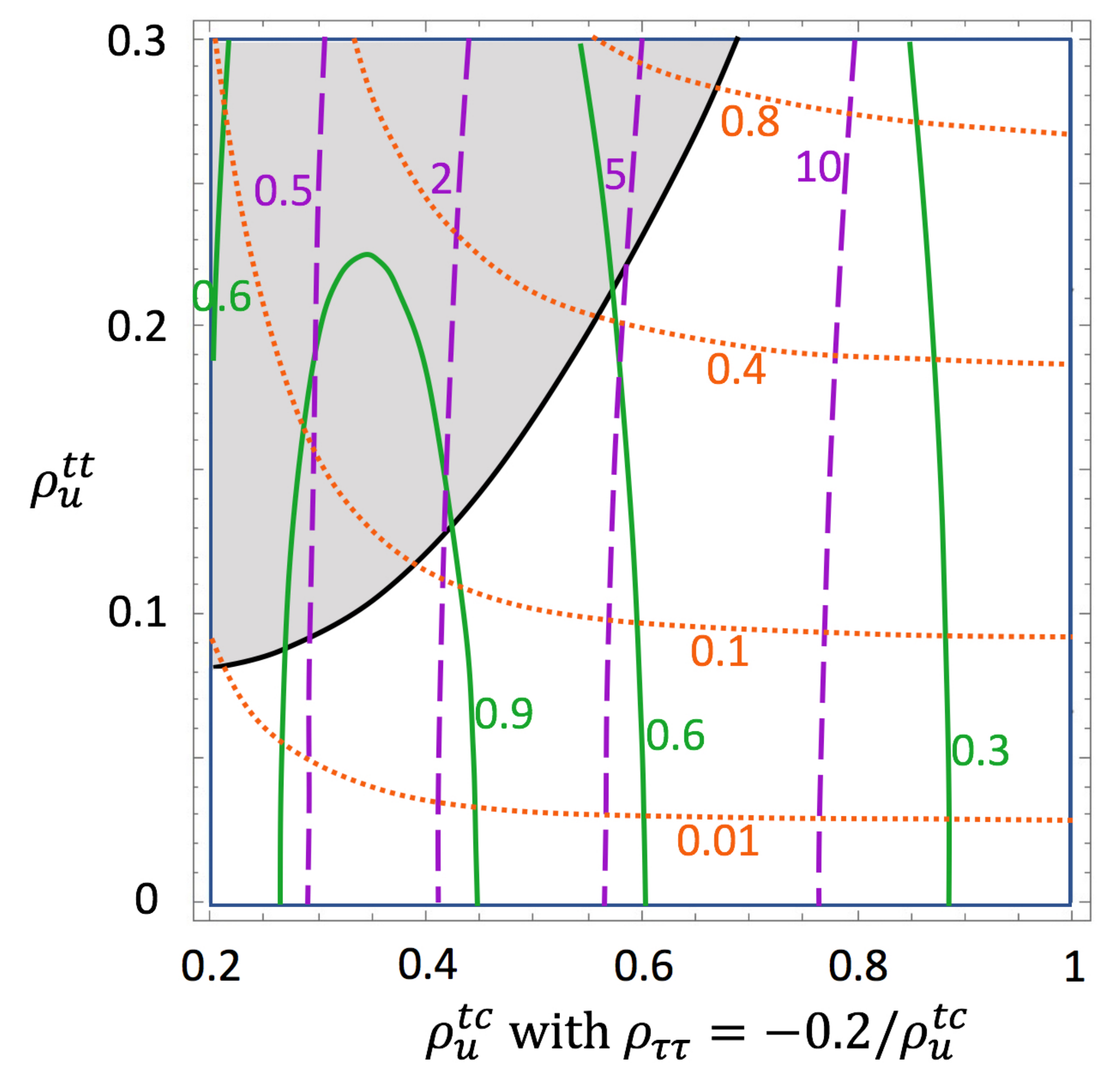}
    \includegraphics[width=0.48\textwidth]{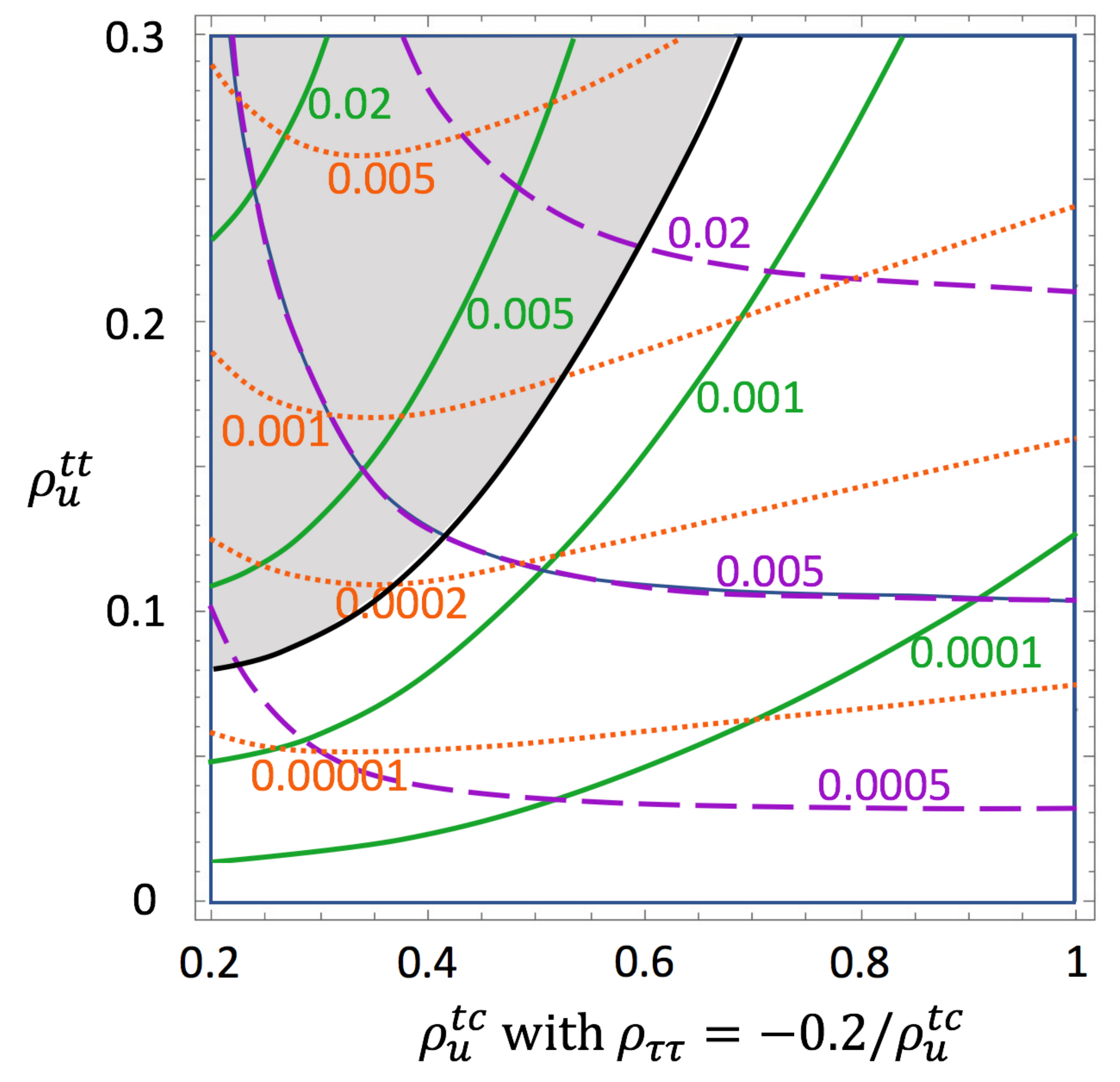}    
    \caption{[{\bf Left}]The cross sections $\sigma(pp\rightarrow c H^-+\bar{c}H^+)\times {\rm BR}(H^\pm)$
      (via $bg\rightarrow c H^-$ process) and [{\bf Right}]
      $\sigma(pp\rightarrow t H^-+\bar{t}H^+)\times {\rm BR}(H^\pm)$
      (via $bg\rightarrow t H^-$ process) at $\sqrt{s}=13$ TeV are shown as a function of
      $\rho_u^{tc}$ and $\rho_u^{tt}$ in cases with
      ${\rm BR}(H^\pm)={\rm BR}(H^\pm\rightarrow \tau^\pm \nu)$ (solid green lines),
      ${\rm BR}(H^\pm)={\rm BR}(H^\pm\rightarrow c\bar{b}, b\bar{c})$ (dashed purple lines)
      and ${\rm BR}(H^\pm)={\rm BR}(H^\pm\rightarrow t\bar{b}, b\bar{t})$ (dotted orange lines), respectively.
      Here the same parameter set is taken as in Figure~\ref{cross_tautau}.}
    \label{production_other_chargedH}
  \end{center}
\end{figure}
In addition to the $cg\rightarrow bH^+$ production process, the production processes $bg\rightarrow c H^-$
and $bg\rightarrow t H^-$ would be also important when we have non-zero $\rho_u^{tc}$ and $\rho_u^{tt}$.
The production cross sections via these processes are computed by the calchep~\cite{Belyaev2012qa},
\begin{align}
  \sigma(pp\rightarrow c H^-+\bar{c}H^+)&=15~|\rho_u^{tc}|^2~[{\rm pb}],\\
  \sigma(pp\rightarrow t H^-+\bar{t}H^+)&=0.46~|\rho_u^{tt}|^2~[{\rm pb}].
\end{align}

In Figure~\ref{production_other_chargedH}, we show the cross sections 
$\sigma(pp\rightarrow c H^-+\bar{c}H^+)\times {\rm BR}(H^\pm)$ [pb] (via $bg\rightarrow c H^-$ process) [Left] and
$\sigma(pp\rightarrow t H^-+\bar{t}H^+)\times {\rm BR}(H^\pm)$ [pb] (via $bg\rightarrow t H^-$ process) [Right]
at $\sqrt{s}=13$ TeV as a function of
$\rho_u^{tc}$ and $\rho_u^{tt}$ in cases with
${\rm BR}(H^\pm)={\rm BR}(H^\pm\rightarrow \tau^\pm \nu)$ (solid green lines),
${\rm BR}(H^\pm)={\rm BR}(H^\pm\rightarrow c\bar{b}, b\bar{c})$ (dashed purple lines)
and ${\rm BR}(H^\pm)={\rm BR}(H^\pm\rightarrow t\bar{b}, b\bar{t})$ (dotted orange lines), respectively.
Here the same parameter set is taken as in Figure~\ref{cross_tautau}.
Although the current LHC experiment has looked for the production via $bg\rightarrow tH^-$
process~\cite{ATLAStaunu7,CMSStaunu7,ATLAStaunu13,CMSStaunu13},
we do not have any constraints from this search mode because the cross section is not
so large in the general 2HDM we have discussed here. On the other hand, the cross sections
via $bg\rightarrow cH^-$ process are also significantly large, compared with one
via $cg\rightarrow bH^-$ process.

The charged Higgs boson production via $c\bar{b}\rightarrow H^+$ would also be significant
when $\rho_u^{tc}$ is large. The production cross section via $c\bar{b}\rightarrow H^+$
at $\sqrt{s}=13$ TeV are calculated by the calchep~\cite{Belyaev2012qa},
\begin{align}
\sigma(pp\rightarrow H^\pm)=36~(16)~|\rho_u^{tc}|^2~[{\rm pb}],
\label{ch_prod}
\end{align}
for $m_{H^+}=500$ (600) GeV. Especially when $H^\pm \rightarrow c\bar{b},b\bar{c}$, this production mode
receives constraints from searches for dijet resonances~\cite{CMS:2017xrr}. The constraints
are only available for a resonance mass larger than 600 GeV, and the CMS limit at the $\sqrt{s}=13$ TeV
using $27~{\rm fb}^{-1}$ data is obtained by
\begin{align}
\sigma\times {\rm BR}\times A\le 10~[{\rm pb}],
\end{align}
for the resonance mass to be 600 GeV. Here $A$ is the acceptance for narrow resonances with
the kinematic requirements $|\Delta \eta_{ij}|<1.3$ and $|\eta|<2.5$ and we take $A=0.6$
as suggested by the CMS collaboration~\cite{CMS:2017xrr}. From Eq.~(\ref{ch_prod}), the dijet production cross
section in our 2HDM is given by
\begin{align}
\sigma(pp\rightarrow H^\pm)\times {\rm BR}(H^\pm \rightarrow c\bar{b}+b\bar{c})\times A\le 10~
~|\rho_u^{tc}|^2~[{\rm pb}],
\label{dijet}
\end{align}
for $m_{H^+}=600$ GeV. Therefore, the regions with $|\rho_u^{tc}|\le 1$ are still consistent with
this limit.\footnote{Other charged Higgs boson production modes such as
$\sigma(pp \rightarrow b H^\pm)\times {\rm BR}(H^\pm\rightarrow c\bar{b},b\bar{c})$
and $\sigma(pp \rightarrow c H^\pm)\times {\rm BR}(H^\pm\rightarrow c\bar{b},b\bar{c})$
also receive a constraint from the dijet resonance searches. However, the limit is
weaker than one in Eq.~(\ref{dijet}), because of the smaller cross section.}
The improvement of the dijet resonance search limit in future would also have important impact on this scenario.

In conclusion, the searches for these exotic production
and decay processes would be crucial to probe the scenarios for the anomaly of $R(D^{(*)})$.
Here we only present interesting theoretical predictions for the production cross sections
of heavy neutral and charged Higgs bosons at $\sqrt{s}=13$ TeV at the LHC. Apparently, 
more realistic detailed analyses for the LHC physics would be important.

\clearpage

\section{Summary}
The current experimental results have indicated a discrepancy between the measured values and the SM predictions of
$R(D^{(*)})$. Since the current knowledge of the SM can not explain the discrepancy, it would be worth studying
the possibilities of the extension of the SM. In this paper, we have studied the $R(D^{(*)})$ in a general 2HDM to
clarify how large deviations from the SM predictions of $R(D^{(*)})$ are possible by taking into account various
flavor physics constraints. We found that
it is possible (impossible) to accommodate the 1$\sigma$ region of the Belle's results
if we adopt a constraint BR$(B_c^-\rightarrow \tau^-\bar{\nu})\le30\%$
(BR$(B_c^-\rightarrow \tau^-\bar{\nu})\le10\%$).
To obtain the large deviations,
the masses of the heavy Higgs bosons in the 2HDM should be less than $O(1)$ TeV if the size of
new Yukawa couplings is less than $O(1)$. Therefore, a study for a direct production of these
heavy Higgs bosons at the LHC experiment is very important.

We have studied the productions and decays of the heavy Higgs bosons at the LHC, and discussed
the constraints from the current LHC results and implications at the current and future searches
at the LHC. Especially we found that the exotic productions such as $cg\rightarrow t+H/A$ and
$cg\rightarrow b H^+$ would be significantly large, and the searches for the
exotic decay modes such as $H/A \rightarrow t\bar{c}+c\bar{t},~\mu^\pm \tau^\mp$ and
$H^+\rightarrow c\bar{b}$ as well as $H/A\rightarrow \tau^+\tau^-$ and $H^+\rightarrow \tau^+\nu$
would be quite important to probe the interesting parameter regions which generate the sizable
deviations from the SM predictions of $R(D^{(*)})$ in the general 2HDM. We have also shown that the $p_T$ distribution
of $\tau$ lepton in signal events of $pp\rightarrow b H^+\rightarrow b\tau^+\nu$ would be
significantly different from those in the background events. Therefore, we expect that even the current
data of the LHC would have a sensitivity to probe the interesting regions for $R(D^{(*)})$.
Therefore, the interplay between the flavor physics and the LHC physics would play a crucial
role to reveal the origin of the anomaly of $R(D^{(*)})$.

\section*{Note added}
After we finished this work, we were aware of new (preliminary) LHCb result on
$R_{J/\Psi}={\rm BR}(B_c\rightarrow J/\Psi \tau\bar{\nu})/{\rm BR}(B_c\rightarrow J/\Psi \mu\bar{\nu})$~\cite{BcJPexp},
which is another interesting measurement of lepton flavor universality in $b\rightarrow c l\bar{\nu}$
process. The result is $R_{J/\Psi}^{\rm LHCb}=0.71\pm 0.17\pm 0.18$ and suggests about $2\sigma$ deviation from the SM prediction
$R_{J/\Psi}^{\rm SM}=0.283\pm 0.048$~\cite{Watanabe:2017mip}.
 Our study on $R_{J/\Psi}$ and
$R_{\eta_c}={\rm BR}(B_c\rightarrow \eta_c \tau\bar{\nu})/{\rm BR}(B_c\rightarrow \eta_c \mu\bar{\nu})$
(which is possibly another measurement of lepton flavor universality in $b\rightarrow c l\bar{\nu}$
process, and the SM prediction is $R_{\eta_c}^{\rm SM}=0.31^{+0.12}_{-0.07}$~\cite{Dutta:2017xmj}) in the general 2HDM
shows that $(R_{J/\Psi},~R_{\eta_c})=(0.287,~0.37)$ at the reference point 1 and $(0.284,~0.39)$ at the
reference point 2 in our Scenario (1) and
$(R_{J/\Psi},~R_{\eta_c})=(0.290,~0.35)$ at the reference point 1 and $(0.286,~0.30)$ at the reference
point 2 in our Scenario (3), and hence in the general 2HDM, the large deviation for $R_{J/\Psi}$ seems
difficult, similar to one for $R(D^*)$, on the other hand, the deviation for $R_{\eta_c}$ might be larger.

\section*{Acknowledgments}
We would like to acknowledge Yu Nakano and Masamichi Sakurai for a collaboration at the early stage.
We also like to thank C.-P. Yuan for providing model files for calchep and useful discussions.
This work was supported in part by JSPS KAKENHI Grant Number JP16K05319.
\appendix
\section{Higgs (Georgi) Basis}
In this Appendix, we show a relation between Higgs (Georgi) basis (which we adopt in our analysis) and
general basis of Higgs boson fields in a two Higgs doublet model.

In order to break the standard model gauge symmetry SU(3)$_C\times$SU(2)$_L\times$U(1)$_Y$
to U(1)$_{\rm em}$, both neutral Higgs boson fields in the SU(2)$_L$ doublets can get vacuum expectation values (vevs)
in general. In the general basis, Yukawa interactions in quarks and leptons are written by
\begin{align}
{\cal L}_{\rm Yukawa}&=-\bar{Q}^{'i}_L \Phi_1 y_{d_1}^{ij} d_{R}^{'j}-\bar{Q}^{'i}_L \Phi_1 y_{d_2}^{ij} d_R^{'j}
-\bar{Q}_L^{'i}\tilde{\Phi}_2 y_{u_1}^{ij} u_R^{'j}-\bar{Q}_L^{'i}\tilde{\Phi}_2 y_{u_2}^{ij} u_R^{'j} \nonumber\\
&-\bar{L}_L^{'i} \Phi_1 y_{e_1}^{ij} e_R^{'j}-\bar{L}_L^{'i} \Phi_2 y_{e_2}^{ij} e_R^{'j},
\label{generalYukawa}
\end{align}
where $i,j$ denote flavor indices and the summations are taken into account.
Here $\tilde{\Phi}_a=i\sigma_2 \Phi_a^*$~($a=1,2$) where $\sigma_2$ is a Pauli matrix. We parameterize Higgs boson doublet
fields as follows:
\begin{align}
\Phi_1=\left(
\begin{array}{c}
\phi_1^+\\
\frac{v_1+\phi_1^0+iG_1^0}{\sqrt{2}}
\end{array} \right),~~~\Phi_2=\left(
\begin{array}{c}
\phi_2^+\\
\frac{v_2+\phi_2^0+iG_2^0}{\sqrt{2}}
\end{array} \right),
\end{align}
where both Higgs boson doublets get vevs. We can always perform the following transformation:
\begin{align}
\left(
\begin{array}{c}
H_1\\
H_2
\end{array}
\right)=\left(
\begin{array}{cc}
\cos\beta & \sin\beta\\
-\sin\beta &\cos\beta
\end{array}
\right)\left(
\begin{array}{c}
\Phi_1\\
\Phi_2
\end{array}
\right),
\label{HiggsTrans}
\end{align}
where we define a mixing angle $\beta$ to satisfy
\begin{align}
\tan\beta=\frac{v_2}{v_1},~~~v=\sqrt{v_1^2+v_2^2},
\end{align}
so that only one of Higgs boson doublets $(H_1)$ receives the vev $(v)$ shown in Eq.~(\ref{HiggsBasis}).
This is called Higgs (Georgi) basis.
Here $\phi_{1,2}^+,~G_{1,2}^0$,~$\phi_{1,2}^0$ in $\Phi_{1,2}$ are related to
$G^+,~H^+,~G,~A,~\phi_{1,2}$ in $H_{1,2}$ (shown in Eq.~(\ref{HiggsBasis}))
via the transformation in Eq.~(\ref{HiggsTrans}).

Under the Higgs basis, we rewrite Yukawa interactions in Eq.~(\ref{generalYukawa}) as follows:
\begin{align}
{\cal L}_{\rm Yukawa}=&-\bar{Q}_L^{'i}H_1 Y_{d_1}^{ij} d_R^{'j}-\bar{Q}_L^{'i}H_2 Y_{d_2}^{ij} d_R^{'j}
-\bar{Q}_L^{'i} \tilde{H}_1 Y_{u_1}^{ij} u_R^{'j}-\bar{Q}_L^{'i} \tilde{H}_2 Y_{u_2}^{ij} u_R^{'j}
\nonumber \\
&-\bar{L}_L^{'i} H_1 Y_{e_1}^{ij} e_R^{'j}-\bar{L}_L^{'i} H_2 Y_{e_2}^{ij} e_R^{'j},
\end{align}
where Yukawa couplings $Y_{f_{1,2}}$ are written in terms of original Yukawa couplings $y_{f_{1,2}}$
for $f=d,u,e$ in Eq.~(\ref{generalYukawa}),
\begin{align}
\left(
\begin{array}{c}
Y_{f_1}\\
Y_{f_2}
\end{array}
\right)=\left(
\begin{array}{cc}
\cos\beta &\sin\beta\\
-\sin\beta & \cos\beta
\end{array}
\right)\left(
\begin{array}{c}
y_{f_1}\\
y_{f_2}
\end{array}
\right).
\end{align}
Since only $H_1$ receives vev, masses of quarks and leptons are
induced from the Yukawa couplings $Y_{f_1}$ $(f=d,u,e)$. Therefore,
we diagonalize the Yukawa couplings
$Y_{f_1}$ by biunitary transformations
\begin{align}
\left(U_{f_L} Y_{f_1} U_{f_R}^\dagger\right)^{ij} =y_f^i\delta^{ij},
\end{align}
for $f=d,u,e$ in order to obtain the fermion's mass eigenbasis.
The mass eigenbasis for the fermions ($f_{L,R}$) is defined by
\begin{align}
f_{L,R}=U_{f_{L,R}} f_{L,R}'.
\end{align}
Note that the CKM $(V)$ and MNS $(V_{\rm MNS})$ matrices are defined by $V=U_{u_L}U_{d_L}^\dagger$
and $V_{\rm MNS}=U_{e_L}U_{\nu_L}^\dagger$ (where we assume that
$U_{\nu_L}$ diagonalizes the $3\times 3$ left-handed Majorana neutrino mass matrix in the low-energy
effective theory), respectively.
Comparing these with the Yukawa interactions in Higgs basis (Eq.~(\ref{HiggsBasis})),
the relation between Yukawa couplings in both bases are obtained by
\begin{align}
y_f &=U_{f_L}\left[ y_{f_1}\cos\beta+y_{f_2}\sin\beta\right]U_{f_R}^\dagger, \nonumber\\
\rho_f &=U_{f_L}\left[ -y_{f_1}\sin\beta+y_{f_1}\cos\beta\right]U_{f_R}^\dagger,
\label{relations}
\end{align}
where $y_f$ and $\rho_f$ ($f=d,u,e$) are defined in Higgs basis (Eq.~(\ref{HiggsBasis})) and $y_f$
is diagonalized and it is related to the fermion mass $y_f^i=\sqrt{2}m_f^i/v$, 
on the other hand, $y_{f_{1,2}}$ are defined in general basis shown in Eq.~(\ref{generalYukawa}).

We stress that any bases can be transferred to the Higgs basis. In general, if there is no symmetry to
distinguish Higgs boson doublets $\Phi_1$ and $\Phi_2$, it is difficult to
define the most natural basis without considering the theory beyond the 2HDM.
However, it would be useful to discuss the relations between the original basis and Higgs basis in some
specific cases. For well-known type I, type II,
type X (lepton-specific) and type Y (flipped) 2HDM, for example,
an extra $Z_2$ symmetry restricts the Yukawa interactions
in the model, 
\begin{itemize}
\item Type I $(y_{d_1}=y_{u_1}=y_{e_1}=0)$,
\item Type II $(y_{d_2}=y_{u_1}=y_{e_2}=0)$,
\item Type X (lepton-specific) $(y_{d_1}=y_{u_1}=y_{e_2}=0)$,
\item Type Y (flipped) $(y_{d_2}=y_{u_1}=y_{e_1}=0)$.
\end{itemize}
As a consequence, the flavor violating Yukawa couplings are not allowed. However, in many cases,
such a $Z_2$ symmetry may not be exact, and hence the symmetry breaking may induce the flavor violating
Yukawa couplings. For example, in the Type II 2HDM, the exact $Z_2$ symmetry forbids
the Yukawa couplings $y_{d_2},~y_{u_1}$ and $y_{e_2}$, as shown above, and hence these Yukawa couplings
may be induced as corrections to the $Z_2$ symmetry breaking. When we write these $Z_2$ breaking Yukawa
couplings as $\Delta y_{A}$ ($A=d_2,~u_1,~e_2$), the relations between the Yukawa couplings in the
original basis and Higgs basis (Eq.~(\ref{relations})) are written by
\begin{align}
y_f &=U_{f_L}\left[ y_{f_1}\cos\beta+\Delta y_{f_2}\sin\beta\right]U_{f_R}^\dagger, \nonumber\\
\rho_f &=U_{f_L}\left[ -y_{f_1}\sin\beta+\Delta y_{f_2}\cos\beta\right]U_{f_R}^\dagger,
\end{align}
for $f=d,~e$ and 
\begin{align}
y_f &=U_{f_L}\left[ \Delta y_{f_1}\cos\beta+y_{f_2}\sin\beta\right]U_{f_R}^\dagger, \nonumber\\
\rho_f &=U_{f_L}\left[ -\Delta y_{f_1}\sin\beta+y_{f_2}\cos\beta\right]U_{f_R}^\dagger,
\end{align}
for $f=u$. Therefore, we can express the flavor violating Yukawa coupling $\rho_f$ in the Higgs
basis in terms of the symmetry breaking Yukawa coupling $\Delta y_A$ as follows:
\begin{align}
\rho_f&=-y_f\tan\beta +U_{f_L}\Delta y_{f_2} U_{f_R}^\dagger \frac{1}{\cos\beta},\nonumber \\
&=-\frac{\sqrt{2} m_f}{v} \tan\beta +U_{f_L}\Delta y_{f_2} U_{f_R}^\dagger \frac{1}{\cos\beta},~~{\rm for}~f=d,~e, \\
\rho_f&=y_f\frac{1}{\tan\beta} -U_{f_L}\Delta y_{f_2} U_{f_R}^\dagger \frac{1}{\sin\beta},\nonumber \\
&=\frac{\sqrt{2} m_f}{v}\frac{1}{\tan\beta} -U_{f_L}\Delta y_{f_2} U_{f_R}^\dagger \frac{1}{\sin\beta},~~{\rm for}~f=u,
\end{align}
where $y_f$ is diagonalized in the Higgs basis and it is related to the fermion mass $y_f=\sqrt{2}m_f/v$.
One can see that if the $Z_2$ symmetry is exact $(\Delta y_A=0)$, the Yukawa couplings $\rho_f$ are
diagonal and expressed by the fermion mass and $\tan\beta$, and hence the flavor violating components
in $\rho_f$ are induced by the $Z_2$ symmetry breaking Yukawa couplings $\Delta y_A$ as well as flavor
structures of the Yukawa couplings $y_{f_{1,2}}$ in the original basis (in other words, the flavor structures of $U_{f_{L,R}}$).
If the mixing matrices $U_{f_{L,R}}$ are very close to the unit matrix, the flavor violating Yukawa couplings
$\rho_f$ in the Higgs basis may be directly related to the $Z_2$ breaking Yukawa couplings $\Delta y_A$
in the original basis. In general, however, the flavor violation in the Higgs basis is realized by the
complex combination of the flavor violations in the original basis.

In our analysis for $R(D^{(*)})$, we adopt the Higgs basis because the flavor violating charged Higgs
interactions are simply parameterized by the flavor violating Yukawa couplings $\rho_f$ in the Higgs
basis. In order to induce the large deviation in $R(D^{(*)})$ effectively, we consider some simple
flavor violation in the Higgs basis. On the other hand, if we consider the simple flavor
violation in the original basis, that may induce the very complex flavor violation in the Higgs basis,
and hence that induces not only the flavor violation for $R(D^{(*)})$ but also other flavor violation
which may generate strong constraints from other processes. In that sense, we expect that our approach is
conservative to see the possibly large effects on $R(D^{(*)})$. Therefore, in our analysis,
using the Higgs basis, we try to clarify how large deviations are possible within the framework
of the 2HDM in general.

\section{Hadronic matrix elements for $\bar{B}\rightarrow D^{(*)}l\bar{\nu}$}
In this Appendix, we summarize hadronic matrix elements for $\bar{B}\rightarrow D^{(*)}l\bar{\nu}$
which we use in our numerical analysis. The formula we use is
taken from Refs.~\cite{Caprini1997mu,FormfactorsforRDRDs}.\footnote{See also Ref.~\cite{Ivanov:2016qtw}.}
The hadronic matrix elements relevant to $\bar{B} \rightarrow D l^-\bar{\nu}$ in the 2HDM are written as
\begin{align}
  \langle D(p_D)| \bar{c}\gamma^\mu b | \bar{B}(p_B)\rangle & =
  \left(p_B^\mu+p_D^\mu -\frac{m_B^2-m_D^2}{q^2} q^\mu \right) f_+(q^2)+\frac{m_B^2-m_D^2}{q^2}q^\mu f_0(q^2), \nonumber \\
  \langle D(p_D)| \bar{c} b |\bar{B}(p_B)\rangle & =
  \frac{m_B^2-m_D^2}{m_b-m_c} f_0(q^2).\nonumber 
\end{align}
where $p_B$ and $p_D$ ($m_B$ and $m_D$) are momenta (masses) of B and D mesons, respectively, and
$q$ is a momentum transfer $q=p_B-p_D$ ($m_l^2\le q^2 \le (m_B-m_D)^2$), and $m_b$ and $m_c$ are $b$ and $c$ quark masses,
respectively. The relevant hadronic matrix elements for $\bar{B}\rightarrow D^* l^-\bar{\nu}$ are given by
\begin{align}
\langle D^*(p_{D^*},\epsilon)|\bar{c}\gamma_\mu b | \bar{B}(p_B)\rangle &=-i \epsilon_{\mu\nu\rho\sigma} \epsilon^{\nu *}
p_B^\rho p_{D^*}^\sigma \frac{2V(q^2)}{m_B+m_{D^*}}, \nonumber\\
\langle D^*(p_{D^*},\epsilon)|\bar{c} \gamma_\mu \gamma_5 b | \bar{B}(p_B)\rangle
& = \epsilon^*_{\mu} (m_B+m_{D^*}) A_1(q^2)\nonumber \\
-(p_B+p_{D^*})_\mu (\epsilon^* \cdot q)&\frac{A_2(q^2)}{m_B+m_{D^*}} 
-q_\mu (\epsilon^*\cdot q)\frac{2m_{D^*}}{q^2}\left\{A_3(q^2)-A_0(q^2)\right\},\nonumber \\
  \langle D^*(p_{D^*},\epsilon)|\bar{c}\gamma_5 b |\bar{B}(p_B)\rangle
  &=-\frac{1}{m_b+m_c} q_\mu \langle D^*(p_{D^*},\epsilon)|\bar{c}\gamma^\mu \gamma_5 b|
  \bar{B}(p_B)\rangle,\nonumber 
\end{align}
where
\begin{align}
A_3(q^2)=\frac{m_B+m_{D^*}}{2m_{D^*}} A_1(q^2)-\frac{m_B-m_{D^*}}{2m_{D^*}} A_2(q^2).\nonumber
\end{align}
Here $p_{D^*}$ and $m_{D^*}$ are momentum and mass of $D^*$, respectively, and $q$ is a momentum transfer
$q=p_B-p_{D^*}$.

Here the form factors $f_0$, $f_+$, $V$ and $A_i$ $(i=1-3)$ are given by
\begin{align}
f_0(q^2)&=\frac{1}{2\sqrt{m_Bm_D}} \left[ \frac{(m_B+m_D)^2-q^2}{m_B+m_D}h_+(q^2)-\frac{(m_B-m_{D})^2-q^2}{m_B-m_D}h_-(q^2)\right], \nonumber \\
f_+(q^2) &=\frac{1}{2\sqrt{m_Bm_D}}\left[(m_B+m_D)h_+(q^2)-(m_B-m_D)h_-(q^2)\right], \nonumber \\
V(q^2) &= \frac{m_B+m_{D^*}}{2\sqrt{m_Bm_{D^*}}} h_V(q^2),\nonumber 
\end{align}
\begin{align}
A_0(q^2) &=\frac{1}{2\sqrt{m_Bm_{D^*}}}\left[\frac{(m_B+m_{D^*})^2-q^2}{2m_{D^*}}h_{A1}(q^2) \right.\nonumber \\
&\left.~~~~~~~~~~~~~~~~~~~-\frac{m_B^2-m_{D^*}^2+q^2}{2m_B}h_{A2}(q^2)-\frac{m_B^2-m_{D^*}^2-q^2}{2m_{D^*}}h_{A3}(q^2)\right],\nonumber \\
A_1(q^2) &=\frac{(m_B+m_{D^*})^2-q^2}{2\sqrt{m_Bm_{D^*}}(m_B+m_{D^*})}h_{A1}(q^2),\nonumber \\
A_2(q^2) &=\frac{(m_B+m_{D^*})}{2\sqrt{m_Bm_{D^*}}}\left[h_{3A}(q^2)+r_{D^*}h_{A2}(q^2)\right],\nonumber 
\end{align}
where $r_D^{(*)}=m_{D^{(*)}}/m_B$ and the heavy quark effective theory (HQET) form factors are given by
\begin{align}
  h_+(q^2)&=\frac{1}{2\left[1+r_D^2-2r_Dw_D(q^2)\right]}\left[-(1+r_D)^2\left\{w_D(q^2)-1\right\} V_1(q^2)\right.\nonumber \\
    &\left.\hspace{3cm}+(1-r_D)^2\left\{w_D(q^2)+1\right\}S_1(q^2)\right],\nonumber \\
  h_-(q^2) &=\frac{(1-r_D^2)\left\{w_D(q^2)+1\right\}}{2\left[1+r_D^2-2r_Dw_D(q^2)\right]}\left[S_1(q^2)-V_1(q^2)\right],\nonumber \\
h_V(q^2) &= R_{1D^{*}}(q^2)h_{A1}(q^2),\nonumber \\
h_{A2}(q^2) &=\frac{R_{2D^{*}}(q^2)-R_{3D^{*}}(q^2)}{2r_{D^{*}}}h_{A1}(q^2),\nonumber \\
h_{A3}(q^2) &=\frac{R_{2D^{*}}(q^2)+R_{3D^{*}}(q^2)}{2}h_{A1}(q^2).\nonumber
\end{align}
The $q^2$ dependence of these form factors comes through $w_{D^{(*)}}(q^2)=\frac{m_B^2+m_{D^{(*)}}^2-q^2}{2m_Bm_{D^{(*)}}}$, and 
\begin{align}
  V_1(q^2) &= V_{1}(1)\left[1-8\rho_{D}^2 Z_D(q^2)+(51\rho_{D}^2-10)Z_D^2(q^2)-(252\rho_{D}^2-84)Z_D^3(q^2)\right], \nonumber \\
S_1(q^2) &= V_1(q^2)\left[1+\Delta\left\{-0.019+0.041\left(w_D(q^2)-1\right)-0.015\left(w_D(q^2)-1\right)^2\right\}\right],\nonumber \\
 h_{A1}(q^2)&= h_{A1}(1)\left[1-8\rho_{D^{*}}^2Z_{D^{*}}(q^2)+(53\rho_{D^{*}}^2-15)Z_{D^{*}}^2(q^2)-(231\rho_{D^{*}}^2-91)Z_{D^{*}}^3(q^2)\right],\nonumber \\
R_{1D^{*}}(q^2)&=R_1(1)-0.12\left\{w_{D^{*}}(q^2)-1\right\}+0.05\left\{w_{D^{*}}(q^2)-1\right\}^2, \nonumber \\
R_{2D^{*}}(q^2)&=R_2(1)+0.11\left\{w_{D^{*}}(q^2)-1\right\}-0.06\left\{w_{D^{*}}(q^2)-1\right\}^2, \nonumber \\
R_{3D^{*}}(q^2)&=1.22-0.052\left\{w_{D^{*}}(q^2)-1\right\}+0.026\left\{w_{D^{*}}(q^2)-1\right\}^2, \nonumber \\
 Z_{D^{(*)}}(q^2) &= \frac{\sqrt{w_{D^{(*)}}(q^2)+1}-\sqrt{2}}{\sqrt{w_{D^{(*)}}(q^2)+1}+\sqrt{2}}.\nonumber
\end{align}
The numerical values for parameters $\rho_{D^{(*)}}^2$, $R_{1,2}(1)$, $\Delta$, $h_{A1}(1)$ and $V_1(1)$ we use in our numerical analysis
are listed in Appendix D.


\section{Various $b\rightarrow s(d)$ transition processes}
\subsection{$B_{d,s}-\bar{B}_{d,s}$ mixing}
In the presence of $\rho_d^{sb,db}$, $b\rightarrow s(d)$ flavor transition occurs at the tree level.
On the other hand, the Yukawa couplings $\rho_u^{t i}~(i=t,c,u)$ can induce $b\rightarrow s(d)$ transition
via a charged Higgs boson mediation at the loop level.
Such tree level and loop level flavor transitions would be strongly constrained from $B_{d,s}-\bar{B}_{d,s}$ mixing, such
as $B_{d,s}$ meson mass differences, $\Delta m_{B_{d,s}}$. The effective Lagrangian relevant to
$B_{d,s}-\bar{B}_{d,s}$ mixing is given by
\begin{align}
  {\cal L}_{\rm eff} &=
    C_{VLL}^i (\bar{d}_i\gamma^\mu P_L b)(\bar{d}_i \gamma_\mu P_L b)
    +C_{SLR}^i (\bar{d}_i P_L b)(\bar{d}_i P_R b) \nonumber \\
    &+C_{SLL}^i (\bar{d}_i P_L b)(\bar{d}_i P_L b)
    +C_{SRR}^i (\bar{d}_i P_R b)(\bar{d}_i P_R b),
\end{align}
where $i=1$ and $2$ for $B_d-\bar{B}_d$ and $B_s-\bar{B}_s$, respectively.
The contributions induced by the neutral Higgs boson mediations at the tree level are
given by
\begin{align}
  (C_{SLR}^i)^{\rm 2HDM} &=-\sum_{\phi=h,H,A}\frac{y_{\phi bd_i}^{d*} y_{\phi d_i b}}{m_{\phi}^2},\\
    (C_{SLL}^i)^{\rm 2HDM} &=-\sum_{\phi=h,H,A}\frac{(y_{\phi b d_i}^{d*})^2}{2m_\phi^2},\\
    (C_{SRR}^i)^{\rm 2HDM} &=-\sum_{\phi=h,H,A}\frac{(y_{\phi d_i b}^{d})^2}{2m_\phi^2},
\end{align}
where $y^f_{\phi ij}$ is shown in Eqs.~(\ref{yukawa}).
The charged Higgs boson generates the contribution to $C_{VLL}$ via $\rho_u$ Yukawa couplings
at the one loop level as follows:
\begin{align}
  (C_{VLL}^i)^{\rm 2HDM}&=\frac{1}{128\pi^2 m_{H^+}^2}
    \sum_{k,l} (V^\dagger \rho_u)^{d_i k} (\rho^\dagger_u V)^{lb}\left[
      (\rho_u^\dagger V)^{kb}(V^\dagger \rho_u)^{d_i l} G_1(x_k,x_l) \right.\nonumber \\
      &\left.-\frac{4g^2m_{u_k} m_{u_l} }{m_{H^+}^2} V_{k b}V^*_{l d_i}G_2(x_k,x_l,x_W)
      +\frac{g^2 m_{u_k}m_{u_l}}{m_W^2} V_{k b} V^*_{l d_i}G_3(x_k,x_l,x_W)
      \right],
\end{align}
where $x_k=m_{u_k}^2/m_{H^+}^2$ and $x_W=m_W^2/m_{H^+}^2$.
Functions $G_x~(x=1,2,3)$ are defined by
\begin{align}
  G_1(x,y) &=\frac{1}{x-y}\left[
    \frac{x^2\log x}{(1-x)^2}+\frac{1}{1-x}-
    \frac{y^2\log y}{(1-y)^2}-\frac{1}{1-y}
    \right],
\label{G1_func}
  \\
  G_2(x,y,z) &=-\frac{1}{(x-y)(1-z)}
  \left[
    \frac{x \log x}{1-x}-\frac{y\log y}{1-y}-\frac{x\log\frac{x}{z}}{z-x}
    +\frac{y\log \frac{y}{z}}{z-y}
    \right],
  \\
  G_3(x,y,z) &=-\frac{1}{x-y}
  \left[
    \frac{1}{1-z}\left(\frac{x\log x}{1-x}-\frac{y\log y}{1-y}\right)
    -\frac{z}{1-z}\left(
    \frac{x\log \frac{x}{z}}{z-x}-\frac{y\log \frac{y}{z}}{z-y}\right)\right].
\end{align}
From the effective Lagrangian, we can obtain $B_{d,s}$ mass differences, $\Delta m_{B_{d,s}}$,
\begin{align}
  \Delta m_{B_{d_i}} &=2{\rm Re}\left[\langle \bar{B}_{d_i} |(-{\cal{L}}_{\rm eff})|B_{d_i}\rangle \right],\nonumber \\
  &=-2 {\rm Re} (C_{VLL})\frac{m_{B_{d_i}} F_{B_{d_i}}^2 B_{B_{d_i}}}{3}
  -2{\rm Re}(C_{SLR})\left(\frac{1}{24}+\frac{R}{4}\right) m_{B_{d_i}} F_{B_{d_i}}^2 B_{B_{d_i}}\nonumber \\  
  &+2{\rm Re}(C_{SLL}+C_{SRR})\frac{5 R m_{B_{d_i}} F_{B_{d_i}}^2 B_{B_{d_i}}}{24},
\end{align}
where $m_{B_{d_i}}$, $F_{B_{d_i}}$ and $B_{B_{d_I}}$ are a mass, a decay constant and Bag parameter
of $B_{d_i}$ meson, respectively, and $R$ is
\begin{align}
R=\left(\frac{m_{B_s}}{m_s+m_b}\right)^2.
\end{align}
The numerical values of the decay parameters are listed in Appendix D.

\subsection{$b\rightarrow s \gamma$ and $b\rightarrow s l^+l^-$ ($l=e$ and $\mu$)}

The effective operators relevant to $b\rightarrow s \gamma$ and $b\rightarrow s l^+l^-$ are given by
\begin{align}  
{\cal L_{\rm eff}}= &\frac{4G_{\rm F}}{\sqrt{2}}V_{tb} V^*_{ts}\left[\frac{e}{16\pi^2}
    C_7 m_b (\bar{s}\sigma_{\mu\nu}P_R b) F^{\mu\nu}
    +\frac{g_3}{16\pi^2} C_8 m_b(\bar{s}\sigma_{\mu\nu} T^a P_R b) G^{a,\mu\nu} \right.\nonumber \\
    &+ \left.\frac{e^2}{16\pi^2}C_{9(l)} (\bar{s}\gamma^\mu P_L b) (\bar{l}\gamma_\mu l)
    +\frac{e^2}{16\pi^2}C_{10(l)} (\bar{s}\gamma^\mu P_L b) (\bar{l}\gamma_\mu\gamma_5 l)
    \right],
\end{align}
where $l$ is a charged lepton $l=e$ or $\mu$.
The charged Higgs boson contributions with non-zero $\rho_u$ are expressed by
\begin{align}
  C_7^{\rm 2HDM} &=\frac{1}{4\sqrt{2}G_{\rm F} m_{H^+}^2 V_{tb}V_{ts}^*}\sum_i (V^\dagger \rho_u)^{si}(\rho_u^\dagger V)^{ib}
  \left[Q_u G_{\sigma 1}(x_i)+Q_{H^+} G_{\sigma 2}(x_i)\right],\\
  C_8^{\rm 2HDM} &=\frac{1}{4\sqrt{2}G_{\rm F} m_{H^+}^2 V_{tb}V_{ts}^*}\sum_i (V^\dagger \rho_u)^{si}(\rho_u^\dagger V)^{ib} G_{\sigma 1}(x_i),\\
    C_{9(l)}^{\rm 2HDM} &=\frac{1}{2\sqrt{2} G_{\rm F}m_{H^+}^2 V_{tb}V_{ts}^*}
    Q_l \sum_i (V^\dagger \rho_u)^{si}(\rho_u^\dagger V)^{ib}
    \left[Q_u G_{\gamma 1}(x_i)+Q_{H^+} G_{\gamma 2}(x_i)\right]\nonumber \\
    &+\frac{1}{4\pi \alpha V_{tb} V_{ts}^*} (T_{3l}-2Q_l s_W^2)\sum_i (V^\dagger \rho_u)^{si}(\rho_u^\dagger V)^{ib} G_Z(x_i),\\
    C_{10(l)}^{\rm 2HDM} &=\frac{1}{4\pi \alpha V_{tb} V_{ts}^*} (-T_{3l})\sum_i (V^\dagger \rho_u)^{si}(\rho_u^\dagger V)^{ib} G_Z(x_i),
    \label{C10_general}
\end{align}
where $Q_{u,H^+,l}$ are electric charges of up-type quark, charged Higgs boson and charged lepton
$(Q_{u,H^+,l}=2/3,+1,-1)$, respectively and $T_{3l}=-1/2$, and various functions are given by
\begin{align}
  G_{\sigma 1}(x)&=-\frac{2+3x-6x^2+x^3+6x\log x}{12(1-x)^4},\\
  G_{\sigma 2}(x)&=-\frac{1-6x+3x^2+2x^3-6x^2\log x}{12(1-x)^4},\\  
  G_{\gamma 1}(x)&=-\frac{16-45x+36x^2-7x^3+6(2-3x)\log x}{36 (1-x)^4},\\
  G_{\gamma 2}(x)&=-\frac{2-9x+18x^2-11x^3+6x^3\log x}{36 (1-x)^4},\\
  G_Z(x) &=\frac{x(1-x+\log x)}{2(1-x)^2}.
\end{align}
We note that a term proportional to $G_{\gamma 1,~\gamma2}$ in $C_{9(l)}^{\rm 2HDM}$ originates from
$\gamma$ penguin contribution, on the other hand, terms proportional to $G_Z$
in $C_{9,10(l)}^{\rm 2HDM}$ come from $Z$ penguin contribution. Therefore, $C_{9,10(l)}^{\rm 2HDM}$
are universal to all lepton flavor $l$. We also note that in $G_{\gamma 1}(x)$ there is a log-enhancement
when $x$ is small.

\subsection{$B_s\rightarrow l^+l^-~(l=\mu~{\rm and}~\tau)$}
In a general 2HDM, the Yukawa couplings $\rho_d^{sb(bs)}$ 
induce $B_s \rightarrow l^+ l^-$ process at the tree level, and furthermore non-zero $\rho_u$
Yukawa couplings also generate $B_s \rightarrow l^+ l^-$ process at the one loop level.

The effective operators for $B_s\rightarrow l^+l^-$ ($l=\mu$ and $\tau$) are
parameterized as follows:
\begin{align}
  {\cal L}_{\rm eff}&=\frac{G_{\rm F}^2 m_W^2}{\pi^2}
  \left[ C_{A(l)}^{bs}(\bar{b} \gamma_\mu P_L s)(\bar{l}\gamma^\mu \gamma_5 l)
    +C_{A(l)}^{bs'}(\bar{b} \gamma_\mu P_R s)(\bar{l}\gamma^\mu \gamma_5 l) \right.\nonumber \\
    &\left.+C_{S(l)}^{bs}(\bar{b} P_L s)(\bar{l}  l)
    +C_{S(l)}^{bs'}(\bar{b} P_R s)(\bar{l}  l)+C_{P(l)}^{bs}(\bar{b} P_L s)(\bar{l}\gamma_5  l)
    +C_{P(l)}^{bs'}(\bar{b} P_R s)(\bar{l}\gamma_5  l)
    \right].
\end{align}
The decay rate is given by
\begin{align}
  \Gamma(B_s\rightarrow l^+ l^-) &=\frac{G_{\rm F}^4 m_W^4}{8\pi^5} f_{B_s}^2 m_{B_s} m^2_l
  \sqrt{1-\frac{4m_l^2}{m_{B_s}^2}} \nonumber \\
  &\times \left[
    \left| C_{A(l)}^{bs}-C_{A(l)}^{bs'}+\frac{m_{B_s}^2}{2m_l(m_b+m_s)}
    \left(C_{P(l)}^{bs}-C_{P(l)}^{bs'}\right)\right|^2
      \right.\nonumber \\
      &\hspace{1cm}+\left.\left| \frac{m_{B_s}^2}{2m_l(m_b+m_s)}
      \left(C_{S(l)}^{bs}-C_{S(l)}^{sb'}\right)\right|^2
      \left(1-\frac{4m_l^2}{m_{B_s}^2}\right)\right].
\end{align}
The 2HDM contributions mediated by the neutral Higgs bosons $\phi=h,~H,~A$ at the tree level are
\begin{align}
  \frac{G_{\rm F}^2 m_W^2}{\pi^2} (C_{S(l)}^{bs})^{\rm 2HDM}&=\sum_\phi\frac{y^{d*}_{\phi sb}{\rm Re}(y^e_{\phi ll})}{m_\phi^2},\\
  \frac{G_{\rm F}^2 m_W^2}{\pi^2} (C_{S(l)}^{bs'})^{\rm 2HDM}&=\sum_\phi\frac{y^{d}_{\phi bs}{\rm Re}(y^e_{\phi ll})}{m_\phi^2},\\
  \frac{G_{\rm F}^2 m_W^2}{\pi^2} (C_{P(l)}^{bs})^{\rm 2HDM}&=\sum_\phi\frac{ i y^{d*}_{\phi sb}{\rm Im}(y^e_{\phi ll})}{m_\phi^2},\\
  \frac{G_{\rm F}^2 m_W^2}{\pi^2} (C_{P(l)}^{bs'})^{\rm 2HDM}&=\sum_\phi\frac{ i y^{d}_{\phi bs}{\rm Im}(y^e_{\phi ll})}{m_\phi^2}. 
\end{align}

At the one loop level, the charged Higgs boson via non-zero $\rho_u$ Yukawa couplings can induce
the contribution to $C_{A(l)}^{bs}$ which is the same as the effective operator proportional to
$C_{10(l)}^{\rm 2HDM}$ discussed in $b\rightarrow s l^+l^-$ process [Eq.~(\ref{C10_general})]:
\begin{align}
  \frac{G_{\rm F}^2 m_W^2}{\pi^2}(C_{A(l)}^{bs})^{\rm 2HDM}&=\frac{4G_{\rm F}}{\sqrt{2}}V_{tb}^* V_{ts}\frac{\alpha}{4\pi}C_{10(l)}^{\rm 2HDM *},
\end{align}
which is induced by $Z$ penguin contribution. We note that the expression in Eq.~(\ref{C10_general}) is correct for $l=\tau$
because it is lepton flavor universal.
In addition, in the presence of $\rho_e^{\tau\tau,\mu\tau}$ Yukawa couplings, the box diagram generates the following
contribution to $C^{bs}_{A(\tau)}$:
\begin{align}
  \frac{G_{\rm F}^2 m_W^2}{\pi^2}(C_{A(\tau)}^{bs})^{\rm 2HDM}&=-  
  \frac{(\rho_e^\dagger \rho_e)^{\tau\tau}(V^\dagger \rho_u)^{bi}(\rho_u^\dagger V)^{is}}{128\pi^2 m_{H^+}^2}B(x_i),\nonumber \\
\end{align}
where $x_i=m_{u_i}^2/m_{H^+}^2$ and the function $B(x)$ is defined by
\begin{align}
B(x)=\frac{1-x+x\log x}{(1-x)^2}.
\end{align}

\newpage
\section{Various parameters for our numerical analysis}
Here we summarize numerical values of various parameters we use in our numerical calculation below.
\begin{table}[h]
\begin{center}
  \begin{tabular}{|c|c|c||c|c|c|} \hline
     Quantity & Value & Reference &  Quantity & Value & Reference   \\ \hline \hline
     \multicolumn{3}{|c||}{CKM parameters}   & \multicolumn{3}{|c|}{parameters for hadronic matrix elements}    \\ \hline
     $\lambda$& 0.22506 &\cite{PDG}&$\rho_{D}^2$&1.128&\cite{Amhis2016xyh}\\ 
     A&0.811~&\cite{PDG}&$\rho_{D^*}^2$&1.205&\cite{Amhis2016xyh}\\ 
     $\bar\rho$&0.124&\cite{PDG}&$R_1(1)$&1.404&\cite{Amhis2016xyh}\\ 
     $\bar\eta$&0.356&\cite{PDG}&$R_2(1)$&0.854&\cite{Amhis2016xyh}\\
     \cline{1-3}
     \multicolumn{3}{|c||}{$B$ and $D$ meson parameters}   &$\Delta$&1& \cite{Tanaka2010se}
  \\ \cline{1-3}
     $m_{Bd}$&5.280 [GeV]&\cite{PDG}&$h_{A1}(1)$&0.908&\cite{FFC1}\\ 
     $m_{B^-}$&5.279 [GeV]&\cite{PDG}&$V_{1}(1)$&1.07&\cite{V10}\\ \cline{4-6}
     $m_{Bs}$&5.367 [GeV]&\cite{PDG}& \multicolumn{3}{|c|}{SM particle masses and $G_{\rm F}$}\\ \cline{4-6}
     $M_{Bc}$&6.275 [GeV]&\cite{PDG}&$m_\mu$&0.105676 [GeV] &\\ 
     $m_{D}$&1.865 [GeV]&\cite{PDG}&$m_\tau$&1.77686 [GeV]&\\
     $m_{D^*}$&2.007 [GeV]&\cite{PDG}&$m_c(m_c)$&1.27 [GeV]&\\
     $\tau_{Bd}$&$\hspace{10pt}2.309\times10^{12}$ [GeV$^{-1}$]$\hspace{10pt}$&\cite{PDG} &  $m_t$&173.21 [GeV]&\\
     $f_{Bd}\sqrt{B_{Bd}}$ &227.7 [MeV]&\cite{MILC} &      $m_d(2{\rm GeV})$&0.0047 [GeV]&\\
     $\tau_{B^-}$&$\hspace{10pt}2.489\times10^{12}$ [GeV$^{-1}$]$\hspace{10pt}$&\cite{PDG} &   $m_s(2{\rm GeV})$&0.096 [GeV]&\cite{PDG}\\
     $f_{B^-}$ &186 [MeV]&\cite{HPQCD1} &     $m_b(m_b)$&4.18 [GeV]&\\
      $\tau_{Bs}$& $\hspace{10pt}$$2.294\times10^{12}$ [GeV$^{-1}$]$\hspace{10pt}$&\cite{PDG} &      $m_W$&80.385 [GeV]&      \\
     $f_{Bs}\sqrt{B_{Bs}}$ &274.6 [MeV] &\cite{MILC} &      $m_Z$&91.188 [GeV]&\\
     $\tau_{Bc}$&$\hspace{10pt}$$7.703\times10^{11}$ [GeV$^{-1}$]$\hspace{10pt}$&\cite{PDG} & $m_h$&125.09 [GeV]&\\
     $f_{Bc}$&0.434 [GeV]&\cite{HPQCD2} & $G_F$&$\hspace{5pt}1.166\times 10^{-5}$ [GeV$^{-2}]\hspace{5pt}$&\\
     \hline
    \end{tabular}
   \end{center}
\end{table}

\end{document}